\documentclass[a4paper,11pt]{article}
\pdfoutput=1 

\usepackage{jheppub} 

\usepackage[T1]{fontenc} 

\usepackage{mathrsfs}
\usepackage{amsmath}
\usepackage{amsfonts}
\usepackage{amssymb}
\usepackage{MnSymbol}
\usepackage[dvipsnames]{xcolor}
\usepackage{ulem}
\usepackage{marginnote}

\usepackage{slashed}

\notoc

\newcommand{\D}{\mathrm{d}}
\def\vec{\mathbf}

\newcommand{\rom}[1]{\uppercase\expandafter{\romannumeral #1\relax}}

\title{\boldmath Functional methods for false-vacuum decay in real time}

\author[]{Wen-Yuan Ai,}
\author[]{Bj\"orn~Garbrecht,}
\author[]{Carlos Tamarit}

\affiliation[]{Physik Department T70, James-Franck-Stra{\ss}e,\\
Technische Universit\"at M\"unchen, 85748 Garching, Germany}

\emailAdd{ai.wenyuan@tum.de}
\emailAdd{garbrecht@tum.de}
\emailAdd{carlos.tamarit@tum.de}

\makeatletter
\gdef\@fpheader{
}
\makeatother

\abstract{We present the calculation of the Feynman path integral in real time for tunneling in quantum mechanics and field theory, including the first quantum corrections. For this purpose, we use the well-known fact that Euclidean saddle points in terms of real fields can be analytically continued to complex saddles of the action in Minkowski space. We also use Picard-Lefschetz theory in order to determine the middle-dimensional steepest-descent surface in the complex field space, constructed from Lefschetz thimbles, on which the path integral is to be performed. As an alternative to extracting the decay rate from the imaginary part of the ground-state energy of the false vacuum, we use the optical theorem in order to derive it from the real-time amplitude for forward scattering. While this amplitude may in principle be obtained by analytic continuation of its Euclidean counterpart, we work out in detail how it can be computed to one-loop order at the level of the path integral, i.e. evaluating the Gau{\ss}ian integrals of fluctuations about the relevant complex saddle points. To that effect, we show how the eigenvalues and eigenfunctions on a thimble can be obtained by analytic continuation of the Euclidean eigensystem, and we determine the path-integral measure on thimbles. This way, using real-time methods, we recover the one-loop result by Callan and Coleman for the decay rate.  We finally demonstrate our real-time methods explicitly, including the construction of the eigensystem of the complex saddle, on the archetypical example of tunneling in a quasi-degenerate quartic potential.}

\begin{document} 
\begin{flushright}                          
{TUM-HEP-1201-19}
\end{flushright}
\vskip-1.2cm

\maketitle
\flushbottom

\newpage

\hrule
\tableofcontents
\vskip.5cm
\hrule

\newpage

\normalem

\section{Introduction}
\label{sec:intro}

Tunneling is one of the signature phenomena of quantum theory. The most prominent example realized
in nuclear physics is alpha decay, but there are also important technical applications such as the tunneling microscope. Vacuum transitions~\cite{Kobzarev:1974cp,Coleman:1977py,Callan:1977pt}
through tunneling play an important role in particle physics models
and for their cosmological implications. Metastable vacua can decay through the nucleation of classical, expanding bubbles, and gravitational waves are produced in their collisions~\cite{Witten:1984rs,Kosowsky:1991ua,
Caprini:2009fx,Caprini:2015zlo}. There are close analogies
between vacuum tunneling and first-order phase transitions at finite
temperature~\cite{Langer:1967ax,Langer:1969bc,Affleck:1980ac,Linde:1980tt,Linde:1981zj}. In extensions of the Standard Model where electroweak symmetry breaking in the early universe occurs through such a first-order transition, bubbles may turn out to be pivotal for generating the cosmic matter-antimatter asymmetry~\cite{Kuzmin:1985mm,Shaposhnikov:1987tw,Morrissey:2012db} (for a recent review, see Ref.~\cite{Garbrecht:2018mrp}).

\begin{figure}[!]
  \centering
  \hspace{10pt}
  \includegraphics[scale=0.6]{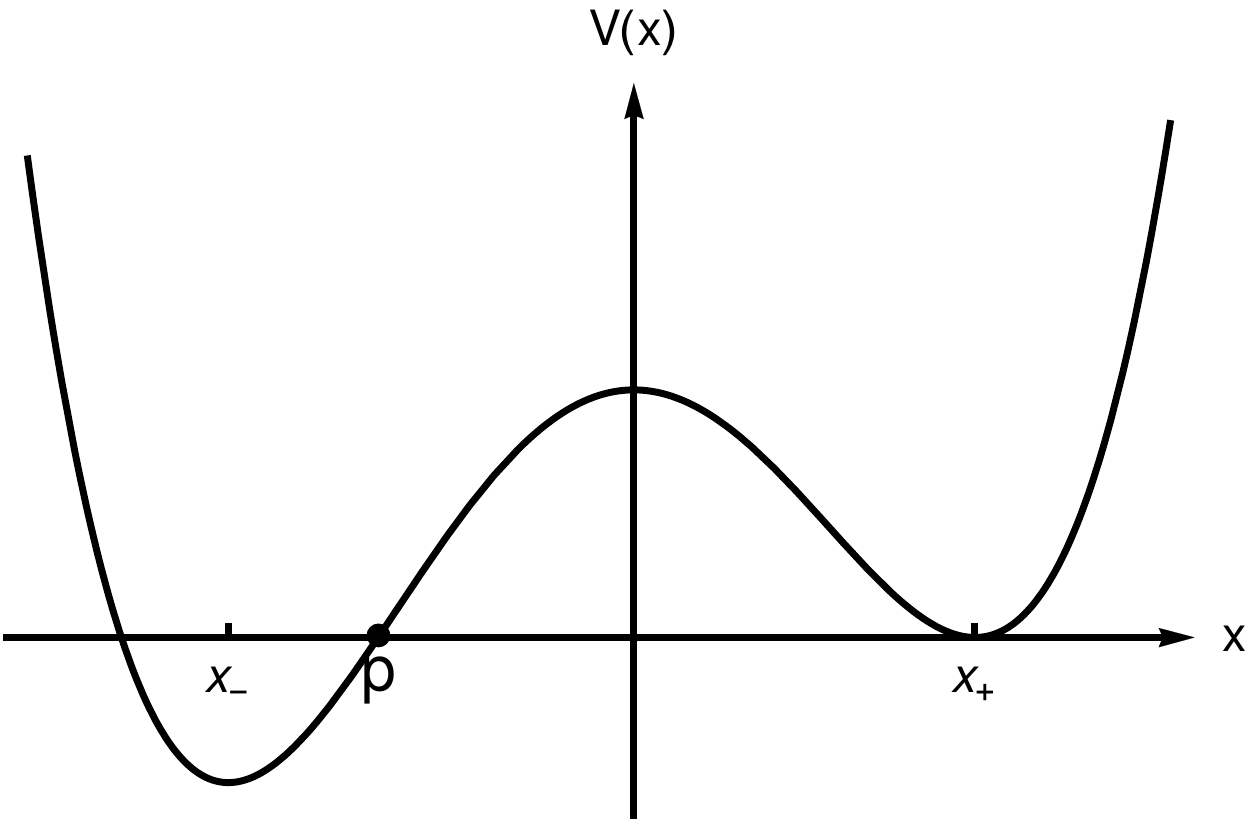}
  \caption{The classical  potential $V(x)$ in a theory with a false vacuum.
  \label{fig:potential}}
\end{figure}

To recall the basic theoretical aspects, we note that quantum-mechanical tunneling occurs in potentials of the form shown in Figure~\ref{fig:potential}. We assume that a particle initially occupies the ground state around the local minimum at $x_+$, which we refer to as the false vacuum (in view of the generalization to field theory, while this term may not be the most fitting choice in quantum mechanics). It is separated from the true vacuum at $x_-$ by a local maximum which we assume here without loss of generality to be located at $x=0$. Through quantum tunneling, the particle can hop from the false vacuum over the barrier to the region around the true vacuum $x_-$, which is energetically forbidden in classical physics. We also indicate the escape point $\mathbf{p}$ beyond which the motion of the particle can be described by a classically allowed trajectory. The theoretical description of vacuum transitions had first been given in the context of statistical physics at finite temperature~\cite{Langer:1967ax,Langer:1969bc}, which was extended to zero-temperature quantum field theory in Refs.~\cite{Kobzarev:1974cp,Coleman:1977py,Callan:1977pt} (see also Ref.~\citep{Coleman:1988}).

Following Callan and Coleman, the tunneling rate can be calculated from a Euclidean transition amplitude~\cite{Coleman:1977py,Callan:1977pt}. We review here the basic picture, while in Appendix~\ref{sec:Callan-Coleman}, a detailed review of the evaluation of the Euclidean path integral is provided.
We begin with the following Euclidean transition amplitude:
\begin{align}
\label{heatkernal}
Z^E[\mathcal{T}]\equiv\langle x_f|e^{-H\mathcal{T}}|x_i\rangle=\int\mathcal{D}x\, e^{-S_E[x]}\,,
\end{align}
where $\mathcal{T}$ and $S_E$ are the Euclidean time and Euclidean action, $x_i$ and $x_f$ are the initial and final positions, respectively. The expression in the middle can be expanded using a complete set of energy eigenstates:
\begin{align}
\label{relation}
\langle x_f|e^{-H\mathcal{T}}|x_i\rangle=\sum_n e^{-E_n\mathcal{T}}\langle x_f|n\rangle\langle n|x_i\rangle\,.
\end{align}
For large $\mathcal{T}$, the lowest-lying state dominates and hence the above Euclidean transition amplitude contains the information of the lowest-energy and its wave function. In the case of quantum tunneling, $x_i$ and $x_f$ are chosen to be the metastable minimum $x_+$.

One can evaluate Eq.~\eqref{heatkernal} through the method of steepest descent, on which the technical details are presented in Appendix~\ref{sec:Callan-Coleman}. We first need to find out all the stationary points. Note that in the Euclidean equations of motion the potential appears flipped upside down. This allows for a solution starting at $x_+$ in the infinite past $\tau\to-\infty$, reaching the turning point $\mathbf p$ at some time $\tau_0$, eventually bouncing back to $x_+$ for $\tau\to\infty$. The soliton thus obtained is called the bounce which,  among the stationary points, is of particular importance for tunneling. In order to relate this solution to the decay rate, one needs to analyze the fluctuations about the bounce~\cite{Callan:1977pt}.
In particular, there is one mode with a negative eigenvalue because of the metastability of the false vacuum as well as a Goldstone zero mode because of the spontaneous breakdown of time-translation invariance. 
In quantum field theory, tunneling proceeds via the nucleation of bubbles containing the true vacuum within the false-vacuum phase. Since the bounce of least action is hyperspherically symmetric, the particular dynamics depends only on the hyperradial coordinate. The discussion therefore proceeds in analogy with the one for quantum mechanics, up to the effect of the additional hyperspherical excitations about the bounce or the bubble.

Next, we need to integrate the fluctuations about the stationary points. Since the fluctuation operator (the generalization of the Hessian matrix) evaluated at the bounce contains a negative eigenvalue, it is shown in Ref.~\cite{Callan:1977pt} that performing the Gau{\ss}ian functional integral around the bounce (as well as multi-bounce stationary points) leads to an imaginary part in the Euclidean transition amplitude~\eqref{heatkernal}. This then implies an imaginary part also for the ground-state energy via Eq.~\eqref{relation}, which can be interpreted as the decay rate. However, Eq.~\eqref{heatkernal} is apparently real (and so is its quantum field theoretical generalization with a path integral defined in terms of real fields). When carrying out the steepest descent evaluation of the path integral
thoroughly, constructing the integration contours using Picard-Lefschetz theory~\cite{Pham:83,Berry:91Hy,Witten:2010cx,Witten:2010zr}, one finds canceling imaginary contributions from two different steepest-descent contours passing through the bounces,  which connect them to the false-vacuum solution in one case and a new stationary point  called shot in the other~\cite{Andreassen:2016cvx} (see Appendix~\ref{sec:Callan-Coleman}). An imaginary part can only arise when restricting the fluctuations 
to the steepest-descent contour that passes through the false vacuum and the bounce but not the shot, which is analogous to imposing boundary conditions connected with the false vacuum.\footnote{The importance of boundary conditions connected with the false vacuum, and how this leads to complex false-vacuum effective actions, was also pointed out in Ref.~\cite{Plascencia:2015pga}.} The extraction of the imaginary part as carried out in Ref.~\cite{Callan:1977pt} proceeds along somewhat different lines, considering deformations of the potential, which leads to the same integration contour in the vicinity of the bounce stationary point as Picard-Lefschetz theory does. Notably, the correct contour leads to a factor of $1/2$ in front of the decay rate, different from what one would expect from the na\"ive steepest descent evaluation of the fluctuation integral about the bounce.
The construction of the integration contour from the flow equations of Picard-Lefschetz theory (which define the relevant steepest-descent directions) may appear somewhat simpler than the argument based on the analytic continuation to another theory with a deformed potential.
Moreover, in Section~\ref{sec:MinkowskiPathIntegral}, we show that one can use the flow equations to compute the path integral for arbitrary complex times interpolating between the Minkowski and the Euclidean cases.

Apart from the above issues, computing the eigenvalue of the Hamiltonian for the ground state of the false vacuum using the Euclidean path integral does not shed light on how tunneling proceeds in the real-time formulation of the path integral. For quantum-mechanical cases, one way to recognize the role of instantons in real-time quantum tunneling is to compare their form with that of the solution to the static Schr\"{o}dinger Equation that can be obtained in the Wentzel-Kramers-Brillouin (WKB) expansion, cf. Appendix~\ref{app:sec:wkb}. The connection between Euclidean instantons and tunneling in the WKB approximation has been further studied in quantum mechanics and field theory in Refs.~\cite{Gervais:1977nv,Bitar:1977wy}. This, however, still tells us little about the real-time picture of the tunneling process
in a functional approach. Therefore, it would be interesting
to relate the decay rate to amplitudes that are calculated in real time, i.e. in Minkowski space:
\begin{align}
Z^M[T]\equiv\langle x_f|e^{-iHT}|x_i\rangle=\int \mathcal{D}x\: e^{iS_M[x]}.
\end{align}
The obstacle is, of course, that the Euclidean instantons do in general not have a correspondence in {\it real} configurations in Minkowski spacetime. And thus we generally have no real classical solutions that could dominate the quantum tunneling process. It is therefore very difficult to evaluate the Minkowski transition amplitude.  
Recently, however, it has been understood that one can analytically continue the path integral over real paths to one over complex paths when applying Picard-Lefschetz theory~\cite{Pham:83,Berry:91Hy,Witten:2010cx,Witten:2010zr}. Therefore, we may attempt to find a deformed but equivalent integration contour that contains {\it complex} stationary points which can be identified as Minkowski correspondences of the Euclidean instantons. Then the expansion of the Minkowski path integral around these complex saddle points\footnote{When the paths are complexified, all the stationary points are saddle points due to the complex structure.} will give the dominant contributions to the Minkowski transition amplitude and should generate the same results as those obtained from an expansion around instantons in the Euclidean path integral. Based on these new developments, some progress in understanding quantum tunneling in the real-time formalism has been reported in Refs.~\cite{Cherman:2014sba,Tanizaki:2014xba}. However, an  understanding of how to carry out the integration along the complexified field paths in Minkowski spacetime has yet been missing. It is thus the aim of this paper to develop the formalism to accomplish this task. In this respect, it should be noted that, when using Picard-Lefschetz theory to construct integration contours defined by flow equations, there is no straightforward analytic continuation relating the Euclidean contours to the Minkowski ones, because the flow equations are not holomorphic. For this reason, performing the path integral in Minkowski space-time is a nontrivial endeavour.

Complex saddle points now have become a very useful concept in the study of 
series expansions around the perturbative vacuum~\cite{Basar:2013eka}. Even when complex saddles are not on the integration contour, they could still encode very important information about physical observables as a consequence of resurgence~\cite{Ecalle,Delabaere,Costin}. Resurgence theory states that the
expansion around the perturbative vacuum encodes the information of all nonperturbative saddles. For additional applications of Picard-Lefschetz theory in quantum field theories and quantum mechanics, see Refs.~\cite{Cristoforetti:2012su,Cristoforetti:2013wha,Alexandru:2016gsd,Serone:2017nmd,Alexandru:2017lqr,Mou:2019tck,Mou:2019gyl,Fukushima:2019iiq} and references therein. In this work, we shall discuss complex saddles which lie on the (deformed) integration contour and directly describe the nonperturbative phenomenon of quantum tunneling. The idea that such complex saddles may recover the results obtained from the instanton techniques was suggested in Ref.~\cite{Cherman:2014sba} for the double-well model. Here, we work out this proposition in a much more detailed and concrete way for the general case, including the integration of fluctuations around the complex saddles. In particular, we transfer the original problem of solving the gradient flow equations which define the deformed integration contour---constructed in terms of steepest-descent surfaces attached to saddle points, or Lefschetz thimbles---
to one of solving proper eigenequations; this allows us to successfully carry out the path integral on the integration contour which passes through the relevant complex saddle points. Using the real-time transition amplitude that we have derived in this way, we shall further show that, under plausible assumptions, the particle tunneling rate and the false-vacuum decay rate can be derived from an optical theorem for tunneling based on the unitarity of the evolution operator. Apart from the potential applications of the techniques that we have developed in order to perform path integrals on Lefschetz thimbles associated with complex saddles, the derivation of the decay rate in the real-time formalism could also shed new insights on
the dynamics of the vacuum transition. Obtaining the decay rate from an optical theorem for metastable vacua suggests that after the decay, there is in principle a superposition of all possible nucleated configurations rather than a unique classical critical bubble. We leave the possible consequences of this picture and the derivation of the quantum state after tunneling to future work.  

The organization of this paper is as follows. We begin our discussion with the construction of an optical theorem for false-vacuum decay in Section~\ref{sec:optical-theorem}. Thus, the decay rate of the metastable vacuum can be related to the real-time false-vacuum to false-vacuum transition amplitude (that we also refer to as forward scattering amplitude, in reference to particle collisions, that the optical theorem is typically applied to), which we will compute from the Minkowski path integral in the subsequent parts of this work. In Section~\ref{sec:MinkowskiPathIntegral}, we apply Picard-Lefschetz theory to the Minkowski path integral, and we discuss a particularly important complex saddle point---the complex bounce. We transfer the problem of solving the gradient flow equations related to the complex bounce to an eigenproblem in the proper sense for the Minkowski fluctuation operator evaluated at the complex bounce. The  determinant of this operator enters into the formula for the decay rate. Based on our expression for the functional determinant, we prove in Section~\ref{sec:continuation} that it is indeed related to its Euclidean counterpart by an inverse Wick rotation. In particular, this implies that it picks up an extra factor of $i$ due to the integration
over the collective coordinate pertaining to time-translation invariance.
Our proof is based on the explicit continuation of the Euclidean eigensystem
of the quadratic fluctuation operator to the Minkowski case, where discrete
modes and the continuum spectrum have to be distinguished. Furthermore,
when calculating the logarithmic determinant, contributions that are
finite have to be separated from those that are proportional to the volume
of spacetime in order to establish the correct behaviour under analytic continuation. As concrete examples, we discuss in
Section~\ref{sec:continuation:examples} a trivial vacuum state
and the archetypical scenario of tunneling between quasi-degenerate vacua in a quartic potential.
Given the analytic continuation, we finally recover the Callan-Coleman result for the decay rate from the real-time amplitude and the optical theorem in Section~\ref{sec:Minkowskian:decay:path:integral}, and we conclude this
paper in Section~\ref{sec:conclusions}. We collect several technical details in the appendices. In Appendix~\ref{sec:Callan-Coleman}, the Gau{\ss}ian approximation to the Euclidean path integral using Picard-Lefschetz theory is reviewed, and in Appendix~\ref{app:methods:determinant},
we summarize various methods of calculating the one-loop functional determinant.
These results can be compared with the decay rate inferred in Appendix~\ref{app:sec:wkb} using the WKB approximation from the imaginary part of the zero-point energy of the
false vacuum, or, more directly, from the probability current
that flows toward the global ground state. This way, we provide a
comprehensive survey of the computation of the first quantum corrections
to tunneling, to which we can relate our results from the functional
approach in real time.
Throughout this article (save Appendix~\ref{app:sec:wkb}) we set $\hbar=c=1$.

\section{Optical theorem for the decay of the false vacuum}

\label{sec:optical-theorem}

In the approach by Callan and Coleman, the decay rate of the false vacuum is attributed to a complex energy. In  Appendix~\ref{sec:Callan-Coleman},
Picard-Lefschetz theory is used in order to explain how the imaginary part can emerge from a purely real Euclidean path integral. This analysis, however, while successfully predicting the decay rate and the evolution of the emerging classical bubbles, does not tell us much about the dynamical picture of tunneling in Minkowski spacetime. Therefore, we aim to formulate false-vacuum decay based on the Minkowski path integral.  The computation of the tunneling amplitude in Minkowski space  has been developed in Refs.~\cite{Turok:2013dfa,Cherman:2014sba,Tanizaki:2014xba}. In the present paper, we go beyond the previous works by calculating the determinant of fluctuations around the complex saddle points, and also by relating these real-time results more directly to the decay rate of the false vacuum. In order to achieve the latter, the rate shall here be obtained from the unitarity of the evolution operator ${\cal U}$. To prepare for that, we first recall the optical theorem in quantum field theory. The latter is most commonly used in the derivation of decay rates and cross sections for perturbative reactions in quantum field theory, while we aim here for an application to processes based on nonperturbative, solitonic solutions.

\subsection{The optical theorem in scattering theory}
The  optical theorem in scattering theory relies on the unitarity of the $S$-matrix, $S^\dagger S=\mathbf{1}$. Inserting $S=\mathbf{1}+iM$ to $S^\dagger S=\mathbf{1}$, we have\footnote{To avoid mix up of notations, we use $M$ instead of $T$ to denote the so-called ``$T$-matrix'' because we reserve $T$ for the real-time period in the amplitudes.}
\begin{align}
\label{unitarity}
-i(M-M^\dagger)=M^\dagger M.
\end{align}
We can take the matrix element of this equation between particle states, say $|{\bf p}_1{\bf p}_2\rangle$ and $|{\bf k}_1{\bf k}_2\rangle$ for a two-by-two scattering for concreteness and simplicity. To evaluate the right-hand side, we insert a complete and normalized set of intermediate states $\{{\bf q}_n\}$:
\begin{align}
\langle {\bf p}_1{\bf p}_2|M^\dagger M|{\bf k}_1{\bf k}_2\rangle=\displaystyle\sum_{n}\langle {\bf p}_1{\bf p}_2|M^\dagger |\{{\bf q}_n\}\rangle \langle \{{\bf q}_n\}|M|{\bf k}_1{\bf k}_2\rangle.
\end{align}
Thus, Eq.~\eqref{unitarity} yields
\begin{align}
-i\left[\langle {\bf p}_1{\bf p}_2|M|{\bf k}_1{\bf k}_2\rangle-\langle {\bf p}_1{\bf p}_2|M^\dagger|{\bf k}_1{\bf k}_2\rangle\right]=\displaystyle\sum_{n}\langle {\bf p}_1{\bf p}_2|M^\dagger |\{{\bf q}_n\}\rangle \langle \{{\bf q}_n\}|M|{\bf k}_1{\bf k}_2\rangle.
\end{align}
Further, letting the initial and final states be the same, i.e. taking ${\bf p}_i={\bf k}_i$, we obtain
\begin{align}
\label{opticaltheorem}
-i\left[\langle {\bf k}_1{\bf k}_2|M|{\bf k}_1{\bf k}_2\rangle-\langle {\bf k}_1{\bf k}_2|M^\dagger|{\bf k}_1{\bf k}_2\rangle\right]=\displaystyle\sum_{n}\langle {\bf k}_1{\bf k}_2|M^\dagger |\{{\bf q}_n\}\rangle \langle \{{\bf q}_n\}|M|{\bf k}_1{\bf k}_2\rangle.
\end{align}
Therefore the imaginary part of the $M$-matrix corresponds to the decay probability of the initial state into all possible intermediate states.

\subsection{Optical theorem for false-vacuum decay}

A crucial ingredient to the optical theorem as discussed above is the unitarity of the $S$-matrix. For the case of vacuum decay, we consider a finite (but still large) time interval $T$ (i.e. $[-T/2,T/2]$). Further, the unstable false-vacuum state is not a true asymptotic state of the free theory. In place of the $S$-matrix\footnote{Note that defining the $S$-matrix in terms of its action on free states rather than scattering states requires in particular a different treatment involving M{\o}ller operators.}, we therefore need to use the unitary time-evolution operator ${\cal U}(T) $ in order to compute amplitudes of the form
\begin{align}
\label{eq:Finout}
\langle {\rm F}|{\cal U}(T) |{\rm I}\rangle,\quad\textnormal{where}\;\quad {\cal U}(T) =e^{-i H T}
\end{align}
$H$ is a Hermitian Hamiltonian. Since ${\cal U}(T)$ is still unitary, the above argument leading to the optical theorem~\eqref{opticaltheorem} can still be applied.

Based on this, we now construct an optical theorem for false-vacuum decay.
For this purpose, we need to specify a false-vacuum state. While the true-vacuum state is stationary with a node-free wave-function, that is the eigenfunction for the lowest eigenvalue
of the Hamiltonian, it is less straightforward to exactly specify the false vacuum. Nonetheless,
the existence of the false-vacuum state is implied in the approach by Callan and Coleman
to vacuum decay~\cite{Callan:1977pt} because the pertaining complex eigenvalue is found.

For the present purpose, we therefore proceed with the approximate description of the false-vacuum state $|{\rm FV}\rangle$
through a wave function which is ground-state-like and node-free in the region of the potential well around $x_+$ (see Fig.~\ref{fig:potential}). Further, rather than being  stationary, it should feature a time-dependent amplitude due to the probability current leaking into the region beyond the potential barrier in conjunction with the total conservation of probability. This setup describes the configuration of interest,
a particle that approximately resides in a local ground state at $x_+$, that is
a metastable configuration however.
While we believe that these assumptions are plausible to this end,
in Section~\ref{sec:bc} and in Appendix~\ref{app:sec:wkb}, we show that a
state of this form indeed exists. 
Of course, the false-vacuum state $|{\rm FV}\rangle$ should be considered as an unstable resonant state. Corresponding states appear in the evaluation of matrix
elements in scattering theory, where unstable external particles are put on the mass
shell, a procedure that is applied e.g. in the computation of production cross
sections for unstable particles, of their decay rate or in the case of the
narrow-width approximation to processes with intermediate states on the mass shell, cf. e.g.
the discussion of production cross sections for the Higgs boson in Ref.~\cite{Dittmaier:2011ti}.

We then consider the following element of ${\cal U}(T)$:
\begin{align}
\label{amplitude}
\langle \text{FV}|{\cal U}(T) |\text{FV}\rangle=\langle {\rm FV}|e^{-iHT}|{\rm FV}\rangle.
\end{align} 
Inserting ${\cal U}(T) =\exp(i\sigma)\mathbf{1}+iM(T)$ into ${\cal U}(T) ^\dagger {\cal U}(T) =\mathbf{1}$ for the matrix element above, we have
\begin{align}
\label{opticaltheoremforfalsevacuum}
-i\left[\langle\text{FV}|e^{-i\sigma} M(T)|\text{FV}\rangle-\langle\text{FV}|e^{i\sigma}M(T)^\dagger|\text{FV}\rangle\right]=\displaystyle\sum_n\langle\text{FV}|M(T)^\dagger|\{{\bf q}_n\}\rangle \langle \{{\bf q}_n\}|M(T)|\text{FV}\rangle,
\end{align}
where the $|\{{\bf q}_n\}\rangle$ are a complete set of operators in the  Heisenberg picture, e.g. eigenstates of position operators 
in the case of quantum-mechanical particle tunneling, or analogous eigenstates of field operators in quantum field theory. The phase $\exp(i\sigma)$ is unobservable and related
to the normalization of the energy of the false-vacuum state. It will be identified
more specifically in Section~\ref{sec:continuation:determinant}.
The left-hand side of Eq.~\eqref{opticaltheoremforfalsevacuum} is simply the imaginary part of the amplitude,
\begin{align}
\label{opticaltheoremforfalsevacuum2}
2\,\text{Im}\langle\text{FV}|e^{-i\sigma} M(T)|\text{FV}\rangle=\displaystyle\sum_n\langle\text{FV}|M(T)^\dagger|\{{\bf q}_n\}\rangle \langle \{{\bf q}_n\}|M(T)|\text{FV}\rangle.
\end{align}
When dividing by the normalization $\langle {\rm FV}|{\rm FV}\rangle$,
the right-hand side is the total probability for the false vacuum to decay into arbitrary states
within the time $T$.

\subsection{Boundary conditions on the path integral, Lefschetz thimbles and the false-vacuum state}
\label{sec:bc}

We are close to our goal when we can find out the imaginary part in Eq.~\eqref{opticaltheoremforfalsevacuum2}. As will be motivated next, computing $\langle\text{FV}|M(T)|\text{FV}\rangle$ with path-integral methods requires implementing appropriate constraints in order to pick the false-vacuum state rather than the true vacuum.

We may start by recalling that, in their seminal work, Callan and Coleman set to extract the false-vacuum decay rate from the transition amplitude between eigenstates $|x_+\rangle$ of the position operators in  the Schr\"odinger picture, with eigenvalues given by the location  $x_+$ of the false vacuum. Using the spectral resolution of the identity in terms of projectors onto eigenstates $|n\rangle$ of the Hamiltonian with eigenvalues $E_n$, one can write
\begin{align}
  \left\langle x_+\Big|e^{-iHT}\Big|x_+\right\rangle=\sum_n e^{-i E_nT}\left|\langle x_+|n\rangle\right|^2.
\end{align}
When a real value of $T$ is approached from the lower complex half-plane (as corresponds to rotating from Euclidean to Minkowski time), one can make the replacement $T\rightarrow T(1-i\epsilon)$, which leads to
\begin{align}
\label{eq:xptransition}
 \lim_{T\rightarrow\infty} \left\langle x_+\Big|e^{-iHT(1-i\epsilon)}\Big|x_+\right\rangle= e^{-i E_0T(1-i\epsilon)}\left|\langle x_+|0\rangle\right|^2,
\end{align}
where $E_0$ is the eigenvalue with the lowest real part.

If $|0\rangle$ could be identified with the false-vacuum state, one would expect its energy $E_0$ to be complex, which would allow one to extract the decay rate as $\varGamma=2\,|{\rm Im}E_0|$.
However, as noted by Callan and Coleman (working directly in Euclidean space),
the complex energy of an unstable state cannot be an eigenvalue  corresponding to a finite-norm eigenstate of the Hamiltonian. In fact, a finite-norm unstable state cannot be an eigenstate and thus cannot have a well-defined energy; considering a complex-energy eigenvalue implies an approximation in which the false vacuum is treated as a non-normalizable state.
Nonetheless, Callan and Coleman show that a deformation of the contour of the path integration leads to a finite result for the transition amplitude which can be understood as arising from a complex value of $E_0$ in the right-hand side of Eq.~\eqref{eq:xptransition}.
From this, they obtain a nonzero decay rate, with the result matching the usual quantum-mechanical estimates of tunneling probabilities.
There seems to be a contradiction because the false-vacuum state is not a normalizable energy
eigenstate and thus does not take part in the spectral resolution of the identity in terms
of the finite-norm states $|n\rangle$. One would need another choice of a complete basis including the false
vacuum. If the latter is treated as having finite norm and thus not  being a true eigenstate, it would mix with other states under the action of
the Hamiltonian. A similar problem would be expected when approximating $|{\rm FV}\rangle$ as a nonnormalizable eigenstate with a complex energy (when acting on infinite-norm states, the Hamiltonian has not the usual Hermiticity properties). In any case, even if these overlaps could be neglected, in the $T\rightarrow\infty$
 limit one would
always pick the contribution from the lowest eigenvalue---the true vacuum---as opposed to
the false vacuum. 
In summary, one then expects the transition amplitude of Eq.~\eqref{eq:xptransition} to be proportional to the true-vacuum transition amplitude:
\begin{align}
\label{ampitude:T:TV}
 \lim_{T\rightarrow\infty} \left\langle x_+\Big|e^{-iHT(1-i\epsilon)}\Big|x_+\right\rangle\sim \lim_{T\rightarrow\infty} \left\langle {\rm TV}\Big|e^{-iHT(1-i\epsilon)}\Big|{\rm TV}\right\rangle.
\end{align}

This matter has been understood recently through the application of
Picard-Lefschetz theory to approximations of the path integral~\cite{Andreassen:2016cvx}, as we review in Appendix~\ref{sec:Callan-Coleman}. In essence, Callan and Coleman's calculation captures the integration over a subset of the field fluctuations of the path integral, which are those that remain close to the false-vacuum configuration at almost all times (i.e. for $T\to\infty$ all but finite times). When restricting to these local fluctuations, one expects that the lowest energy state accessible to the system is the false vacuum so that the constrained path integral will be related to the false-vacuum transition-amplitude rather than that of the true vacuum. The restriction to a subset of field fluctuations can be made more precise with Picard-Lefschetz theory, in which the integration contour of the path integral is deformed into a sum of complex steepest-descent paths constructed from downward flows from saddle points, or Lefschetz thimbles.  Each downward flow can be interpreted as describing the local dynamics about a saddle, while summing over all the contours should capture the full path integral and hence the dynamics of the true vacuum. By restricting the path integral to certain combinations of Lefschetz thimbles (or subsets thereof) connected with the false vacuum~\cite{Andreassen:2016cvx},
contributions from the global extremum of the action at the true vacuum are excluded.
In particular, it is thus avoided that these dominate the amplitude as
in Eq.~\eqref{ampitude:T:TV}, so that we may write
\begin{align}
\label{amplitudexx}
&\langle {\rm FV}|e^{-iHT}|{\rm FV}\rangle=\mathcal{N}^2\langle x_+|e^{-iHT(1-i\epsilon)}|x_+\rangle_{\rm c.L.t.}\ {\rm\ for\ large\ }T,
\end{align}
where we indicate that the path integral is constrained to be evaluated on a set of Lefschetz thimbles by the subscript ${\rm c.L.t.}$
We will specify the particular choice the Lefschetz thimbles that isolates the relevant contributions to false-vacuum decay in Section~\ref{sec:complexified:path:integral}. Once the false vacuum to false vacuum transition amplitude~\eqref{amplitudexx} is computed, we can use it to obtain the decay rate via the optical-theorem relation~\eqref{opticaltheoremforfalsevacuum2} as we will show in Section~\ref{sec:Minkowskian:decay:path:integral}. That derivation of the decay rate does not make use of the interpretation of the false-vacuum state as an eigenstate with complex energy.
On the other hand, we can still follow Callan and Coleman and relate the
amplitude to the decay rate as
\begin{align}
\label{eq:GammaCC}
 \varGamma= -2\,{\rm Im}E_0 =-2\, {\rm Im}\,\left( \lim_{T\rightarrow\infty} \frac{i}{T(1-i\epsilon)}\log \left\langle x_+\Big|e^{-iHT(1-i\epsilon)}\Big|x_+\right\rangle_{\rm c.L.t.}\right),
\end{align}
or we may employ the Euclidean version of this relation as in their original work.
In the remainder of this paper, we drop the subscript ${\rm c.L.t.}$ and imply that
all amplitudes are evaluated on thimbles constituting contours of path integration that contain the false vacuum but omit the true one.

In principle, the selection of contributions from certain saddle points only can be understood  in terms of boundary conditions for the path integral. In the context of effective actions---of relevance to tunneling because   
the false-vacuum transition-amplitude can be connected to an effective action evaluated at an extremum~\cite{Garbrecht:2015cla,Garbrecht:2015oea,Plascencia:2015pga}---these issues have been discussed in Refs.~\cite{Weinberg:1987vp,Plascencia:2015pga}. To understand the role of boundary conditions, we may go back to the false-vacuum transition-amplitude and consider the spectral resolution of the identity in terms of  eigenstates $|x,t\rangle$ of the Heisenberg-picture position operators:
\begin{align}\label{eq:FVtoFV}\begin{aligned}
 \langle {\rm FV}|e^{-iHT}|{\rm FV}\rangle&=\int \D x\, \D y \, \left\langle {\rm FV}\bigg|x,\frac{T}{2}\right\rangle\left\langle x,\frac{T}{2}\bigg|e^{-iHT}\bigg|y,-\frac{T}{2}\right\rangle\left\langle y,-\frac{T}{2}\bigg|{\rm FV}\right\rangle
 \approx\,{\cal N}^2 \left\langle x_+|e^{-iHT}|x_+\right\rangle.
\end{aligned}\end{align}
In this equation, we have  used the fact that the false-vacuum wave-functions $\langle x,t|{\rm FV}\rangle$ are dominated by their contribution near $x=x_+$, independently of time as they correspond to a quasi-stationary state: We assume $T$ to be large, but below the time in which the probability density around $x_+$ no longer dominates over its value close to the true vacuum. 
The right-hand side of Eq.~\eqref{eq:FVtoFV} is equivalent to a path integral with boundary conditions fixed by the false vacuum localized at $x_+$, demanding  that $x(t)$ should approach the constant value  $x_+$ for $|t|=T/2$. The fact that such boundary values are to be considered independently of $T$ further implies the boundary condition $\dot x\rightarrow0$. The added requirement can be understood as the reason behind the restriction to particular Lefschetz thimbles in Eq.~\eqref{amplitudexx}. As will be discussed in Section~\ref{sec:complex:saddles}, the discarded steepest descent contours are those for which the boundary condition $\dot{x}(t)\rightarrow0$ is not met. As saddle points, they involve the shot and the true vacuum and are thus not capturing the transition amplitude from the quasi-stationary state $|{\rm FV}\rangle$ onto itself.

For the discussion in this section, we have made some assumptions about the state $|{\rm FV}\rangle$
and its wave function that yet need to be justified. In particular, we have assumed
that the wave function takes an approximately Gau{\ss}ian shape about the location
$x_+$ of the false vacuum (provided the potential is quadratic to leading order at this point), and a probability current leaking into the region of the true
vacuum that also leads to a decaying amplitude as imposed by probability conservation.
The previous features are confirmed quantitatively by finding false-vacuum solutions for the wave function using the WKB method to solve the time-independent Schr\"odinger equation, as we carry out in Appendix~\ref{app:sec:wkb}. 
Since the complex energy-eigenvalue of the WKB solution 
agrees with the one inferred per Eqs.~\eqref{amplitudexx} and~\eqref{eq:GammaCC} from the path-integral approach and since there
is only one unstable mode in the fluctuation operator, the WKB result
indeed approximates the wave function of the state $|{\rm FV}\rangle$.
Of course, the WKB solution with complex energy cannot be normalizable because of
its divergent amplitude when taking time to $-\infty$. (Otherwise, with the Hamiltonian being Hermitian, there would be real values for the energy.) Therefore, it can only be an approximation to physical situations in which a particle is
placed in the local ground state close to $x_+$ at some point in the finite past.
Furthermore, the WKB approximation will not be applicable once
the wave function develops as sizable reflux from the true-vacuum back to the false-vacuum region, cf. the numerical example in Ref.~\cite{Andreassen:2016cvx}.
It will turn out that
in observable quantities, the normalizations of $|{\rm FV}\rangle$ cancel such that
this does not lead to a practical issue.

\section{Complex saddle points in the path integral with complex time and the Minkowski case}

\label{sec:MinkowskiPathIntegral}

One notoriously subtle point about the Callan-Coleman theory of tunneling~\cite{Coleman:1977py,Callan:1977pt} is due to the perturbative expansion around a saddle point of the Euclidean action which is not an extremum. Rather, it exhibits one negative, unstable mode that is of crucial relevance for tunneling but also requires careful treatment of the path integral by appropriate methods. As an alternative to the original approach, it has been pointed out in Refs.~\cite{Tanizaki:2014xba,Andreassen:2016cvx} that Picard-Lefschetz theory as reviewed in Refs.~\cite{Witten:2010cx,Witten:2010zr} is particularly suitable to address this issue. In order to keep the present work self-contained, in Appendix~\ref{sec:Callan-Coleman} we review the application of this method to the Euclidean path integral describing tunneling processes. In the following, we carry out some developments such as to apply Picard-Lefschetz theory to Minkowski path integrals, as well as to those with a general complexified time coordinate that interpolates between the Minkowski and Euclidean cases. Perturbation theory is then based on the expansion about saddle points that can be found in terms of complex rather than purely real field configurations~\cite{Turok:2013dfa,Cherman:2014sba}. We discuss the complex saddles in quantum mechanics in Section \ref{sec:complex:saddles}, while Section \ref{sec:complexified:path:integral} summarizes the properties of their associated thimbles and the integration on them. Technical details on the flow equations that define the thimbles and on the definition of the integral measure are given in Section \ref{app:sec:Flow-Jacobian}. A generalization of the results to the case of quantum field theory is given in Section \ref{app:sec:QFT}.

\subsection{Complex saddles}
\label{sec:complex:saddles}

Given certain boundary conditions, the equations of motion in Minkowski space or for a generalized, complexified time variable may not have solutions in terms of real field configurations. However, it is possible to obtain solutions from complex field configurations.
As we discuss in the present section, this is precisely the situation of relevance for quantum tunneling. The solutions then correspond to complex saddle points of the action, and pertaining to these are Lefschetz thimbles generated by the downward gradient flows, in analogy with the discussion of the Euclidean case in Appendix~\ref{sec:Callan-Coleman}. Throughout this article, ``flow'' will refer to a one-dimensional steepest descent path in complexified field space which passes through one (or more) saddle points. In the general case, the thimbles that collect the flows can be seen as complex integration contours that start and end in convergence regions in which the integrand becomes exponentially suppressed. They can be viewed as cycles in a relative homology, and thus we will use ``cycle'' in the following  for any integration path that links regions of convergence. In general, a deformed integration contour of the path integral (which leaves the result of the integration unchanged) can be expressed as a linear combination of thimbles with integer coefficients. In case that the downward  flows from the saddle points end up linking several of these saddles, individual thimbles might not be valid integration cycles, but one can still construct cycles from combinations of thimbles or subsets of thimbles, and the deformed integration contour will again be a linear combination of cycles. This will be the case for the tunneling problem.

\begin{figure}[h]
  \centering
  \hspace{10pt}
  \includegraphics[scale=0.8]{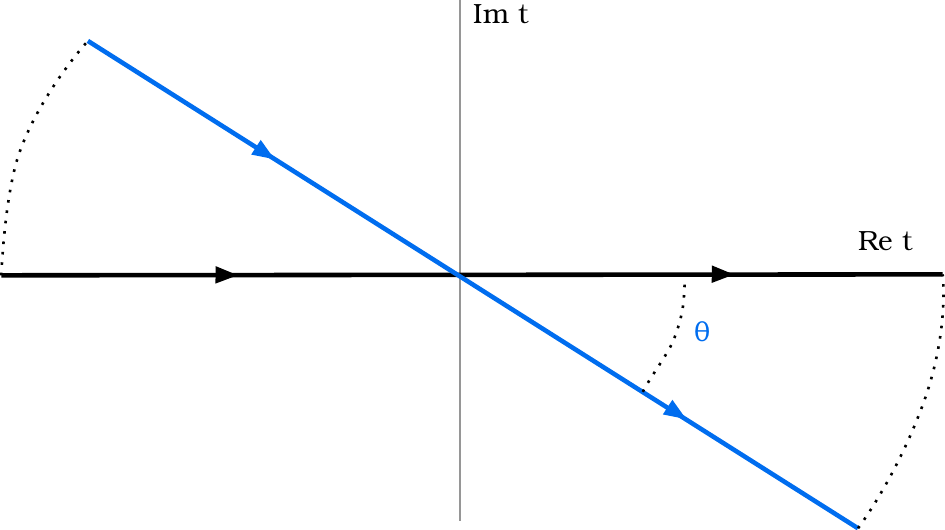}
  \caption{Time contour rotated by an arbitrary angle $\theta$.
  \label{fig:time}}
\end{figure}

In order to find out the relevant saddle points and the corresponding Lefschetz thimbles, we begin with the transition amplitude in complexified time, with a time contour rotated as in Figure~\ref{fig:time}:
\begin{align}
\label{MinkowskiPropagator}
U_{\theta}(x_+,T/2;x_+,-T/2)\equiv \langle x_+|e^{-iHT e^{-i\theta}}|x_+\rangle=\int\mathcal{D}x(t)\,e^{i S_{\theta}[x(t)]}\equiv Z^{\theta}[T],
\end{align}   
where
\begin{align}
\label{action:Minkowski}
S_{\theta}[x]=e^{-i\theta}\int_{-T/2}^{T/2} \D t\left[\frac{1}{2}\left(\frac{\D x}{\D t}\right)^2\cdot e^{2i\theta}-V(x)\right]
\end{align}
and where $\theta\in[\epsilon,2\pi]$ with $\epsilon$ being a positive infinitesimal. The Euclidean and Minkowski transition amplitudes and path integrals are recovered for $\theta=\pi/2$ and $\theta=\epsilon$, respectively.  All the paths in the path integral in Eq.~\eqref{MinkowskiPropagator} are understood to observe the Dirichlet boundary conditions $x(\pm T/2)=x_+$, where $x_+$ is the location of the metastable minimum. The stable minimum is at $x_-$ and in between these vacua, at $x=0$, there is a local maximum of the potential, cf. Figure~\ref{fig:potential}.

The saddle points are given by the solutions to the equation of motion
\begin{align}
\label{MinkowskiEoM}
e^{2i\theta}\cdot\frac{\D^2 x(t)}{\D t^2}+V'(x)=0
\end{align}
subject to the above Dirichlet conditions. When we work in real paths and take the limit $\theta\rightarrow 0^+$ and $T\rightarrow \infty$, we can deduce from the potential that we may have two solutions; the first one is the trivial false-vacuum solution $x_{F}(t)\equiv x_+$, and the second one is similar to the so-called shot configuration in Euclidean case \cite{Andreassen:2016cff,Andreassen:2016cvx} (cf. Appendix~\ref{sec:Callan-Coleman}), with the particle starting in the false vacuum with nonzero velocity and arriving at the top of the potential barrier at $t=0$. The expansion around it cannot describe quantum tunneling, where the initial energy is the one of the false-vacuum ground state. Furthermore, as discussed at the end of Section~\ref{sec:optical-theorem}, the computation of the false-vacuum transition-amplitude by path integral methods requires imposing the boundary condition $\dot{x}(t)\rightarrow0$ for large times, which is not satisfied by the latter solution. To extract the information about false-vacuum decay, one therefore needs to complexify the paths $x(t)$ to $z(t)$ and find nontrivial saddle points whose neighbouring configurations capture quantum fluctuations of the false-vacuum state. As will be seen, these complex saddle points are related to the Euclidean bounce configurations \cite{Coleman:1977py} and correspond to fields bouncing back and forth from the false vacuum any number of times.
 
Indeed, in the limit $T\rightarrow\infty$, the complex solutions to Eq.~\eqref{MinkowskiEoM}, denoted as $x_a^\theta(t)$, can be found from substituting $\tau\rightarrow ie^{-i\theta} t$ into the Euclidean
solutions $x_a(\tau)$,
\begin{align}
\label{solutioncont}
{x}^\theta_a(t)=x_a(\tau=ie^{-i\theta}t),
\end{align}
where $a=F, B_n,S$ labels the saddles which are the false vacuum, the multi-bounces and the shot (cf. Appendix~\ref{sec:Callan-Coleman}).
Comparing with the Euclidean equation of motion Eq.~\eqref{EuclideanEoM}, we
see that $x^\theta_a(t)$ solves Eq.~(\ref{MinkowskiEoM}). As noted by Callan and Coleman \cite{Callan:1977pt}, the effect of fluctuations around multi-bounce configurations can be recovered from the results for the single bounce. The same arguments can be used here, and thus in the following, we will mostly restrict the discussion to the single bounce, referring to it simply as the bounce.

While allowing for general values of $\theta$, we are ultimately interested in
Minkowski spacetime that corresponds to $\theta=\epsilon$.
Apparently, when applying Eq.~\eqref{solutioncont} to the trivial false-vacuum solution $x_F(\tau)$, we still obtain an identical trajectory ${{x}}_F^\theta(t)\equiv x_+={\rm const.}$ For the additional Euclidean saddles that are subject to above Dirichlet conditions, i.e. the bounce and the shot, ${x}_a^\theta(t)$ is a holomorphic function at infinity with $\epsilon\leq \theta\leq \pi/2$, such that these still converge to $x_+$ as $t\rightarrow\pm \infty$, thus satisfying the same boundary conditions. In Ref.~\cite{Cherman:2014sba}, it is shown explicitly that this is indeed the case for the kink solution to the quantum-mechanics problem given by the potential~(\ref{eq:pot}) for $g=0$. The bounce solution in field theory, as an instanton, actually takes the form of the kink solution in the thin-wall limit where the vacua become quasi-degenerate (see e.g. Refs.~\cite{Garbrecht:2015oea,Garbrecht:2015yza,Ai:2018guc}). We also recall that in the quantum-mechanics case, the bounce can be viewed as a kink--antikink pair in the quasi-degenerate limit. In general, when applying Eq.~\eqref{solutioncont} to the Euclidean bounce, we therefore obtain a complex saddle point ${x}_B^\theta(t)$ that observes the Dirichlet conditions above, and we refer to it as the {\it complex bounce}.

Remarkably, the complex bounce gives the same exponential suppression of the tunneling amplitude as in the Euclidean formalism, as first noted for the kink instanton in Ref.~\cite{Cherman:2014sba}. To see this, we write the action~(\ref{action:Minkowski}) as
\begin{align}
\label{complexbounceAction}
S_{\theta}[z]=e^{-i\theta}\int_{-T/2}^{T/2} \D t \left(\frac{1}{2}\left[\left(\frac{\D z}{\D t}\right)e^{i\theta}\pm i\sqrt{2V(z)}\right]^2\mp i\left(\frac{\D z}{\D t}e^{i\theta}\sqrt{2V(z)}\right)\right).
\end{align}
The complex bounce ${x}_B^\theta$ is simply the solution of
\begin{subequations}
\label{complexbounceEoM}
\begin{align}
\left(\frac{\D z}{\D t}\right)e^{i\theta}+i\sqrt{2V(z)}&=0,\ {\rm for } -T/2\leq t \leq 0, \\
\left(\frac{\D z}{\D t}\right)e^{i\theta}-i\sqrt{2V(z)}&=0,\ {\rm for }\  0\leq t \leq T/2,
\end{align}
\end{subequations}
which solve the second-order equation of motion~\eqref{MinkowskiEoM}. For $\theta=\pi/2$, Eqs.~\eqref{complexbounceEoM} and~\eqref{MinkowskiEoM} are equivalent to the equation of motion of the Euclidean bounce~\cite{Coleman:1977py}. Substituting Eq.~\eqref{complexbounceEoM} into Eq.~\eqref{complexbounceAction}, we obtain
\begin{align}
\label{Bounceaction}
\mathcal{I}[{x}_B^\theta]:= iS_{\theta}[{x}_B^\theta]=-i e^{-i\theta}\int_{-T/2}^{T/2}\D t\; 2V({x}_B^\theta(t))=-S_E[x_B].
\end{align}
Now, the potential $V$ is polynomial at the tree level and hence holomorphic. Also, ${x}_B^\theta(t)$ is analytic because it is the continuation of the Euclidean bounce $x_B(\tau)$. Then starting from the expression of $iS_{\theta}$ above and rotating the integration  contour via $t\rightarrow -i e^{i\theta}\tau$ (note $\tau\in\mathbb{R}$), one gets {\it minus} the Euclidean bounce action, $-S_E[x_B]$. Note here, since at $t=\pm\infty$, the complex bounce always converges to the false vacuum $x_+$ at which the potential is zero, the integral~\eqref{Bounceaction} at the infinite boundaries does not contribute when we deform the contour from $t$ to $-i e^{i\theta} \tau$.

We thus note that the bounce action is independent of $\theta$ provided the false vacuum
is normalized to lie at \emph{zero} energy. If we were to assign a finite energy to the false vacuum, then an extra contribution proportional to the volume of spacetime would arise. The latter is proportional to the phase $i e^{-i\theta}$ because of the temporal
integration measure $i e^{-i\theta}\D t$. In Section~\ref{sec:continuation:determinant},
we will recover further $\theta$-dependent contributions arising from the one-loop integration of the fluctuations around the bounce.  Some of these contributions correspond to corrections to the Coleman-Weinberg potential and are thus again proportional to the spacetime volume. These volume- and $\theta$-dependent factors are however unphysical, and, as will be seen in Section~\ref{sec:Minkowskian:decay:path:integral}, they do not contribute to decay rates as they are related to the normalization of the false-vacuum state. 


\subsection{Complexified path integral and Gau{\ss}ian approximation}
\label{sec:complexified:path:integral}

\begin{figure}
  \centering
  \includegraphics[scale=.75]{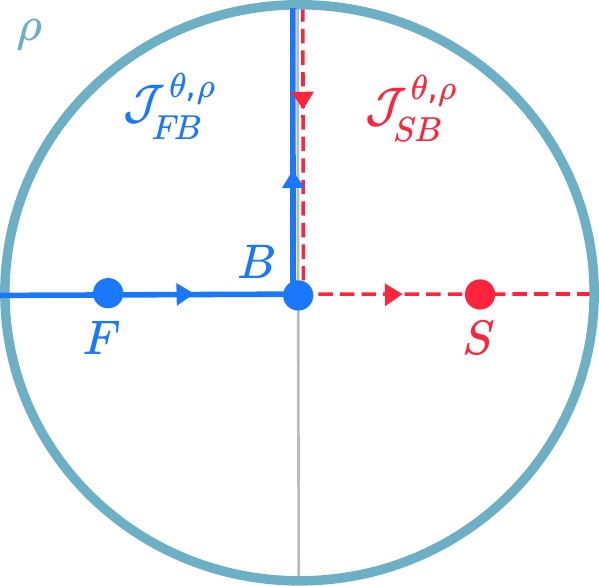}\hskip50pt\includegraphics[scale=.55]{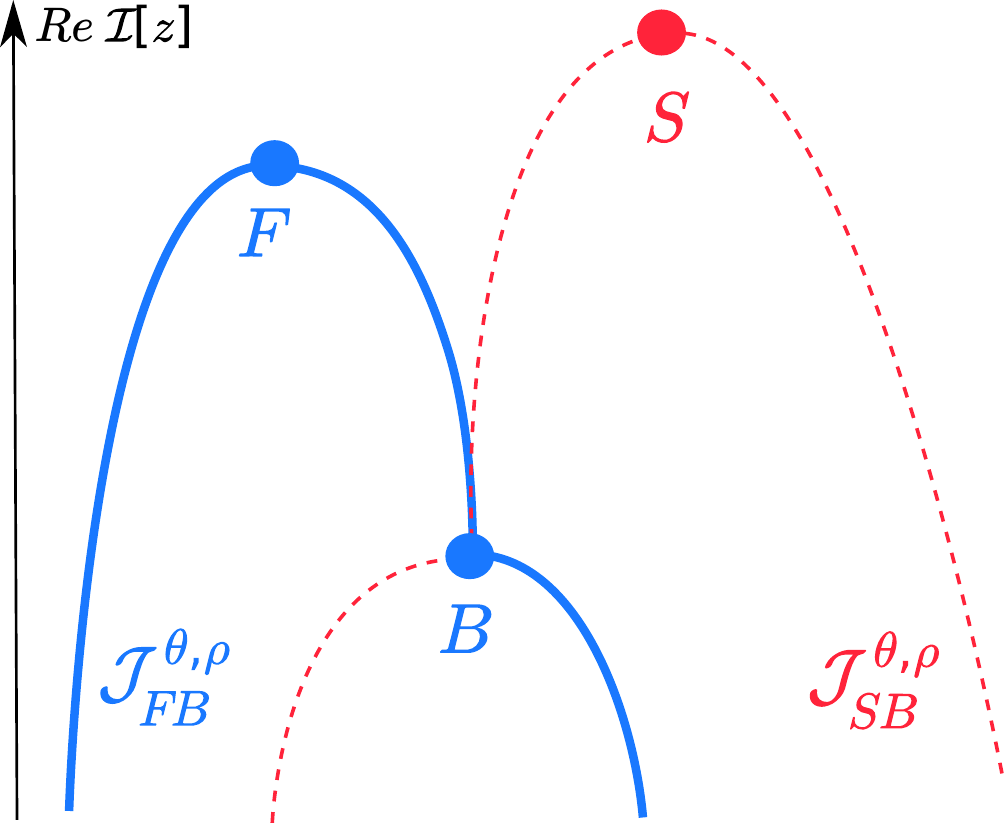}
  \caption{Left: Schematic representation of the 
special one-dimensional complex flows, associated with the negative mode for the bounce, that connect different saddles. The false vacuum, single bounce and shot saddle points are denoted by $F,B,S$ denote, respectively; $\mathcal{J}^{\theta,\rho}_{FB}$ (in blue) and $\mathcal{J}^{\theta,\rho}_{SB}$ (in dashed red) are the 
one-dimensional {flows} 
that connect the false vacuum and the bounce, and the shot and the bounce, respectively. The circle represents infinity. The path integral acquires opposite imaginary parts from the vertical segments of the 
flows. Right: Schematic illustration of the value of the function ${\rm Re}({\cal I}[z])$ along the 
{flows} $\mathcal{J}^{\theta,\rho}_{FB}$ and $\mathcal{J}^{\theta,\rho}_{SB}$, with the same notation and colour code as before. The 
flow $\mathcal{J}^{\theta,\rho}_{FB}$ 
{branches out} at the bounce. {Note that at each saddle point, there are infinitely many additional flows that do not link different saddles, that we do not show in the plot.}
  \label{fig:JFBrho}}
\end{figure}

The parameter $\theta$ in ${x}_a^\theta(t)$ leads to a continuous deformation of the trajectories starting from the original Euclidean saddle points $x_a(\tau)$. We similarly expect that the thimbles for varying $\theta$ will be a continuous deformation of the  Euclidean ones. However, in contrast to the case of the saddle points, this continuous deformation will not be related to a straightforward analytic continuation because, as will be seen below, the equations defining the thimbles are not holomorphic. This makes the integration along the deformed  thimbles nontrivial. In Appendix~\ref{sec:Callan-Coleman} it is argued  that in Euclidean space, despite the presence of the three types of saddle points with their associated thimbles (false-vacuum, bounces, and shot), it is more convenient to define two relevant integration cycles. This is a consequence of the fact that some of the different saddles are connected by one-dimensional steepest-descent flows, and, as mentioned earlier, in such cases there is no direct relation between cycles and thimbles. The relevant cycles are ${\mathcal{J}}_{FB}$--- constructed from the thimbles associated with the false vacuum and the bounces---and ${\mathcal{J}}_{SB}$, which combines the thimbles of the shot and the bounces. 
For arbitrary $\theta$ we expect then deformed integration cycles ${\mathcal{J}}_{FB}^\theta$ and ${\mathcal{J}}_{SB}^\theta$.
Although we  cannot prove that the saddle points remain connected when we deform the Euclidean $\mathcal{J}_{FB}$ to general values of $\theta$, we may note that the necessary condition~\cite{Witten:2010cx} ${\rm Im}\mathcal{I}[{x}_F^\theta]={\rm Im}\mathcal{I}[{x}_{B_n}^\theta]=0$
for the critical points $F$ and $B_n$ to be connected by this flow is satisfied because of Eq.~(\ref{Bounceaction}), which implies that all the saddles have a real value of the functional ${\cal I}=iS_\theta$.\footnote{Had we not fixed $V(x_+)=0$, we would have $iS_{\theta}[{x}^\theta_F]\neq 0$. But then $iS_{\theta}[{x}^\theta_{B_n}]$ would be shifted by the same amount, leading to no physical consequences after normalization.}  A schematic representation of the special flows that connect different saddles
is given in Figure~\ref{fig:JFBrho}.

While we have noted in Eq.~(\ref{Bounceaction})
that the action is invariant under the continuation in the variable $\theta$, we shall now show explicitly that when evaluating the path integral on the deformed cycles, even though the latter are not obtained by analytic continuation of their Euclidean counterparts, at the end of the day one obtains expressions that are straightforwardly related by analytic continuation of the time interval such that the results for arbitrary $\theta$ can be obtained from those in terms of the Euclidean interval $\mathcal T$ by the continuation $\mathcal T\rightarrow ie^{-i\theta}T$. As will  be seen later in Section~\ref{sec:Minkowskian:decay:path:integral}, this gives rise to the same NLO (next-to-leading order, i.e. one-loop here) result for the decay rate as from the Euclidean path integral. 

As a starting point, from Eq.~\eqref{amplitudexx} generalized to arbitrary $\theta$ using Eq.~\eqref{MinkowskiPropagator} we propose
\begin{align}
\label{eq:FVtheta}
&\langle {\rm FV}|e^{-iHTe^{-i\theta}}|{\rm FV}\rangle=\mathcal{N}^2\int_{{\mathcal{J}}_{FB}^{\theta}}\mathcal{D}z\, e^{iS_\theta[z]}\equiv\mathcal{N}^2Z^{\theta}_{FB}[T]\,.
\end{align}
That is, we specify the integration cycle ${\mathcal{J}}_{FB}^\theta$ as the relevant set of Lefschetz thimbles for arbitrary $\theta$, with the Minkowski limit corresponding to $\theta=\epsilon$. The cycle ${\mathcal{J}}_{SB}^\theta$ is discarded because the boundary conditions associated with the shot, with nonzero $\dot{x}$ at large times, do not capture the dynamics of the pseudo-stationary false-vacuum state, as discussed at the end of Section~\ref{sec:optical-theorem}. The cycle ${\mathcal{J}}_{FB}^\epsilon$ is obtained from the deformation of $\mathcal{J}_{FB}$ via a continuous change of $\theta$ from $\theta=\pi/2$ to $\epsilon$, where 
 ${\mathcal{J}}_{FB}^\theta$ is generated by the downward flow from  $x_F^\theta(t)\equiv x_+$ and ${x}_B^{\theta}(t)$ when
substituting $\mathcal{I}[z]=iS_{\theta}[z]$ into Eq.~(\ref{flow}), giving
\begin{align}
\label{Minkowskiflow}
\frac{\partial z(t;u)}{\partial u}=e^{i(\theta-\pi/2)}\left(\frac{\partial^2\overline{z}(t;u)}{\partial t^2}\cdot e^{-2i\theta}+V'(\overline{z}(t;u))\right),
\end{align}
where $z(t;u\to-\infty)=x^\theta_a$ with $a=F,B$. Note that, as emphasized several times, the flow equation is not holomorphic, so that we cannot generate solutions for Minkowski space by analytic continuation of the Euclidean ones.

We aim for a perturbative evaluation of the path integral around the saddles, which requires solving the flow equations in their neighbourhood. As 
${\cal J}^\theta_{FB}$ passes through the false-vacuum and (multi-)bounce saddles, we expect that we can evaluate the path integral $Z^\theta_{FB}$ as a sum of Gau{\ss}ian contributions near the saddles:
\begin{align}
 Z^\theta_{FB}\approx Z^\theta_{F,\text{Gau{\ss}ian} }+\sum_n Z^\theta_{B_n,\text{Gau{\ss}ian}}.
\end{align}
Since the multi-bounces correspond to infinitely separated bounces, the integral of their fluctuations factorizes, and one can use the Euclidean arguments of Ref.~\cite{Callan:1977pt}  to express $Z^\theta_{FB}$ in terms of the single-bounce contribution $Z^\theta_{B,\text{Gau{\ss}ian}}$:\footnote{The same exponentiation arguments were applied for effective actions evaluated at  Minkowski saddles in Ref.~\cite{Plascencia:2015pga}.}
\begin{align}
\label{eq:ZthetaFBexp}
 Z^\theta_{FB}\approx Z^\theta_{F,\text{Gau{\ss}ian} }\exp\left(\frac{ Z^\theta_{B,\text{Gau{\ss}ian}}}{Z^\theta_{F,\text{Gau{\ss}ian}}}\right).
\end{align}
Thus, we just need to estimate $Z^\theta_{F,\text{Gau{\ss}ian}}$ and $Z^\theta_{B,\text{Gau{\ss}ian}}$ by obtaining the downward flows near the corresponding saddle points, and carrying out the integration of fluctuations. 

Expanding $z(t;u)={x}_{a}^\theta(t)+\Delta{z}_a(t;u)$ with $a=F,B$, we obtain the linearized flow equation
\begin{align}
\label{linearizedMinkFlow}
\frac{\partial \Delta{z}_{a}(t;u)}{\partial u}=e^{i(\theta-\pi/2)}\left( e^{-2i\theta}\cdot \frac{\partial^2}{\partial t^2}+V''(\overline{{x}_a^\theta}(t))\right)\overline{\Delta{z}}_a(t;u),
\end{align} 
subject to the boundary conditions
\begin{align}
\label{bc:Delta:z:u}
\Delta z_a(t;u\to-\infty)=0.
\end{align}
The $\Delta z_a$ span the thimble at the saddle points. Therefore, the saddle point
expansion of the path integral is
\begin{align}
\label{saddleExMin}
Z^{\theta}_a&=e^{{\cal I}[{x}_a^\theta]}\notag\\
&\times\int\mathcal{D}\Delta z_a\, e^{ie^{-i\theta}\int_{-T/2}^{T/2}\D t\left[-\frac{1}{2} \Delta z_a(t)\left(e^{2i\theta}\cdot\frac{\D^2}{\D t^2}+V''({x}_a^\theta(t))\right)\Delta z_a(t)-\frac{1}{3!}(g+\lambda {x}_a^\theta(t))\Delta z_a^3(t)-\frac{1}{4!}\lambda\Delta z_a^4(t)\right]}.
\end{align}

To obtain $\Delta z_a(t)$, we write~\cite{Tanizaki:2014xba}
\begin{align}
\label{decomposition:Delta:z}
\Delta z_a=\sum\limits_n \sqrt{-i}e^{i\theta/2}g_n^a(u)\chi_n^a(t),
\end{align}
with $g_n^a(u)\in \mathbb{R}$.
Substituting this separation ansatz into Eq.~\eqref{linearizedMinkFlow}, we obtain
the flow eigenequation
\begin{align}
\label{flow-eigenequationMin}
{\cal M}^{\theta*}_{a}\,\overline{\chi_n^a}(t)\equiv\left( e^{-2i\theta}\cdot \frac{\D^2}{\D t^2}+V''(\overline{{x}_a^\theta}(t))\right)\overline{\chi_n^a}(t)={\kappa}_n^a \chi_n^a(t)
\end{align}
and
\begin{align}
\label{tildega}
{\kappa}_n^a g_n^a(u)=\frac{\D{{g}_n^{a}}(u)}{\D u}.
\end{align}
Equation~(\ref{flow-eigenequationMin}) can be combined with its complex conjugate
such as to form an eigenvalue equation with a Hermitian operator, cf. Section~\ref{app:sec:Flow-Jacobian}. Therefore, we can impose the orthonormality relation
\begin{align}
\label{app:norm-relations}
\int_{-T/2}^{T/2}\D t\; \overline{{\chi}_m^a}(t)\chi_n^a(t)=\delta_{mn}.
\end{align}
Repeating further the analysis of Appendix~\ref{sec:Callan-Coleman}, we find $g_n^a(u)={a}_n^a \exp( {{\kappa}_n^a u})$ with ${a}_n^a\in \mathbb{R}$ and ${\kappa}_n^a\in \mathbb{R}^+$ as required by Eq.~(\ref{bc:Delta:z:u}). Using the above orthonormalization and Eq.~\eqref{flow-eigenequationMin}, one can check that the quadratic term in the exponential of the integrand of the path integral becomes negative definite (except for the zero mode), such that we are dealing with a Wiener integration. From the decomposition~(\ref{decomposition:Delta:z}), we define the path integral measure as
\begin{align}
\label{path:integral:measure}
\mathcal{D}\Delta z_a=\delta_a J_a\prod_{n}\frac{\D g^a_n}{\sqrt{2\pi}}\,.
\end{align}
The factor $\delta_a$ that will be specified below accounts for the fact that the integration cycle 
${\cal J}^\theta_{FB}$ does not include the entirety of all the downward flows starting form the bounce saddle points. As we will discuss below, the false-vacuum and (multi-)bounce saddle points are joined by flows that branch out at the bounce saddles, and the integration contour only picks half of these branches.


The Jacobian $J_a$ appears here because the path integral is originally defined in terms of field fluctuations in real directions, whereas the $\sqrt{-i}e^{i\theta/2}\chi^a_n(t)$ are in general complex. One may view the $g^a_n$ as real parameters for the integration on the thimble but also the direction
of integration in the complex field space must be accounted for, which is achieved by the factor of $J_a$ that we derive in Section~\ref{app:sec:Flow-Jacobian}.  Note the zero mode will be handled separately and $J_a$ we defined does not include the possible phase from the zero mode. At the Gau{\ss}ian level, the path integral~(\ref{saddleExMin}) gives (where the zero mode is omitted, which we indicate with a prime)
\begin{align}
\label{Gaunon-zero}
Z'^{\theta}_{a,\text{Gau{\ss}ian}}=e^{\mathcal{I}[{x}^{\theta}_a]}\delta_a
J_a\prod_{n\neq 1}\frac{1}{\sqrt{{\kappa}^a_n}}.
\end{align}
Here we are using Callan's and Coleman's notation, in which the negative mode around the bounce is assigned a subscript ``0'', and the zero mode a subscript ``1''.

Regarding the evaluation of Eq.~\eqref{Gaunon-zero}, one might attempt to obtain the solutions to Eq.~\eqref{flow-eigenequationMin} and ${\kappa}_n^a$ from the analytical continuation $\tau\rightarrow ie^{-i\theta}t, \mathcal{T}\rightarrow ie^{-i\theta}T$ of the Euclidean flow eigenequation~\eqref{flow-eigenEq}. But this is impossible due to the complex conjugates appearing in the flow eigenequation. However, as we show in Section~\ref{app:sec:Flow-Jacobian}, the infinite product of the flow eigenvalues can be expressed as
\begin{align}
\label{eq:detkappa}
\prod_{n\neq 1}{\kappa}_{n}^a =\left|{\det}^\prime\left(e^{2i\theta}\cdot \frac{\D^2}{\D t^2}+V''({{x}_a^\theta}(t))\right)\right|,
\end{align}
and the Jacobian $J_a$ is related to the phase of the above determinant (see  Section~\ref{app:sec:Flow-Jacobian}),
\begin{align}
\label{eq:Jphase}
 J_a =\left(\prod_{n\neq1} \sqrt{-i}e^{i\theta/2}\right)\exp\left(-\frac{1}{2}{\rm Arg}\,{\det}^\prime\left(e^{2i\theta}\cdot \frac{\D^2}{\D t^2}+V''({{x}_a^\theta}(t))\right)\right).
\end{align}

Thanks to Eq.~\eqref{eq:detkappa}, we can reduce the problem of solving the flow eigenequation to that of solving the proper eigenvalue problem
\begin{align}
\label{Min-eigen}
{\cal M}^\theta_a f^a_n(t)=\left(e^{2i\theta}\cdot\frac{\D^2}{\D t^2}+V''({x}_a^\theta(t))\right){f}_n^a(t)={\lambda}_n^a{f}_n^a(t).
\end{align}
We use the term \emph{proper} eigenvalue equation to emphasize that in
contrast to the flow eigenequation~\eqref{flow-eigenequationMin}, no complex conjugation of the eigenvector appears here.
Now given this standard form of an eigenequation, it turns out that there is no obstacle in the way of analytic continuation so that the determinant in Eq.~\eqref{eq:detkappa} can be related to its Euclidean counterpart. This is shown in Section~\ref{sec:continuation} in which we explain how to construct the analytical continuation of the eigenmodes and eigenvalues from the Euclidean solutions
to arbitrary values of $\theta$, and we apply this procedure to the specific example of the kink solution in
the archetypical double-well potential~(\ref{potential}) in Section~\ref{sec:continuation:examples}. We therefore obtain
\begin{align}
\label{Min-Gau}
J_a\prod_{n\neq1}\frac{1}{\sqrt{{\kappa}^a_n}}=\left(\prod_{n\neq1} \sqrt{-i}e^{i\theta/2}\right)\left[\left.\left({{\det}^\prime}(-\partial_\tau^2+V''(x_a))\right)\right|_{\mathcal{T}\rightarrow i e^{-i\theta}T}\right]^{-1/2}.
\end{align}

We next need to handle the zero mode in the flow equation around the complex bounce, which is given by
\begin{align}
\label{zeromodeMin}
{\chi}^B_1(t)=\frac{1}{\sqrt{-i}e^{i\theta/2}\sqrt{S_E[x_B]}}\left.\frac{\D x_B(\tau)}{\D \tau}\right|_{\tau\rightarrow ie^{-i\theta}t}=-ie^{i\theta}\frac{1}{\sqrt{-i}e^{i\theta/2}\sqrt{S_E[x_B]}}\frac{\D {x}_B^\theta(t)}{\D t}.
\end{align}
Using Eqs.~\eqref{complexbounceEoM},~\eqref{Bounceaction}, one can verify that
\begin{align}
\int_{-T/2}^{T/2} \D t\, {\chi}_1^B(t) {\chi}_1^B(t) =1.
\end{align}
From the path integral measure defined in terms of the $g^B_n$, using the decomposition \eqref{decomposition:Delta:z} we can relate the change in coordinate $g^B_1$ associated with the zero mode to an infinitesimal time translation of the bounce, $t\rightarrow t+\D t_0$:
\begin{align}
\label{Zeromodemeasure}
\sqrt{-i}e^{i\theta/2}{\chi}_1^B(t)\,\D {g}^B_1=\D\Delta z_B=\frac{\D {x}^\theta_B}{\D t}\D t_0\Rightarrow\frac{\sqrt{-i}e^{i\theta/2}}{\sqrt{2\pi}}\,\D {g}^B_1\rightarrow \sqrt{-i}e^{i\theta/2} \sqrt{\frac{S_E[x_B]}{2\pi}}(ie^{-i\theta}\D t). 
\end{align}
The integration over the zero mode then gives a factor
\begin{align}
\label{Zeromodeintegral}
\sqrt{-i}e^{i\theta/2} \sqrt{\frac{S_E[x_B]}{2\pi}}(ie^{-i\theta}T).
\end{align}

Before putting everything together, we need to identify the fractions $\delta_a$ introduced in the measure of Eq.~\eqref{path:integral:measure}. As argued earlier, $\mathcal{J}^\theta_{FB}$ along which we approximate $Z^\theta_{FB}$ links the false-vacuum saddle point $a=F$ with the bounce saddles $a=B_n$. Near the false vacuum, all steepest-descent directions are part of $\mathcal{J}^\theta_{FB}$, so that $\delta_F=1$. However, the same is not true for the (multi-)bounces. Let us consider first the single bounce. In Section~\ref{sec:continuation}, we show that
the discrete eigenvalues of Eq.~(\ref{Min-eigen}) remain invariant under the analytic continuation. Therefore, just as in the Euclidean case, there is one negative mode $\chi^B_0$ about the complex bounce $B$. It is expected that there is a flow $\mathcal{J}^{\theta,\rho}_{FB}$ in $\mathcal{J}^\theta_{FB}$ that links $F$ and $B$, approaching $B$ along the direction of the negative mode. The latter is thus a direction of steepest ascent relative to $B$, while the flow must continue down to a steepest descent direction.  These two ``in'' and ``out'' directions are related by a relative
factor of $i$ for the flow eigenmodes corresponding to the negative mode
of Eq.~(\ref{Min-eigen}), as can be seen by inspection of Eq.~(\ref{flow-eigenequationMin}), according to which a multiplication of the eigenfunctions by
a factor of $i$ implies a sign change of the eigenvalue. Thus, the flow $\mathcal{J}^{\theta,\rho}_{FB}$ approaches $B$ from a steepest ascent direction, and leaves along one of the two possible steepest-descent directions associated with $i\chi^B_0$ (the two directions corresponding to either positive or negative coefficients $g^B_0$ in the expansion of Eq.~\eqref{decomposition:Delta:z}). The picture is illustrated in Figure~\ref{fig:JFBrho}. The flow associated with the false vacuum thus branches out at $B$, and $\mathcal{J}^{\theta,\rho}_{FB}$ picks only one branch, which is the one that gives the correct sign for the imaginary part of $Z^\theta_{FB}$, which determines the decay rate as a consequence of the optical theorem, as discussed in Section \ref{sec:optical-theorem}. The fact that only one branch of the steepest-descent flow from the bounce is relevant implies $\delta_B=1/2$. Returning to the multi-bounce, in this case, we expect $n$ negative modes for an $n$-bounce, corresponding to each of the single bounces. Generalizing the discussion for the single bounce, the expectation is that the downward flow from $F$ should reach the multi-bounce $B_n$ along with their $n$ steepest ascent directions. At $B_n$ the flow is expected to divide into $2n$ steepest-descent branches (corresponding to two imaginary directions per negative mode), with the flow $\mathcal{J}^{\theta,\rho}_{F{B_n}}$ in the integration cycle picking only half of them. Hence one expects a factor of $\delta_{B_n}=(1/2)^n$ for the Gau{\ss}ian integration near $B_n$, which is already accounted for by the exponentiation formula \eqref{eq:ZthetaFBexp}, as is clear when expanding it in terms of $Z^\theta_B$ with $\delta_B=1/2$ and identifying the $B_n$ contributions as those proportional to $(Z^\theta_B)^n$.

Finally putting the pieces together, from Eqs.~\eqref{eq:ZthetaFBexp},~\eqref{Gaunon-zero},~\eqref{Min-Gau}, and using the zero-mode factor for $Z^\theta_{B,\text{Gau{\ss}ian}}$ given in Eq.~\eqref{Zeromodeintegral}, one arrives at
\begin{align}\label{ZFB-Theta}\begin{aligned}
\frac{Z^\theta_{FB}[T]}{Z^\theta_{F}[T]}\approx&\,\exp\left(\frac{Z^\theta_{B,\text{Gau{\ss}ian}}[T]}{Z^\theta_{F,\text{Gau{\ss}ian}}[T]}\right)\\
=&\,{\exp}\left(\frac{{ie^{-i\theta}T}}{2}\sqrt{\frac{S_E[x_B]}{2\pi}}e^{-S_E[x_B]}\left(\frac{{\det}'[e^{2i\theta}\partial_t^2+V''( x^\theta_B)]}{\det[e^{2i\theta}\partial_t^2+V''(x^\theta_F)]}\right)^{-1/2}\right)\\
=&\left.{\exp}\left(\frac{{\mathcal T}}{2}\sqrt{\frac{S_E[x_B]}{2\pi}}e^{-S_E[x_B]}\left(\frac{{\det}'[-\partial_\tau^2+V''(x_B)]}{\det[-\partial_\tau^2+V''(x_F)]}\right)^{-1/2}\right)\right|_{\mathcal{T}\rightarrow i e^{-i\theta}T}.
\end{aligned}\end{align}
Comparing with the Euclidean formula in the second line of Eq.~\eqref{ZFB-Euc}, it follows that the result for arbitrary $\theta$ is given by the straightforward analytic continuation of the Euclidean time interval $\mathcal T$ to its rotated counterpart $T$. This is in keeping with the expectations coming from the fact that one can formally write the partition functions $Z^\theta_{FB}$ as in Eq.~\eqref{eq:FVtheta}, which formally implies
\begin{align}
 Z^\theta_{FB}[T]=\left.Z^E_{FB}[\cal T]\right|_{\mathcal{T}\rightarrow i e^{-i\theta}T}.
\end{align}

%
%

Although we postpone a more detailed discussion of the tunneling rate until Section~\ref{sec:Minkowskian:decay:path:integral}, it should be noted that according to the optical theorem discussed in Section~\ref{sec:optical-theorem}---see Eq.~(\ref{opticaltheoremforfalsevacuum2})---a nonzero decay rate requires an imaginary part in the transition amplitude ${M}$. The latter is proportional to $-i Z^\epsilon_{FB}$---as follows from identifying $\langle {\rm FV}|\mathsf{1}+iM|{\rm FV}\rangle=Z^\epsilon_{FB}$. This means that $Z^\epsilon_{FB}$ has 
to contain a real part, which in turn actually requires the quotient of determinants in \eqref{ZFB-Theta} to be a negative real number, as it is in the Euclidean case $\theta=\pi/2$ due to the presence of a discrete negative eigenvalue. Since the result for arbitrary $\theta$ is related to the Euclidean one by analytic continuation ${\cal T}\rightarrow ie^{-i\theta} T$, a negative real value for the quotient of determinants for arbitrary $\theta$ requires the quotient to become  $T$-independent in the large $ T$ limit. As will be seen in Section~\ref{sec:continuation}, this is indeed the case since with appropriate regularizations $T$ only appears in contributions from the continuum spectrum which are common for the bounce and the false-vacuum saddle points. The discrete spectrum of the operators is preserved under rotations of the time contour, and thus there is always a negative mode which ensures that the quotient of determinants is a negative real number. Using this in equation \eqref{ZFB-Theta} finally gives
 \begin{align}\label{ZFB-Theta2}\begin{aligned}
\frac{Z^\theta_{FB}[T]}{Z^\theta_{F}[T]}\approx&\,\exp\left(\frac{Z^\theta_{B,\text{Gau{\ss}ian}}[T]}{Z^\theta_{F,\text{Gau{\ss}ian}}[T]}\right)\\
=&\,{\exp}\left(\frac{{-e^{-i\theta}T}}{2}\sqrt{\frac{S_E[x_B]}{2\pi}}e^{-S_E[x_B]}\left|\frac{{\det}'[e^{2i\theta}\partial_t^2+V''( x^\theta_B)]}{\det[e^{2i\theta}\partial_t^2+V''(x^{\theta}_F)]}\right|^{-1/2}\right)\\
=&\left.{\exp}\left(\frac{{i\mathcal T}}{2}\sqrt{\frac{S_E[x_B]}{2\pi}}e^{-S_E[x_B]}\left|\frac{{\det}'[-\partial_\tau^2+V''(x_B)]}{\det[-\partial_\tau^2+V''(x_F)]}\right|^{-1/2}\right)\right|_{\mathcal{T}\rightarrow i e^{-i\theta}T}.
\end{aligned}\end{align}

\subsection{Flow equations and Jacobian}
\label{app:sec:Flow-Jacobian}

In this section, we show how to relate the flow eigenequations to the proper eigenvalue equations in order to derive the Jacobian induced when the path integral is performed on a Lefschetz thimble. This is necessary because we choose to parametrize the path integral by real numbers $g_n^a$, while the flow eigenfunctions span the thimble in complex directions in general.

We start by considering the linearized flow equation about the saddle $z_a$
\begin{align}
\label{app:general-flow}
\frac{\partial\Delta z_a(t;u)}{\partial u}={-ie^{i\theta}{\cal M}^\theta_a}^*\,\overline{\Delta z_a}(t;u),
\end{align}
where 
\begin{align}
{\cal M}^\theta_a=ie^{i\theta}\int \D t' \left.\frac{\delta^2\mathcal{I}[z]}{\delta z(t')\delta z(t)}\right|_{z_a}\equiv ie^{i\theta}{\left.\frac{\delta^2\mathcal{I}}{\delta  z^2}\right|_{z_a(t)}}=e^{-2i\theta}\cdot \frac{\D^2}{\D t^2}+V''({{x}_a^\theta}(t)).
\end{align} 
Following the analysis of Appendix~\ref{sec:Callan-Coleman}, we make the separation ansatz (see Section~\ref{sec:complexified:path:integral})
\begin{align}
\label{app:decomp:flow}
\Delta z_a(t;u)
=\sum\limits_n {\sqrt{-i}e^{i\theta/2}} g^a_n(u) \chi_n^a(t)
=\sum\limits_n {\sqrt{-i}e^{i\theta/2}} {a}_n^a\exp({\kappa}_n^a u) \chi_n^a(t),
\end{align}
where ${a}_n^a\in\mathbb{R}$ and ${\kappa}_n^a\in \mathbb{R}^+$. The last property ensures that the steepest-descent flow  reaches the saddle-point only at the limiting value $u\rightarrow-\infty$.
When there is a zero mode with ${\kappa}_n^a=0$, it needs to be handled separately,
as it is carried out for the mode pertaining to time translations in Section~\ref{sec:complexified:path:integral}. We therefore restrict
the following discussion to the nonzero modes.

For the flow eigenmodes $\chi_n^a$, we then obtain the flow eigenequation
\begin{align}
\label{app:flow-eigenEq}
{{{\cal M}^\theta_a}^*}\,\overline{\chi_n^a}(t)={\kappa}_n^a \chi_n^a(t)
\end{align}
with Dirichlet boundary conditions $\chi_n^{a}(t=\pm T/2)=0$.
This equation can be combined with its complex conjugate as
\begin{align}
\label{app:eigen1}
\begin{pmatrix}
\mathbf{0} & {{{\cal M}^\theta_a}^*}\\
{{\cal M}^\theta_a} & \mathbf{0}
\end{pmatrix}
\begin{pmatrix}
\chi_n^a(t)\\ \overline{\chi_n^a}(t)
\end{pmatrix}
={\kappa}_n^a\begin{pmatrix}
\chi_n^a(t)\\ \overline{\chi_n^a}(t)
\end{pmatrix}.
\end{align}
The operator on the left-hand side is Hermitian, such that
we can impose orthonormalization as in Eq.~\eqref{app:norm-relations}. Furthermore, the Hermiticity property implies that the $\kappa_n$ are real, as was assumed earlier.
We use $\{{f}_n^a(t)\}$ to denote the eigenfunctions with corresponding eigenvalues ${\lambda}_n^a$ satisfying
\begin{align}
\label{app:eigenEq}
{{\cal M}^\theta_a}\,{f}_n^a(t)={\lambda}_n^a {f}_n^a(t).
\end{align}

Substituting $z(t;u)=z_a(t)+\sum_n {\sqrt{-i}e^{i\theta/2}} g_n^a(u)\chi_n^a(t)$ into ${\mathcal I}[z]$ and making use of the complex conjugate of Eq.~\eqref{app:flow-eigenEq} and Eq.~\eqref{app:norm-relations}, one obtains up to $\mathcal{O}(\Delta z^2)$
\begin{align}
\label{app:expan}
{\mathcal I}[z]={\mathcal I}[z_a]-{\frac{ie^{-i\theta}}{2}}\int\D t\,\Delta z_a(t;u)\,{{\cal M}_a^\theta}\,\Delta z_a(t;u)
={\mathcal I}[z_a]-\frac{1}{2}\sum_n{{\kappa}_n^a}\,(g_n^a(u))^2.
\end{align}

Now, the goal is to compute the saddle point approximation to the path integral
on the thimble ${\cal J}_a$
\begin{align}
\label{app:pathint1}
Z_a=\int{\cal D}\Delta z_a\; e^{{\cal I}[z]}
= e^{{\cal I}[z_a]}\int{\cal D}\Delta z_a \; e^{{-\frac{ie^{-i\theta}}{2}} \int \D t\; \Delta z_a(t){{\cal  M}_a^\theta } \Delta z_a(t)+\cdots}.
\end{align}
The decomposition~(\ref{app:decomp:flow}) leads us to work with the path integral measure (cf. Eq.~(\ref{path:integral:measure}))
\begin{align}
\label{app:path:integral:measure}
{\mathcal{D}\Delta z_a=J_a\prod_n\frac{\D g_n^a}{\sqrt{2\pi}},}
\end{align}
where $J_a$ is the Jacobian. We have dropped here the argument $u$ of the $g_n^a$ because they now assume the role of integration variables.
Then,
in Gau{\ss}ian approximation, we obtain {(in case the  mode $n=1$
is not a zero mode, it should not be removed from the product, and the prime can be ignored)}
\begin{align}
\label{app:Gau}
{Z'_a=J_a\prod_{n\neq 1}\int\ \frac{\D g_n^a}{\sqrt{2\pi}}\ e^{-\frac{1}{2}\sum_n{\kappa}_n^a({g}_n^a )^2}=J_a\prod_{n\neq 1}\frac{1}{\sqrt{{\kappa}_n^a }}.}
\end{align}

To relate the infinite product of the flow eigenvalues to the determinant of ${\cal M}^\theta_a$, we first note that
Eq.~(\ref{app:eigen1}) is associated with another equation:
\begin{align}
\label{app:eigen2}
\begin{pmatrix}
\mathbf{0} & {{{\cal M}^\theta_a}^*}\\
{{{\cal M}^\theta_a}} & \mathbf{0}
\end{pmatrix}
\begin{pmatrix}
i \chi_n^a(t)\\ -i\overline{\chi_n^a}(t)
\end{pmatrix}
=-{\kappa}_n^a\begin{pmatrix}
i\chi_n^a(t)\\ -i\overline{\chi_n^a}(t)
\end{pmatrix}.
\end{align}
Therefore, ${\kappa}_n^a$ and $-{\kappa}_n^a$ are eigenvalues for the
operator on the left-hand side of Eq.~\eqref{app:eigen1}.
It follows that
\begin{align}
\prod_{n\neq 1} \left[-({\kappa}_n^a)^2\right]={\det}'
\begin{pmatrix}
\mathbf{0} &  {{{\cal M}^\theta_a}^*}\\
{{{\cal M}^\theta_a}} & \mathbf{0}
\end{pmatrix},
\end{align} 
which gives $\prod_{n\neq 1} ({\kappa}_n^a)^2=|{\det}'({{{\cal M}^\theta_a}^*{\cal M}^\theta_a})|$ for the particular block structure. We therefore arrive at 
\begin{align}
\label{app:flowVSeigen}
\prod_{n\neq 1}{{\kappa}_n^a}=|{\det}'({{{\cal M}^\theta_a}})|=\left|\prod_{n\neq 1}{\lambda}^a_n\right|,
\end{align}
where we recall ${\kappa}_{n\neq 1}^a>0$.

In order to work out the Jacobian $J_a$,
we first pick an arbitrary, complete orthonormal basis of \emph{real} functions $\{\varphi_n(t)\}$ such that
\begin{align}
\label{app:real:basis}
\int\D t\; \varphi_m(t)\varphi_n(t)=:( \varphi_m,\varphi_n )=\delta_{mn},
\end{align}
where we have also defined a shorthand notation for the real inner product. We carry out the following analysis for discrete modes,
as it would apply for finite $T$. Taking $T\to \infty$ will lead in general to spectra that have a continuum part, which is the case of interest. We therefore need to assume that the present arguments remain valid in that limit. 
The real basis allows us to decompose $\Delta z_a$ into its components as
\begin{align}
\label{app:Delta:z:decomp:comps}
\Delta z_{a,n}:=(\varphi_n,\Delta z_a )=\sum\limits_m(\varphi_n, \sqrt{-i}e^{i\theta/2} g_m^a {\chi}^a_m)
=:\sum\limits_m R_{nm} g_m^a,
\end{align}
where we have used the decomposition~(\ref{app:decomp:flow}) into the flow eigenmodes
${\chi}_m^a$. It follows that
\begin{align}
\Delta z_a(t)= \sum\limits_n\Delta z_{a,n} \varphi_n(t)
\end{align}
and, consequently, infinitesimally
\begin{align}
\label{app:trafo:gz}
\D \Delta z_{a,n}:=(\varphi_n,\D \Delta z_a)=\sum\limits_m(\varphi_n, \sqrt{-i}e^{i\theta/2}\D g_m^a {\chi}_m^a)
=\sum\limits_m R_{nm} \D g_m^a.
\end{align}
Note how the operator $R$ maps the real coefficients $g_m^a$ in the decomposition~(\ref{app:decomp:flow})  onto the complex $\Delta z_{a,n}$ that parameterize the thimble ${\cal J}_a$ in terms of the real basis $\{\varphi_n(t)\}$.

Given these constructions, as an alternative to the measure~(\ref{app:path:integral:measure}),
we may now express the path integral~(\ref{app:pathint1}) in the component form
\begin{align}
Z'_a=\int_{{\cal J}_a} {\prod_{n\neq1}\left(\frac{{\mathrm{d}} \Delta z_{a,n}}{\sqrt{2\pi}}\right) }\,
e^{{\cal I}[z]},
\end{align}
{where we omit the $n=1$ term if it is a zero mode.}
Through Eq.~(\ref{app:Delta:z:decomp:comps}), the
latter is given as a hypersurface parametrized in terms of the real parameters $g_n^a$. Note
that here, in the given linearized expansion around the saddle, this  hypersurface is
thus approximated by a hyperplane. We can therefore express the path integral as
\begin{align}
Z'_a=\int_{{\cal J}_a}{\left(\prod_{n\neq 1} \frac{\D g^a_{n}}{\sqrt{2\pi}} \right)} \
{{\det}'} R\ e^{{\cal I}[z]},
\end{align}
where Eq.~(\ref{app:trafo:gz}) gives us the Jacobian
\begin{align}
J_a={\det}' R
\end{align}
for the transformation from
the $\{z_{a,n}\}$ to the $\{g_n^a\}$. The prime indicates that the Jacobian $J_a$ defined does not include the contribution from the zero mode which is isolated.

Next, we multiply the complex conjugate of Eq.~(\ref{app:flow-eigenEq}) by ${\chi}_m$ from the left,
such that
{\begin{align}
{(  {\chi}_m, {\cal M}^\theta_a\chi_n)={\kappa}_n \delta_{mn}\;\ \textnormal{(no sum over $n$)},}
\end{align}}
where we have used Eq.~(\ref{app:norm-relations}). Inserting complete
sets in the $\varphi$-basis leads to
{
\begin{align}
{(i e^{-i\theta} )({\sqrt{-i}e^{i\theta/2}}{\chi}_m, \varphi_i)(\varphi_i,{ \cal M}^\theta_a \varphi_j)(\varphi_j,{\sqrt{-i}e^{i\theta/2}}\chi_n)={\kappa}_{n}\delta_{mn}\;\ \textnormal{(no sum over $n$)},}
\end{align}
}
or, using the definition of $R$ in Eq.~(\ref{app:Delta:z:decomp:comps}),
\begin{align}
(i e^{-i\theta} )(R^T)_{mi}{\cal M}^\theta_{a,ij} R_{jn}={\kappa}_{n}\delta_{mn}\;\ \textnormal{(no sum over $n$)},
\end{align}
where ${\cal M}^\theta_{a,ij}$ are the components of ${\cal M}^\theta_{a}$ in the $\varphi$-basis.
Promoting this component equation to one for matrices, taking the determinant on both sides and using Eq.~(\ref{app:flowVSeigen}) yields
\begin{align}
{({\det}' R)^2 {\det}'(i e^{-i\theta} {\cal M}^\theta_a)=\prod_{n\neq 1}{\kappa}_n^a=|{\det}' {\cal M}^\theta_a|}
\end{align}
and, eventually,
\begin{align}
J_a={\det}' R=\left(\prod_{n\neq 1} \sqrt{-i}e^{i\theta/2}\right)\sqrt{\frac{|{\det}' {\cal M}^\theta_a|}{{\det}' {\cal M}^\theta_a}}.
\end{align}
We therefore conclude that the Jacobian $J_a$ is proportional to the   phase of $1/\sqrt{({\det}' {\cal{M}}_a^{\theta})}$.

This together with relation~\eqref{app:flowVSeigen} is particularly useful because we can transfer the original flow eigenproblem to the proper eigenproblem. And the proper eigenequation~\eqref{app:eigenEq} can be conveniently analytically continued between the Euclidean formalism and the Minkowski formalism as we will show in Section~\ref{sec:continuation}.


\subsection{Generalization to quantum field theory}

\label{app:sec:QFT}

The previous treatment of the quantum-mechanical path integral can be easily generalized to quantum field theory, as summarized next. Assuming for simplicity a theory involving a real scalar field $\phi$ with a false vacuum at $\phi=\phi_+$ and a true vacuum at $\phi=\phi_-$,  the relevant transition amplitude for a rotated time contour (the quantum field theoretical generalization of \eqref{MinkowskiPropagator}) is given by
\begin{align}
\label{eq:PropagatorQFT}
Z^\theta[T]=U_{\theta}(\phi_+,T/2;\phi_+,-T/2)\equiv \langle \phi_+|e^{-iHTe^{i\theta}}|\phi_+\rangle=\int\mathcal{D}\phi\,e^{i S_{\theta}[\phi]},
\end{align}   
where the action now is
\begin{align}
\label{action:MinkowskiQFT}
S_{\theta}[\phi]=e^{-i\theta}\int_{-T/2}^{T/2}\D t\int \D^3x \left[\frac{1}{2}\left(\frac{\partial \phi}{\partial t}\right)^2\cdot e^{2i\theta}-\frac{1}{2}\left(\vec{\nabla}\phi\right)^2-V(\phi)\right].
\end{align}
As before, the classical action for a rotated time contour admits complex saddle-point solutions with Dirichlet boundary conditions related to the false vacuum, which for $\theta=\pi/2$ reproduce the Euclidean bounce. The equation for the saddle points is
\begin{align}
 e^{2i\theta}\,\frac{\partial^2\phi}{\partial t^2}-\vec{\nabla}^2\phi+V'(\phi)=0.
\end{align}
Given a Euclidean solution $\phi_a(\tau,\vec{x})$ solving the above for $\theta=\pi/2$, one can construct solutions for arbitrary $\theta$ through analytic continuation,
\begin{align}
 \phi_a^\theta(t,\vec{x}) = \phi_a(\tau=ie^{-i\theta}t,\vec{x}).
\end{align}
As long as $\theta\geq0^+$, the rotated solutions satisfy the same Dirichlet boundary conditions as the Euclidean one.
Moreover, with an appropriate normalization of the potential ensuring $V(\phi_+)=0$, the values of the action at the complex saddle points tending to $\phi_+$ for $t\rightarrow\pm\infty$ coincide with their Euclidean counterparts, as follows from applying the Cauchy
theorem to $S_\theta$ and relating the contour of the time integration to the Euclidean one (as was argued for the quantum-mechanical case below Eq.~\eqref{Bounceaction}).

Using Picard-Lefschetz theory, the path integral can be again approximated by a sum of integrations over some Lefschetz thimbles (or subspaces thereof) passing through the saddle points. The results of the previous sections carry over to the field theoretical case, with the main differences coming from the fact that the fluctuation operators appearing in the flow equations involve now spatial derivatives. In particular, denoting the spacetime coordinates as $x$, one has that the linearized flow equations near a saddle point $\phi^\theta_a(x)$---with $\phi$ along the flow written as $\phi(x;u)=\phi^{\theta}_a(x)+\Delta\phi_a(x;u)$---take now a form analogous to Eq.~\eqref{app:general-flow},
\begin{align}
\label{app:general-flowQFT}
\frac{\partial}{\partial u}\,\Delta \phi_a(x;u)=-ie^{i\theta}{{\cal M}^\theta_a}^*\,\overline{\Delta \phi_a}(x;u),
\end{align}
where the fluctuation operator ${\cal M}^\theta_a$  is now given by
\begin{align}
{\cal M}^\theta_a=e^{2i\theta}\frac{\partial^2}{\partial t^2}-\vec{\nabla}^2+V''(\phi^\theta_a).
\end{align} 
The ansatz
\begin{align}
 \Delta\phi_a(x;u)=\sum_n \sqrt{-i}e^{i\theta/2} g^a_n(u) \chi^a_n(x)
\end{align}
allows again to express the path integration on the thimble
in terms of the (primed if there are zero modes) determinant of ${\cal{M}}^\theta_a$; the derivation goes as in Section \ref{app:sec:Flow-Jacobian}. As there are no zero modes for the false-vacuum saddle point, this gives directly
\begin{align}
\label{eq:ZthetaF}
 Z^\theta_{F}=\left(\prod_n\sqrt{-i}e^{i
\theta/2}\right)(\det \ {\cal M}^\theta_F)^{-1/2}=(\det ie^{-i\theta}{\cal M}^\theta_F)^{-1/2},
\end{align}
where we have used the fact that the Euclidean action evaluated at the constant false-vacuum saddle point is zero, given the choice of normalization of the potential $V(\phi_+)=0$.

For the bounce saddle point, one has to deal separately with the zero-mode integration. There are four zero modes in quantum field theory, related to the invariance of the theory under temporal and spatial translations. The zero modes are related to derivatives of the bounce solution with respect to its spacetime coordinates:
\begin{align}
\label{eq:zeroQFT}
 \chi_{1,(\mu)}^{B}=\frac{\partial_\mu\phi^{\theta}_B}{\sqrt{\int d^4 x (\partial_{\mu}\phi^{\theta}_B)^2}}\Rightarrow\Delta\phi_B^{(\mu)}=\sqrt{-i}e^{i\theta/2} g^B_{1,(\mu)} \frac{\partial_\mu\phi_B^\theta}{\sqrt{\int d^4 x (\partial_{\mu}\phi_B^\theta)^2}},
\end{align}
where  there is no summation in $\mu=0,\dots, 3$. With the path integral measure defined as before in terms of the $g^a_n$ (see Eq.~\eqref{app:path:integral:measure}),  we can connect  $g^B_{1,(\mu)}$ in Eq.~\eqref{eq:zeroQFT} with a coordinate translation $x^\mu\rightarrow x^\mu+y^\mu$, as in the quantum-mechanical example:
\begin{align}
 \D\Delta\phi^{(\mu)}_B = \D y^{(\mu)} \partial_{\mu} \phi^{\theta}_B \Rightarrow \prod_{\mu=1}^4\sqrt{-i}e^{i\theta/2}\frac{\D g_1^{(\mu)}}{\sqrt{2\pi}}= \prod_{\mu=1}^4 \D y^{(\mu)}\sqrt{\frac{\int d^4 x(\partial_{\mu}\phi^{\theta}_B)^2}{2\pi}}.
\end{align}
Since the bounce is obtained from an analytic continuation of the Euclidean bounce, one can relate the spacetime integrals above to their Euclidean counterparts by rotating the time contour. Then one can apply Callan's and Coleman's arguments of Ref.~\cite{Callan:1977pt} for the Euclidean bounce,  which, relying on its O(4) invariance and the fact that its action is stationary under dilatation transformations, imply
\begin{align}
- e^{2i\theta} \int \D^4 x\,(\partial_{0}\phi_B^\theta)^2= \int \D^4 x\,(\partial_{i}\phi_B^\theta)^2= -i e^{i\theta}S_E[\phi_B^\theta], \quad i=1,2,3.
\end{align}
Putting everything together, the path integration along the thimble corresponding to the bounce saddle-point gives
\begin{align}
 Z_{B}^\theta=\frac{ie^{-i\theta}V^{(3)}T}{2}\,e^{-S_E[\phi_B]}\left(\frac{S_E[\phi_B]}{2\pi}\right)^2({\rm det}' {\cal M}^\theta_B)^{-1/2}\left(\prod_n\sqrt{-i}e^{i
\theta/2}\right),
\end{align}
where $V^{(3)}$ is the three-dimensional volume factor arising from the integration of the spatial zero modes, while the factor $1/2$ arises as in the quantum-mechanical case from choosing a particular steepest-descent path that passes through both the false-vacuum and bounce saddle points, and which around the latter picks only one of the two steepest-descent branches associated with the negative mode.

The quantum field theoretical generalization of Eq.~\eqref{ZFB-Theta2} becomes now:
\begin{align}\label{ZFB-Theta-QFT}\begin{aligned}
\frac{Z^\theta_{FB}[T]}{Z^\theta_{F}[T]}\approx&\,{\exp}\left(\frac{{-e^{-i\theta}TV^{(3)}}}{2}\left(\frac{S_E[\phi_B]}{2\pi}\right)^2e^{-S_E[\phi_B]}\left|\frac{{\det}' {\cal M}^\theta_B}{\det {\cal M}^\theta_F}\right|^{-1/2}\right)\\
=&\left.{\exp}\left(\frac{{i\mathcal TV^{(3)}}}{2}\left(\frac{S_E[\phi_B]}{2\pi}\right)^2e^{-S_E[\phi_B]}\left|\frac{{\det}' {\cal M}^E_B}{\det {\cal M}^E_F}\right|^{-1/2}\right)\right|_{\mathcal{T}\rightarrow i e^{-i\theta}T}.
\end{aligned}\end{align}
Again, the ratio of determinants will be shown to be independent of $T$, and the last line shows that the result for arbitrary $\theta$ can be simply obtained from the Euclidean result by analytic continuation of the Euclidean time interval $\cal T$.

\section{Analytic continuation of the fluctuation spectrum}
\label{sec:continuation}

We now work out how the solutions to the proper eigenvalue problem transform under rotations of the time variable within the complex plane, i.e. changes of $\theta$. As advertised earlier, we will see that the determinants  of fluctuations for arbitrary $\theta$ are related to their Euclidean counterparts by the analytic continuation $\mathcal T\rightarrow ie^{-i\theta}T$. Moreover, the quotients of determinants over the bounce and false-vacuum saddle points appearing in Eqs.~\eqref{ZFB-Theta},~\eqref{ZFB-Theta-QFT} will be shown to be $T$-independent, as needed for $\theta$-independent decay rates. We present the arguments for the field-theoretical case, assuming a spatially homogeneous geometry for the vacuum transition. For tunneling problems, this corresponds to the thin-wall limit, where the bubble wall is approximated as planar compared to its radial profile. In Section~\ref{sec:spherical:geom}, we outline how the analytic continuation can also be applied to tunneling transitions when the thin-wall limit does not apply and the spherical geometry of the bubble has to be taken into account. 
The present discussion can be easily reduced to the quantum-mechanics examples that are discussed in other
parts of this work.

In order to derive the analytic continuation, we consider a fluctuation determinant in a scalar theory around a background configuration ${\phi}_a^\theta$ that extremizes the effective action (after complexification in field space, if necessary) and interpolates between two vacuum configurations.
For a time contour analytically continued in the complex plane by a rotation of the angle $\theta$
from the Euclidean case $\theta=\pi/2$ in the clockwise direction, the eigenvalue problem for the fluctuation operator is given by
\begin{align}
\label{eq:flukes}
{\cal M}^\theta_a\Delta \phi^\theta_{\{\iota\}}(t,\vec{x})\equiv \left[e^{2i\theta}\frac{\partial^2}{\partial t^2}-\vec{\nabla}^2+V''(\phi_a^\theta(t,\vec{x}))\right]\Delta \phi^\theta_{\{\iota\}}(t,\vec{x})=\lambda \Delta \phi^\theta_{\{\iota\}}(t,\vec{x}),
\end{align}
where the background $\phi^\theta_a$ satisfies 
\begin{align}
\label{eq:bg}
 \left[e^{2i\theta}\frac{\partial^2}{\partial t^2}-\vec{\nabla}^2+V'(\phi^\theta_a(t,\vec{x}))\right]\phi^\theta_a=0.
\end{align}
The $\{\iota\}$ are labels that determine the eigenvalue $\lambda=\lambda(\{\iota\})$ and uniquely specify the eigenstate, i.e.  the eigenfunction. The labels are more or less
directly related to properties of the eigenfunctions, and below we identify these based on their asymptotic behavior. Note that for $\theta\to 0^+$, we recover the fluctuation equation in Minkowski space. 


\subsection{Eigenmodes and eigenvalues}

We proceed with identifying the eigenmodes from their form for large $|t|$, where the background field configuration $\phi^\theta_a$ has the following asymptotic behaviour
\begin{align}
\phi^\theta_a(t,\vec{x})\rightarrow \phi_{\pm}, {\rm\ as\ } t\rightarrow\pm\infty.
\end{align}
Here $\phi_\pm$ denotes the vacua with $V'(\phi_\pm)=0$.
The potential near the vacua is approximately parabolic
\begin{align}
 V_\pm(\phi)=\frac{m^2_\pm}{2}(\phi-\phi_\pm)^2,
\end{align}
where $m^2_\pm$ are the effective masses of fluctuations around the vacua. We take $\phi_+$ to be the initial vacuum. For the tunneling problem, we have $\phi_-=\phi_+$ since the bounce $\phi_B^\theta$ approaches the false vacuum both at $t=\pm\infty$. Thus we have $m_+^2=m_-^2\equiv m^2$. This is also true for the kink soliton (in the time direction) when $\phi_-\neq \phi_+$ but still $V''(\phi_-)=V''(\phi_+)$ because of the $Z_2$ symmetry between the two vacua in the double-well potential. Our following analysis therefore applies to the bounce as well as to the kink soliton.

Inserting this into the equation of motion for the background~\eqref{eq:bg} and assuming spatial homogeneity, one can see that the extremal solution approaches the vacua $\phi_\pm$ exponentially fast, as long as $\theta>0$:
\begin{align}
\phi^\theta_a(t,\vec x)\rightarrow\phi_\pm+ C e^{\mp (i\cos\theta +\sin\theta) m t}\quad \text{for}\quad t\rightarrow\pm\infty.
\end{align}
Within Eq.~(\ref{eq:flukes}) for the fluctuations, for large $|t|$ the background will then sit at the vacua up to exponentially suppressed corrections, such that we may replace $V''(\phi^\theta_a)\rightarrow m^2$, and we are left with equations of the linear form
\begin{align}
 \left[e^{2i\theta}\frac{\partial^2}{\partial t^2}-\vec{\nabla}^2+m^2\right]\Delta \phi^\theta_{\{\iota\}}(t,\vec{x})=\lambda \Delta \phi^\theta_{\{\iota\}}(t,\vec{x})
 \;\ \textnormal{ for large $|t|$}.
\end{align}
We construct the asymptotic solutions $\Delta\phi^\theta_{\{\iota\}}$
in terms of an expansion in a complete and orthonormal set of functions on three-dimensional space.\footnote{One may as well turn this around, consider the differential operator
for large $|\vec x|$ and construct solutions from a complete set of functions of time $t$. However, we proceed as we do because we are ultimately interested in the analytic dependence on the parameter $\theta$.} Therefore, the labels $\{\iota\}$ have
to be chosen according to the symmetries of the background $\phi^\theta_a$.
This is particularly simple for the planar-wall geometry, while we
discuss more general situations in Section~\ref{sec:spherical:geom}.


For the planar wall, we can work with eigenfunctions of the three-dimensional Laplacian $\vec{\nabla}^2$. For these, we choose exponential functions $\exp(i\vec{k}\cdot\vec{x})$, which are eigenfunctions with eigenvalue $-\vec{k}^2$, and thus, we use $\vec{k}$ as labels that characterize the spatial behaviour. Note that because of the spatial translation symmetries in the planar-wall limit, $\vec k$ is a conserved quantity over the evolution in Euclidean time. Thus for either the bounce or the kink soliton, we will have the same asymptotic quantum numbers at $t=\pm \infty$. We can then separate the eigenfunctions as
\begin{align}
\label{eq:separation:kspace}
 \Delta\phi^\theta_{\{\eta,\vec k\}}(t,\vec{x})\sim \varphi_{\{\eta,\vec k\}}(t)e^{i\vec{k}\cdot\vec{x}}, \,\,t\rightarrow \pm\infty,
\end{align}
where
\begin{align}
 e^{2i\theta}\frac{\D^2}{\D t^2}\varphi_{\{\eta,\vec k\}}(t) = (\lambda-\vec{k}^2-m^2)\varphi_{\{\eta,\vec k\}}(t),
\end{align}
and $\eta$ is an additional label characterizing the asymptotic temporal behaviour.
The asymptotic solutions are then of the form
\begin{align}
 \Delta\phi^\theta_{\{\eta,\vec k\}}(t,\vec{x})\sim \exp\left[ \pm e^{-i\theta}\sqrt{\lambda-\vec{k}^2-m^2}\,\,t\right]e^{i\vec{k}\cdot\vec{x}}.
\end{align}
For some of the values of $\lambda$, the full $\Delta\phi^\theta_{\{\iota\}}(t,\vec{x})$ as a solution to Eq.~(\ref{eq:flukes}) is a normalizable eigenvector, either in the proper or improper sense. Solutions that are improperly normalizable have an oscillatory asymptotic behaviour in the temporal direction. This happens for $\lambda$ such that { $e^{-i\theta}\sqrt{\lambda-\vec{k}^2-m^2}$} is purely imaginary. Properly normalizable solutions correspond to functions that decay at infinity in {the} time direction. This occurs when { $e^{-i\theta}\sqrt{\lambda-\vec{k}^2-m^2}$} has a real part. According to the node theorem, the spectrum of properly normalizable solutions is discrete.

For either case, it is useful to characterize the asymptotic oscillation
frequency, or, respectively, the decay rate by real parameters that
we introduce here as
\begin{align}
\begin{aligned}
e^{-i\theta}\sqrt{\lambda-\vec{k}^2-m^2}=&i\kappa_{\theta},\,\kappa_{\theta}\in\mathbb{R}, &&\text{ (for continuous  } \lambda \text{, oscillating solution)},
\\
\sqrt{\lambda-\vec{k}^2-m^2}=&{ -i\beta,  \, \beta\in\mathbb{R}^+,}
&&\text{ (for discrete  } \lambda\text{, decaying solution)}.
\label{eq:oscdec}
\end{aligned}
\end{align}
We can thus choose these parameters to take the place of $\eta$, such that we use ${\{\iota\}}=\{\kappa_{\theta},\vec k\}$ or, respectively, ${\{\iota\}}=\{\beta,\vec k\}$ as the labels of the eigenstate. We will see shortly why there is no necessity to add a subscript ``$\theta$'' to $\beta$. An important aspect is that, given solutions to the eigenvalue equations (not necessarily normalizable) for some value of $\theta$, one can obtain solutions for another value of $\theta$ by analytic continuation in the time variable. This follows from the fact that both the background and fluctuation equations for arbitrary $\theta$ can be obtained by analytic continuation from Euclidean time. Thus we may construct all solutions by rotating from Euclidean time: $\tau\rightarrow ie^{-i\theta}t$. A key concern is that solutions that are (im)properly normalizable for one value of $\theta$ are not necessarily so for another value. However, we show next that one can construct (im)properly normalizable solutions for arbitrary $\theta$ by supplementing the analytic continuation of the time variable with complex rotations of the parameters of the solutions. This analysis will also explain that the phases in Eq.~(\ref{eq:oscdec}) are indeed chosen such as to maintain the parameters $\kappa_{\theta}$, $\beta$ real for all values of $\theta\in (0,\pi/2]$. 

First, we show that Euclidean solutions which are temporally
decaying will continue to be so after analytic continuation, without modification of the parameters $\beta$. For $\theta=\pi/2$, the operator ${\cal M}^\theta_a$ in Eq.~(\ref{eq:flukes})
is Hermitian. The eigenvalues are therefore real, and it is possible to choose real eigenfunctions. A temporally decaying Euclidean solution therefore should have the asymptotic form
\begin{align}
 \Delta\phi^E_{\{\beta_{E},\vec k\}}(\tau,\vec{x})\propto \exp\left( \mp \beta_{E}\tau\right),\,\,\tau\rightarrow \pm\infty,\beta_{E}\in\mathbb{R}^+
.
\end{align}
(Here and in the following, we occasionally replace superscripts or subscripts $\theta=\pi/2$ with $E$. Note however that this does not apply to the action, where Eq.~\eqref{Bounceaction} holds.) The analytic continuation to arbitrary $\theta$, obtained by substituting $\tau\rightarrow i e^{-i\theta}t$ gives 
\begin{align}
 \Delta\phi^\theta_{\{\beta_{E},\vec k\}}(\tau,\vec{x})\propto \exp\left( \mp ie^{-i\theta}\beta_{E} t\right)=\exp(\mp (\sin\theta) \beta_{E} t)\exp(\mp i(\cos\theta) \beta_{E} t),\,\,t\rightarrow \pm\infty
 .
\end{align}
As long as $\theta>0$, the rotated solution will still be temporally decaying. Therefore the straightforward rotation of Euclidean decaying solutions in the temporal argument gives acceptable eigenfunctions for a rotated time contour as well. And this means that the real decay parameters $\beta_E$ remain unchanged and real for all values
of $\theta$, what explains the phase choice in Eq.~(\ref{eq:oscdec}), and we therefore suppress the subscript $E$ or $\theta$ on $\beta$.
Another important consequence is that the discrete Euclidean eigenvalues $\lambda$, that are a function of $\beta$
as per Eq.~(\ref{eq:oscdec}), are preserved under analytic continuation in time---in particular the discrete zero\footnote{The zero mode will be traded for a collective coordinate and hence introduce a dependence on $i e^{-i\theta}T$ as per Eq.~(\ref{Zeromodeintegral}).} and negative modes crucial in tunneling computations. To summarize these results,
given a discrete Euclidean mode with eigenvalue $\lambda$, its analytic continuation to general values of $\theta$ is
\begin{align}
\label{continuation:disc:lambda}
\Delta\phi^\theta_{\{\beta,\vec k\}}(t,\vec{x})
=
\sqrt{i}e^{-i\theta/2} \Delta\phi^E_{\{\beta,\vec k\}}(i e^{-i\theta} t,\vec{x}),
\end{align}
with the same eigenvalue $\lambda$. Below, when we discuss the scalar
product, we will derive the normalizing factor.

Second, as for temporally oscillating Euclidean solutions, these go as
\begin{align}
\label{continuum:asymp:E}
 \Delta\phi^E_{\{\kappa_{E},\vec k\}}(\tau,\vec{x})\propto \exp\left(i \kappa_{E}\tau\right),\,\,\tau\rightarrow \pm\infty,
\end{align}
where $\kappa_{E}\in\mathbb{R}$ depends on the free parameters that define the Euclidean solutions. The eigenvalue $\lambda$ now is part of a continuum spectrum. For convenience, we work with complex continuum eigenfunctions.
Note however that since ${\cal M}_E$ is Hermitian, we could have chosen also a
real basis in terms of eigenfunctions that asymptotically behave like sine and
cosine. This remark will be of importance when we discuss the normalization and completeness of the eigenmodes in Sections~\ref{sec:normal} and~\ref{sec:complete}.
The analytic continuation of the continuum mode ~\eqref{continuum:asymp:E} in the temporal variable only gives
\begin{align}
 \Delta\phi^E_{\{ {\kappa_{E}},\vec k\}}(ie^{-i\theta}t,\vec{x})\propto \exp\left(-e^{-i\theta}\kappa_{E} t\right),\,\,t \rightarrow \pm\infty.
\end{align}
This is however not an acceptable eigenfunction, as it grows exponentially in one of the time directions. However, we can also complexify the parameter of the asymptotic oscillations of the original Euclidean solution, replacing $\kappa_{E}\rightarrow- ie^{i\theta}\kappa_{\theta}$ (recall that both $\kappa_{E}$ and $\kappa_{\theta}$ are defined to be real) and arrive at a new solution that is normalizable in the improper sense. Further, since the eigenvalues are related to the exponents in the asymptotic solutions through Eq.~\eqref{eq:oscdec}, this implies that the continuum eigenvalues for arbitrary $\theta$  are obtained from the Euclidean eigenvalues by the appropriate analytic continuation. In the Euclidean case, one would have e.g.
\begin{align}
\label{eq:lambdacontE}
  \lambda\equiv\lambda^E(\kappa_{E},\vec k)=\kappa^2_{E}+\vec{k}^2+m^2,\quad \kappa_{E},\vec{k}\in\mathbb{R},
\end{align}
while, upon the continuation given by Eq.~\eqref{eq:oscdec}, the corresponding eigenvalue for arbitrary $\theta$ is
\begin{align}
\label{eq:lambdaosc}
  \lambda\equiv\lambda(\theta,\kappa_{\theta},\vec k)=\lambda^E(-i e^{i\theta}\kappa_{\theta},\vec k)=- e^{2i\theta}\kappa^2_{\theta}+\vec{k}^2+m^2, \quad\kappa_{\theta},\vec{k}\in\mathbb{R},
\end{align}
i.e. it is different from the Euclidean one and depends on $\theta$, in contrast to the discrete modes.
The analytically continued, improperly normalizable mode is obtained when
making the replacement $\tau\rightarrow i e^{-i\theta}t$ as well as $\kappa_{E}\rightarrow -ie^{i\theta}\kappa_{\theta}$ in the Euclidean solution as
\begin{align}
\label{continuum:eigenfunction:theta}
\Delta\phi_{\{\kappa_{\theta},\vec k\}}^\theta(t,\vec{x})=\Delta\phi^E_{\{-ie^{i\theta}\kappa_{\theta},\vec{k}\}}(ie^{-i\theta}t,\vec{x}),
\end{align}
where we have again stipulated a normalization that we will confirm below.

\subsection{Normalization of the eigenmodes}
\label{sec:normal}

In order to prepare for the calculation of the fluctuation determinant, we next need to verify that the analytically continued modes constitute, just as the Euclidean eigensystem, a complete orthonormal basis. First, we note that the continuation of the Euclidean differential operator in Eq.~(\ref{eq:flukes}) to arbitrary values of $\theta$ is not Hermitian.
As a consequence, we have found above that the continuum eigenvalues are
generally complex, while the discrete eigenvalues remain real. In either case,
the eigenfunctions, which can be chosen real in the Euclidean case (even though, for convenience, we have chosen a complex basis for the continuum spectrum), become complex. Nonetheless, since the eigenfunctions are obtained by analytic rotations of Euclidean ones, the real scalar product as in Eq.~(\ref{app:real:basis}) remains invariant. In particular, different eigenfunctions remain orthogonal with respect { to} this scalar product that does not involve complex conjugation.

For discrete eigenfunctions, we have shown thus far that
\begin{align}
\Delta\phi^\theta_{\{\beta,\vec k\}}(t,\vec{x})=\sqrt{N}\Delta\phi^E_{\{\beta,\vec k\}}(ie^{-i\theta}t,\vec{x}),
\notag
\end{align}
where $N$ accounts for possible differences in normalization.
The scalar product of the eigenfunctions is:
\begin{align}
 \int \D t\, \D^3 \vec{x}\, \Delta\phi_{\{\beta,\vec k\}}^\theta(t,\vec{x})\Delta\phi_{\{\beta^\prime,\vec k'\}}^\theta(t,\vec{x})= N\int \D t\, \D^3 \vec{x}\, \Delta\phi^E_{\{\beta,\vec k\}}(ie^{-i\theta}t,\vec{x})\Delta\phi_{\{\beta^\prime,\vec k'\}}^E(ie^{-i\theta}t,\vec{x}).
\end{align}
Given the analytic time dependence of the Euclidean eigenfunctions, using the Cauchy theorem, the integration contour can be rotated without changing the value of the integral. (Note that the discrete eigenfunctions vanish for large values of complex time along the arcs that join the rotated axes.) A change of the contour by replacing $t \rightarrow -i e^{i\theta}\tau$ therefore gives\footnote{Recall that $\tau$ and $t$ are both defined to be real.}
\begin{align}
\int \D t\, \D^3 \vec{x}\, \Delta\phi^\theta_{\{\beta,\vec k\}}(t,\vec{x})\Delta\phi^\theta_{\{\beta^\prime,\vec k'\}}(t,\vec{x})
=& -i e^{i\theta}N\int \D\tau\, \D^3 \vec{x}\, \Delta\phi_{\{\beta,\vec k\}}^E(\tau,\vec{x})\Delta\phi^E_{\{\beta',\vec k'\}}(\tau,\vec{x})
\notag
\\
=&-i e^{i\theta} N \delta_{\beta\beta'}
(2\pi)^3\delta(\vec k-\vec k')
,
 \label{eq:discretenorm}
\end{align}
where we have used the standard orthonormality relation for the Euclidean eigenfunctions. Thus, using
\begin{align}
\label{normalization:continued:disc}
N=i e^{-i\theta},
\end{align}
which is the normalization appearing in Eq.~\eqref{continuation:disc:lambda},
we obtain orthogonal eigenfunctions with the proper norm.


For the temporally oscillating solutions, we have to consider in addition the effect of the analytic continuation of the parameter $\kappa$. We first note that we normalize the Euclidean scalar product as
\begin{align}
\label{eq:scalar:product:continuum}
 \int \D\tau\, \D^3\vec{x}\, \widetilde{\Delta\phi}_{\{\kappa_E,\vec{k}\}}^{E}(\tau,\vec{x})\Delta\phi_{\{\kappa'_{E},\vec{k'}\}}^{E}(\tau,\vec{x})= 2\pi \delta(\kappa_E-{\kappa'_{E}})(2\pi)^3\delta^3(\vec{k}-\vec{k'}).
\end{align}
Here, the tilde indicates the reciprocal eigenfunction, which is different from the eigenfunction
in case we choose to work with a complex basis for convenience. Since ${\cal M}_E$ is Hermitian,
such a complex basis can however be decomposed into a real basis.
It is then understood that the two real basis functions constituting a complex one are individually
continued analytically, both for the eigenfunction and its reciprocal.
For a rotated time contour, the corresponding eigenfunctions are obtained by analytic continuation of $\tau$ and $\kappa_E$ according to Eq.~\eqref{continuum:eigenfunction:theta}.
The inner product is a function of $\kappa_\theta$, and it is fixed for the Euclidean case $\theta=\pi/2$ in Eq.~(\ref{eq:scalar:product:continuum}).
In order to continue to general values of $\theta$,
\begin{align}
&\int \D t\, \D^3 \vec{x}\, \widetilde{\Delta\phi}_{\{\kappa_\theta,\vec{k}\}}^{\theta}(t,\vec{x})\Delta\phi_{\{\kappa'_{\theta},\vec{k'}\}}^{\theta}(t,\vec{x})
\notag\\
&=\int  \D t\, \D^3 \vec{x}\, \widetilde{\Delta\phi}_{\{-ie^{i\theta}\kappa_\theta,\vec{k}\}}^{E}(ie^{-i\theta}t,\vec{x})\Delta\phi_{\{-ie^{i\theta}\kappa'_\theta,\vec{k'}\}}^{E}(ie^{-i\theta}t,\vec{x}),
\end{align}
we evaluate the right-hand side by shifting the integration contour
through the replacement $t\to -i e^{i\theta}t$
such as to maintain the integration manifestly convergent. Then, the phase that appears as a prefactor
cancels the phase from the continuation of the $\delta$-function, and we obtain
\begin{align}
&\int \D t\, \D^3 \vec{x}\, \widetilde{\Delta\phi}_{\{\kappa_\theta,\vec{k}\}}^{\theta}(t,\vec{x})\Delta\phi_{\{\kappa'_{\theta},\vec{k'}\}}^{\theta}(t,\vec{x})
\notag\\
=& -i e^{i\theta}
2\pi \delta(-ie^{i\theta}(\kappa_\theta-{\kappa'_{\theta}}))(2\pi)^3\delta^3(\vec{k}-\vec{k'})
=2\pi \delta(\kappa_\theta-\kappa_\theta')(2\pi)^3\delta^3(\vec{k}-\vec{k'})\label{eq:scalar:product:continuum:theta}
\end{align}
as the generalization of Eq.~(\ref{eq:scalar:product:continuum}). {The $\delta$-function of a complex argument
is understood here as the analytic continuation of some real representation}.
This fixes the normalization $N'=1$ that has been implied in Eq.~\eqref{continuum:eigenfunction:theta}.
Note that when rotating the integration  in order to apply the Cauchy theorem, the contributions from the integration along the  arcs at infinite complex time are also expected to vanish in this case because of the oscillating nature of the solution---as opposed to the exponential decay for the discrete eigenfunctions.


\subsection{Completeness of the eigenmodes}
\label{sec:complete}

Consider the sum over projection operators 
\begin{align}
\label{eq:Itheta}
 {\cal I}_\theta=\int  \frac{\D^3 \vec{k}}{ (2\pi)^3} \sum_{ \beta} \Delta\phi_{\{\beta,\vec k\}}^\theta(x) \Delta\phi_{\{\beta,\vec k\}}^\theta(x')
 &+\int \frac{\D \kappa_\theta}{2\pi} \frac{\D^3\vec{k}}{(2\pi)^3}\,\widetilde{\Delta\phi}_{\{\kappa_\theta,\vec{k}\}}^{\theta}(x) \Delta\phi_{(\kappa_\theta,\vec{k})}^{\theta}(x').
\end{align}
Using the relation to the Euclidean eigenfunctions with the appropriate normalization obtained above, we have:
\begin{subequations}
\begin{align}
 \Delta\phi_{\{\beta,\vec k\}}^\theta (t,\vec{x}) = &\,\sqrt{ie^{-i\theta}} \Delta\phi_{\{\beta,\vec k\}}^E(ie^{-i\theta}t,\vec{x})\quad&&\text{(decaying, discrete)},\\
 \Delta\phi_{(\kappa_\theta,\vec{k})}^\theta (x) = &\,  \Delta\phi_{\{-ie^{i\theta}\kappa_\theta,\vec{k}\}}^E(ie^{-i\theta}t,\vec{x})\quad&&\text{(oscillating, continuum)}.
\end{align}
\end{subequations}
The above implies, after analytic continuation of the integral in $\kappa_\theta$ in Eq.~\eqref{eq:Itheta}  to an integral over $ie^{-i\theta}\kappa_E$ (where $\kappa_E\in\mathbb{R}$),
\begin{equation}
\begin{aligned}
 {\cal I}_\theta=\,ie^{-i\theta}\bigg(&\int \frac{\D^3 \vec{k}}{(2\pi)^3} \sum_{\beta} \Delta\phi_{\{\beta,\vec k \}}^E(ie^{-i\theta}t,\vec{x}) \Delta\phi_{\{\beta,\vec k\}}^E(ie^{-i\theta}t',\vec{x'})
 \notag\\
 +&\int\frac{\D \kappa_E}{2\pi} \frac{\D^3\vec{k}}{(2\pi)^3}\,\widetilde{\Delta\phi}_{\{\kappa_E,\vec{k}\}}^{E}(ie^{-i\theta}t,\vec{x}) {\Delta\phi_{(\kappa_E,\vec{k})}^{E}}(ie^{-i\theta}t,\vec{x})\bigg).
\end{aligned}
\end{equation}
The term in parentheses is the sum over projectors over the Euclidean eigenfunctions, analytically continued in time. Assuming a complete Euclidean basis, this is nothing but $\delta(ie^{-i\theta}(t-t'))\delta^3(\vec{x}-\vec{x'})$. Thus,
\begin{equation}\begin{aligned}
{\cal I}_\theta=ie^{-i\theta}  \delta(ie^{-i\theta}(t-t'))\delta^3(\vec{x}-\vec{x'}) = \delta(t-t')\delta^3(\vec{x}-\vec{x'}).
\end{aligned}\end{equation}
In summary, the sum over the rotated projectors onto the rotated eigenfunctions is equal to the identity operator, which shows that the rotated basis is complete.

\subsection{Fluctuation determinant}
\label{sec:continuation:determinant}

Given the orthonormal eigenfunctions for arbitrary $\theta$, the differential operator ${\cal M}^\theta_a$ can be expanded in a basis of orthogonal projectors:
\begin{align}
 {\cal M}^\theta_a(x,x') =& \int \frac{\D^3\vec{k}}{(2\pi)^3}\sum_{\beta} \lambda(\beta,\vec k)\Delta\phi_{\{\beta,\vec k\}}^\theta(x) \Delta\phi_{\{\beta,\vec k\}}^\theta(x')
 \notag\\
 +&\int \frac{\D \kappa_\theta}{2\pi} \frac{\D^3\vec{k}}{(2\pi)^3}\, \lambda(\theta,\kappa_\theta,\vec{k}) \,\widetilde{\Delta\phi}_{\{\kappa_\theta,\vec{k}\}}^{\theta}(x) \Delta\phi_{\{\kappa_\theta,\vec{k}\}}^{\theta}(x').
\end{align}
In the continuum integral,  we made explicit the dependence of the eigenvalues on $\theta$.
The determinant can be calculated as
\begin{align}
\label{eq:log:det}
\log\det {\cal M}^\theta_a=&{\rm tr}\,\log {\cal M}^\theta_a=\int \D t\, \D^3 \vec{x}
\sum_{\beta}\int\frac{\D^3\vec{k}}{(2\pi)^3}
\Delta\phi^{\theta}_{\{\beta,\vec k\}}(t,\vec x) \Delta\phi^{\theta}_{\{\beta,\vec k\}}(t,\vec x)
\log\lambda(\beta,\vec k)
\notag\\
+&\int \D t\, \D^3 \vec{x}
\int\frac{\D\kappa_\theta}{2\pi} \frac{\D^3\vec{k}}{(2\pi)^3}
\widetilde{\Delta\phi}^{\theta}_{\{\kappa_\theta,\vec k\}}(t,\vec x) \Delta\phi^{\theta}_{\{\kappa_\theta,\vec k\}}(t,\vec x)
\log\lambda(\theta,\kappa_\theta,\vec k)
\notag\\
=&\underset{\rm disc}{\rm tr}\,\log{\cal M}^\theta_a+\underset{\rm cont}{\rm tr}\,\log{\cal M}^\theta_a,\hspace{-8cm}&&
\end{align}
what we have decomposed into a discrete and a continuum piece.

The contribution from the discrete modes can be readily evaluated using
the orthonormality relation~(\ref{eq:discretenorm}) and replacing
$(2\pi)^3\delta^3(\vec k -\vec k)\to V^{(3)}$, as it is appropriate
for planar wave vectors, where $V^{(3)}$
is the volume of the three-dimensional space in which the vacuum transition
occurs. This leads to the sum
over the logarithms of the discrete eigenvalues, which matches the Euclidean result and remains independent of $\cal T$:
\begin{align}
\label{tr:log:discrete}
\underset{\rm disc}{\rm tr}\,\log{\cal M}^\theta_a= V^{(3)}\int\frac{\D^3\vec{k}}{(2\pi)^3}\sum\limits_{\beta}\log\lambda(\beta,\vec k)= V^{(3)}\int\frac{\D^3\vec{k}}{(2\pi)^3}\sum\limits_{\beta}\log\lambda_E(\beta,\vec k).
\end{align}

In order to make proper sense
of the continuum piece, we separate contributions that can be attributed to the solitonic background $\phi^\theta_a$ from those that belong to the vacuum that is approached asymptotically.
The latter give rise to
a term that is proportional to the volume of spacetime. Since in the vacuum one has constant $V''(\phi_+)$ in Eq.~\eqref{eq:flukes}, the temporal part of the vacuum modes with the decomposition \eqref{eq:separation:kspace} is just an exponential function of time.
From the orthonormality~(\ref{eq:scalar:product:continuum}), we see that
the vacuum modes (indicated by an extra superscript $F$) are
\begin{align}
\label{eq:Deltaphivac}
\Delta\phi^{F,\theta}_{\{\kappa_\theta,\vec k\}}
=\exp(i\kappa_\theta t)\exp(i \vec k\cdot \vec x)\,,
\quad\text{ such that }\quad
\widetilde{\Delta\phi}^{F,\theta}_{\{\kappa_\theta,\vec k\}}\Delta\phi^{F,\theta}_{\{\kappa_\theta,\vec k\}}\equiv 1.
\end{align}
Note that for a given $\kappa_\theta$, the eigenvalue for the vacuum mode as well as for the mode
around the background $\phi^\theta_a$ is given by Eq.~(\ref{eq:lambdaosc}). In view of Eq.~\eqref{eq:Deltaphivac}, it is useful to define
\begin{align} 
\label{beta:bounce}
{\cal B}^\theta_{\{\kappa_\theta,\vec k\}}(t,\vec x)=\widetilde{\Delta\phi}^{\theta}_{\{\kappa_\theta,\vec k\}}\Delta\phi^{\theta}_{\{\kappa_\theta,\vec k\}}-1
\end{align}
as the factor that isolates the solitonic background contribution from that of the vacuum
in the integrand of Eq.~(\ref{eq:log:det}).
In the ultraviolet, where $\kappa_\theta\to \pm\infty$,
we expect that $\left|{\cal B}^\theta_{\{\kappa_\theta,\vec k\}}(t,\vec x)\right|\sim |1/\kappa_\theta|^n$, where $n$ is a positive integer in a gradient expansion.
Explicitly, we then decompose the continuum contribution into
\begin{align}
\label{eq:detcon}
\underset{\rm cont}{\rm tr}\,\log{\cal M}^\theta_a=&
\int \D t \int \D^3 \vec{x}
\int\limits\frac{\D\kappa_\theta}{2\pi}\int\frac{\D^3\vec{k}}{(2\pi)^3}\,
\widetilde{\Delta\phi}^{\theta}_{\{\kappa_\theta,\vec k\}}(t,\vec x) \Delta\phi^{\theta}_{\{\kappa_\theta,\vec k\}}(t,\vec x)
\log\lambda(\theta,\kappa_\theta,\vec k)\notag\\
=&\underset{{\rm cont},{\cal F}}{\rm tr}\,\log{\cal M}^\theta_a+\underset{{\rm cont},{\cal B}}{\rm tr}\,\log{\cal M}^\theta_a,
\end{align}
where $\lambda$ are the continuum eigenvalues given by Eq.~(\ref{eq:lambdaosc}).

The first term  yields
\begin{align}
\underset{{\rm cont},{\cal F}}{\rm tr}\,\log{\cal M}^\theta_a=&
\int \D t \int \D^3\vec{x} \int \frac{\D \kappa_\theta \D^3 \vec{k}}{(2\pi)^4}\log\lambda(\theta,\kappa_\theta,\vec k)
\notag\\
=&V^{(4)}\int \frac{\D \kappa_\theta \D^3\vec{k}}{(2\pi)^4} \log \lambda(\theta,\kappa_\theta,\vec{k})= V^{(4)}\int \frac{\D\kappa_\theta \D^3 \vec{k}}{(2\pi)^4} \log (-e^{2i\theta}\kappa_\theta^2+\vec{k}^2+m^2)
\label{eq:logdetFV}
\end{align}
where $V^{(4)}\equiv\int \D t \int \D^3x$ is the real, four-dimensional volume of spacetime,
and where we recall that $m^2$ is effective mass in the asymptotic vacuum:
\begin{align}
 m^2=V''(\phi_+)=V''(\phi_-).
\end{align}
We recall again that $\phi_+=\phi_-$ for the bounce, such that the above equation is trivially satisfied. For the kink soliton, the effective mass at $\phi_+$ and $\phi_-$ coincides because of the $Z_2$ symmetry between the two vacua. The contribution in Eq.~\eqref{eq:logdetFV} is the same as the logarithm of the determinant of the fluctuation operator for the vacuum. In expressions involving ratios of determinants such as Eqs.~\eqref{ZFB-Theta}, \eqref{ZFB-Theta-QFT}, \eqref{ZFB-Euc}, \eqref{eq:GammaQM} or~\eqref{eq:GammaQFT} for the Euclidean or Minkowskian amplitudes (in fact for normalized amplitudes for
any value of $\theta$), these contributions will  cancel out. Being proportional to the four-dimensional volume $V^{(4)}\propto T$, they are not invariant under rotations of the time contour, so that their cancellation is crucial for well-defined physical observables, as discussed at the end of Section~\ref{sec:complexified:path:integral}. In case we calculate
one amplitude without normalizing it by another, the integral~\eqref{eq:logdetFV} must be regularized. A convenient method in the present context may be
Pauli-Villars regularization by a field with mass $M$
because this leads to a vanishing integrand when
$|\kappa_\theta|\to\infty$. We could then proceed evaluating
Eq.~\eqref{eq:logdetFV} by rotating the integration contour in $\kappa_\theta$, such that
\begin{align}
\underset{{\rm cont,reg},{\cal F}}{\rm tr}\,\log{\cal M}^\theta_a
=& V^{(4)}
\int\frac{\D \kappa_\theta \D^3 \vec{k}}{(2\pi)^4}
\log \left(\frac{-e^{2i\theta}\kappa_\theta^2+\vec{k}^2+m^2}{-e^{2i\theta}\kappa_\theta^2+\vec{k}^2+M^2}\right)
\notag\\
=& V^{(4)} ie^{-i\theta} \int\frac{\D \kappa_\theta \D^3 \vec{k}}{(2\pi)^4}
\log \left(\frac{\kappa_\theta^2+\vec{k}^2+m^2}{\kappa_\theta^2+\vec{k}^2+M^2}\right)
\notag\\
\equiv&2ie^{-i\theta} V^{(4)} V_{{\rm CW},E,{\rm reg}}.
\label{eq:logdetFV2}
\end{align}
Here, we have applied the Cauchy theorem and the fact that the contribution from
the arcs at infinity of this regularized integral are vanishing. The
quantity $V_{{\rm CW},E,{\rm reg}}$ is the regularized effective potential that one would obtain in Euclidean space, such that we see that for the false-vacuum contribution the analytic continuation only incurs a phase from the rotation of the infinite time interval, in accordance with an analytic continuation of the Euclidean result by the substitution ${\cal T}\rightarrow ie^{-i\theta}T$. Noticing that the continuum eigenvalues are of the form \eqref{eq:lambdaosc}, the piece~(\ref{eq:logdetFV}) is nothing but the generalization to arbitrary $\theta$ of the usual Coleman-Weinberg potential
evaluated at the false vacuum. For the problem of vacuum decay, this is expected
because for large values of $|t|$ and $|\vec x|$, i.e. almost everywhere in spacetime, the background sits at the false vacuum. Note also that
for $\theta=\epsilon$, expression~\eqref{eq:logdetFV2} corresponds
to $-2 i \sigma$, where $\sigma$ is the phase appearing in Eq.~\eqref{amplitude}
for the false-vacuum amplitude.

The piece in Eq.~\eqref{eq:detcon} that isolates the one-loop terms due to the solitonic background in the effective action is given by
\begin{align}
\underset{{\rm cont},{\cal B}}{\rm tr}\,\log{\cal M}^\theta_a=&
\int \D t \int \D^3 \vec{x}
\int\limits\frac{\D\kappa_\theta}{2\pi}\int\frac{\D^3\vec{k}}{(2\pi)^3}\,
{\cal B}^\theta_{\{\kappa_\theta,\vec k\}}(t,\vec x)
\log\lambda(\theta,\kappa_\theta,\vec k)\notag\\
=&\int \D t \int \D^3 \vec{x}
\int\limits\frac{\D \kappa_\theta}{2\pi}\int\frac{\D^3\vec{k}}{(2\pi)^3}\,
{\cal B}^E_{\{-ie^{i\theta}\kappa_\theta,\vec k\}}(ie^{-i\theta}t,\vec x)
\log\lambda(\theta,\kappa_\theta,\vec k)
\notag\\
=&-ie^{i\theta}\int \D t \int \D^3 \vec{x}
\int\limits\frac{\D\kappa_\theta}{2\pi}\int\frac{\D^3\vec{k}}{(2\pi)^3}\,
{\cal B}^E_{\{-ie^{i\theta}\kappa_\theta,\vec k\}}(t,\vec x)
\log\lambda^E((-ie^{i\theta}\kappa_\theta,\vec k)
\notag\\
=&\int \D t \int \D^3 \vec{x}
\int\limits\frac{\D\kappa_E}{2\pi}\int\frac{\D^3\vec{k}}{(2\pi)^3}\,
{\cal B}^E_{\{\kappa_\theta,\vec k\}}(t,\vec x)
\log\lambda^E(\kappa_E,\vec k).
\label{tracelog:cont:B}
\end{align}
Here, we have again used the Cauchy theorem in order to rotate the contour of the $\kappa_\theta$-integration.
Therefore, the contributions from the arcs at infinity
must vanish. In field-theoretical settings, this generally requires
a regularization of the integral, as discussed above.
In the quantum-mechanical example of Section~\ref{sec:flukes:kink},
we find that ${\cal B}^E_{\{\kappa_\theta,\vec k\}}\sim 1/\kappa_\theta^2$ (cf. Eq.~\eqref{calB:kink})
such that the integrals over the arcs vanish without further ado.
Since in the planar-wall limit, the modes separate according to Eq.~\eqref{eq:separation:kspace},  equation~\eqref{beta:bounce} further simplifies to
\begin{align}
{\cal B}^\theta_{\{\kappa_\theta,\vec k\}}
=\widetilde{\varphi}_{\{\kappa_\theta,\vec k\}}(t) \varphi_{\{\kappa_\theta,\vec k\}}(t)-1,
\end{align}
i.e. this function is independent of $\vec x$.
We can therefore integrate over three-space and obtain
\begin{align}
\underset{{\rm cont},{\cal B}}{\rm tr}\,\log{\cal M}^\theta_a=
V^{(3)}\int \D t
\int\limits\frac{\D\kappa_E}{2\pi}\int\frac{\D^3\vec{k}}{(2\pi)^3}\,
{\cal B}^E_{\{\kappa_\theta,\vec k\}}(t,\vec x)
\log\lambda^E(\kappa_E,\vec k).
\end{align}
This shows that the solitonic contributions to the determinant over the continuum modes are the same as in Euclidean space, and independent of $\cal T$. Together with the result~\eqref{tr:log:discrete} from the discrete modes, this gives the contribution from the bounce to the effective action, which is notably independent of $T$ and $\theta$, in accordance with the classical contribution~\eqref{Bounceaction}. The additional contribution of Eq.~\eqref{tracelog:cont:B} from the continuum modes is identical to the false-vacuum result, and will cancel out when taking ratios of determinants. In summary, our results imply
\begin{align}
\label{eq:ratiodet}
\det{\cal M}^\theta_a =&\, \left.\det{\cal M}^E_a \right|_{{\cal T}\rightarrow ie^{-i\theta}T}, &\hskip-2cm
 \frac{\det {\cal M}^\theta_B}{\det {\cal M}^\theta_F}=&\,\frac{\det {\cal M}^E_B}{\det {\cal M}^E_F}.
\end{align}
As discussed at the end of Sections~\ref{sec:complex:saddles} and~\ref{sec:complexified:path:integral}, physical observables must be independent of  $T$ and $\theta$, which is  achieved if the observables involve ratios as in Eq.~\eqref{eq:ratiodet}. In the tunneling problem, such a ratio of determinants appears for $Z^\theta_{B}[T]/Z^\theta_F[T]$, as given in Eq.~\eqref{ZFB-Theta-QFT}. As we have just shown that the ratio is ${\cal T}$-independent, we can now use the fact that the real Euclidean determinant has a single discrete negative mode, so that we can write (denoting $\partial_E^2\equiv\partial_\tau^2+\vec{\nabla}^2$)
\begin{align}
\label{ZBoverZF:theta}
\frac{Z^\theta_{B}[T]}{Z^\theta_{F}[T]}=\frac{-e^{-i\theta}TV^{(3)}}{2}
\left(\frac{S_E[\phi_B]}{2\pi}\right)^2
e^{-S_E[\phi_B]}\left|\frac{{\det}'[-\partial_E^2+V''(\phi_B)]}{\det[-\partial_E^2+V''(\phi_+)]}\right|^{-1/2}\,.
\end{align}

This matches as expected the Euclidean result under analytic continuation ${\cal T}\rightarrow ie^{-i\theta}T$, when assuming the determinants remain  ${\cal T}$-independent.  However, the result is nontrivial since, as we have seen, the determinants have a dependence on $T$ that is usually hidden by taking the limit $T\to\infty$. Therefore, the explicit derivation of each piece contributing to Eq.~\eqref{ZBoverZF:theta}
presented here corresponds to a more rigorous proof of this expression. As a byproduct, we have gained explicit insight into the fluctuation spectrum on the Lefschetz thimble for saddle points in complex and real time, and we have clarified how the contributions from the continuum spectrum are related to the usual Coleman-Weinberg potential.


\subsection{Spherical geometry}
\label{sec:spherical:geom}

Thus far, in Section~\ref{sec:continuation} we have developed the arguments for
the analytic continuation of the fluctuation modes in the planar-wall limit. In this setting,
temporal and spatial dependencies of the eigenfunctions naturally separate,
what facilitates the analytic continuation into the plane of complex time. Bubbles are, after all, spherical, such that it is in order at least to outline how to carry out the analytic continuation for the important cases where the planar-wall limit does not apply.


First, we still choose labels  $\{\iota\}$ for the eigenfunctions, such that one of them characterizes the asymptotic behaviour in the time direction (decaying or oscillating),
i.e. it can be identified with
$\beta$ or $\kappa_E$ in the previous discussion. Further,
the $\Delta\phi^E_{\{\iota\}}(x)$
must again constitute a complete and orthonormal set of Euclidean eigenfunctions. Note that
only for the planar-wall geometry we can use $\{\kappa_E,\vec k\}$ for this purpose.
For a spherical geometry, we may therefore choose the hyperspherical
angular momenta $\{j,l_1,l_2\}$ (in the notation of e.g. Ref.~\cite{Garbrecht:2015oea})
in addition to $\kappa_E$ or $\beta$. (In that case,  $\kappa_E$ and $\beta$ characterize
the radial oscillations or the decay of the modes about the Euclidean bubble.)
In case this procedure were to be applied to a concrete problem,
the Euclidean modes, that are initially expressed in hyperspherical coordinates,
would have to be written such as to exhibit the explicit dependence on $\tau$. It may then be further advantageous to transform the $\Delta\phi^E_{\{\iota\}}(x)$ to a basis where the time-dependent factor manifestly separates.

Second, the modes $\Delta\phi^E_{\{\iota\}}(x)$ are to be continued according to the above procedure,
i.e. the discrete modes by an analytic continuation in the time variable $\tau\to i e^{-i\theta}t$ only and the
continuum modes by a simultaneous continuation in time and the parameter $\kappa_E\to -i e^{i\theta}\kappa_E$.

Because of the aforementioned complications due to the change from hyperspherical
coordinates to those with an explicit time variable, we have chosen
above to consider the planar-wall limit, where the discussion is simpler
but nonetheless shows the key points about the analytic continuation. Furthermore,
the archetypical example in a quasi-degenerate, quartic potential~\cite{Callan:1977pt}
is of the thin-wall type and it is
the only one known to us where the Euclidean eigenvalue problem
can be fully solved analytically (cf. Ref.~\cite{Garbrecht:2018rqx} for an extensive discussion of the Fubini-Lipatov instanton, which is perhaps the simplest example with
spherical geometry and where only the Green's function but not the spectrum
is known analytically). For these reasons, we also use the archetypical model in
order to  exemplify the analytic continuation of the modes and the
spectrum in the upcoming section.

\section{Examples for the analytic continuation of the fluctuation spectrum}
\label{sec:continuation:examples}

\subsection{Effective action evaluated at a constant  vacuum configuration}

Consider the case where the background $\phi_a^\theta$ is constant, such that the eigenmodes are identical with their asymptotic forms shown in Section~\ref{sec:continuation}.
This means that they are simple plane waves and the spectrum is continuous. 
In Euclidean space in four dimensions, the eigenfunctions are 
\begin{align}
\label{eq:Euclfree}
 \Delta \phi^E_{\{\kappa_E,\vec{k}\}}(\tau,\vec{x})=
 e^{i \hat{k} \cdot \hat{x}},
\end{align}
where $\hat{x}=(\tau,\vec{x})$ is the Euclidean position four-vector and $\hat{k}=(\kappa_E,\vec{k})$. The reciprocal eigenfunctions are simply obtained by substituting $\hat{k}$ with $-\hat{k}$, so that the orthonormality relation is
\begin{align}
\label{eq:normal}
 \int \D^4\hat{x}\,
\widetilde{  \Delta \phi}^E_{\{\kappa_E,\vec{k}\}}(\tau,\vec{x})
 \Delta \phi^E_{\{\kappa'_E,\vec{k}'\}}(\tau,\vec{x})
 =(2\pi)^4 \delta^4(\hat{k}-\hat{k'}).
\end{align}
The eigenvalues are given by $\delta_{mn}\hat{k}^m\hat{k}^n+m^2\equiv \hat{k}^2+m^2$, and the 
logarithmic determinant is
\begin{align}
 \log\det {\cal M}^E_F=V_E^{(4)} \int \frac{\D^4 \hat{k}}{(2\pi)^4}\log (\hat{k}^2+m^2)=\int \D^4\hat{x}\, V_{\rm CW},
\end{align}
where $V_E^{(4)}\equiv\int\D\tau\int\D^3\vec{x}$. That is, it is given by the spacetime integral of the Coleman-Weinberg potential evaluated at the vacuum.

In Minkowski space the eigenfunctions are also improperly normalizable plane waves. In Lorentzian notation $x^\mu=(t,\vec{x})$, $k^\mu=(k^0,\vec{k})$, these are:
\begin{align}
 \Delta \phi^{\theta=0}_{\{k\}}(x)=e^{ik\cdot x}.
\end{align}
Note how the solutions can be obtained from the Euclidean ones in Eq.~\eqref{eq:Euclfree} by replacing $\tau\rightarrow ie^{-i\theta}t$, $\kappa_E\rightarrow -i e^{i\theta}k^0$, with $\theta=0$.
Again, the above Minkowski solutions are normalized as in Eq.~\eqref{eq:normal} and are eigenfunctions of ${\cal M}^{\theta=0}_F$ with eigenvalues $-k^2+m^2$. The determinant is then
\begin{align}
 \log\det {\cal M}^{\theta=0}_F {=} V^{(4)} {ie^{-i\theta}} \int \frac{\D^4 k}{(2\pi)^4}\log (-k^2+m^2),
\end{align}
which, upon regularization of the ultraviolet divergences, can be obtained from the Euclidean result by analytic continuation of $\kappa_E\rightarrow -ie^{i\theta}k^0$ for $\theta=0$.

\subsection{Fluctuation spectrum about an instanton in the double-well potential
and a quantum-mechanical kink}
\label{sec:flukes:kink}



We will apply some of the general developments of this work to the perhaps simplest example of quantum-mechanical tunneling, i.e. the archetypical model from  Ref.~\cite{Callan:1977pt} based on a quartic potential
\begin{align}
\label{potential}
V(x)=-\frac{1}{2}\,\mu^2\,x^2+\frac{g}{3!}\,x^3+\frac{\lambda}{4!}\,x^4+V_0,
\end{align}
where $\mu,\, g,\, \lambda>0$ and $V_0$ is a constant to ensure $V(x_+)=0$ as shown in Figure~\ref{fig:potential}. In the limit $g\to 0$ the vacua become quasi-degenerate. This leaves us in the so-called thin-wall limit~\cite{Coleman:1977py}, where, in the field theoretical case, the size of the bubble wall is small compared to its radius. In the quantum-mechanical case, the quasi-degenerate limit implies that the bounce approximately corresponds to a pair of a kink and an anti-kink. The kink solution is well-known analytically (see Eq.~(\ref{eq:kink})), as it is also true for the fluctuation spectrum about it (Section~\ref{sec:flukes:kink}). This setup is therefore suitable to illustrate the more general aspects of vacuum decay in real time on a case that is analytically tractable.

The kink in the quantum-mechanical quartic potential as well as the bounce
in the archetypical example are described by the same background.
It follows from the Euclidean equation of motion (i.e. Eq.~\eqref{eq:bg} for $\theta=\pi/2$ and vanishing
spatial gradients)
\begin{align}
 \left[-\frac{\D^2}{\D\tau^2}+V'(\phi_a)\right]\phi_a=0
\end{align}
with the potential
\begin{align}
\label{eq:pot}
 V(\phi)=-\frac{1}{2}\mu^2\phi^2+\frac{\lambda}{4!}\phi^4.
\end{align}
The solution asymptotically approaching the two minima is the kink
\begin{align}
\label{eq:kink}
\bar\phi = v\tanh\gamma (\tau-\tau_0),
\end{align}
where $\gamma=\mu/\sqrt{2}$.
It can be considered as a Euclidean saddle point in quantum mechanics, cf. the potential~\eqref{potential}. In field theory, $\tau$ corresponds to the
radial coordinate of a solitonic bubble with a thin wall~\cite{Coleman:1977py}, and it describes bubble nucleation.

Compared to the quantum-mechanical problem, the thin-wall limit for
tunneling in field theory requires the integration over the space of
fluctuations parallel to the wall
(i.e. the $\vec k$-modes in Section~\ref{sec:continuation}).
This leads to ultraviolet divergences that can however be renormalized~\cite{Garbrecht:2015oea}.
To keep this issue aside and to concentrate on the analytic continuation,
we consider in this section the quantum-mechanical kink.
In either case, quantum mechanics or field theory, one may also derive
analytic expressions for Green's functions about the kink that allow for
as systematic perturbation expansion~\cite{Garbrecht:2015oea,Bezuglov:2018qpq,Bezuglov:2019uxg}.

Writing $u=\tanh\gamma(\tau-\tau_0)$, the eigenvalue equation
\begin{align}
\label{eq:fluctQM}
 \left[-\frac{\D^2}{\D\tau^2}+V''(\bar\phi)\right]\Delta\phi^E_{\{\kappa_E,\vec k\}}(\tau)=\lambda\Delta\phi^E_{\{\kappa_E,\vec k\}}(\tau)
\end{align}
becomes
\begin{align}
\label{eq:Legendre}
 \left[\frac{\D}{\D u}(1-u^2)\frac{\D}{\D u}-\frac{\varpi^2}{1-u^2}+6\right]\Delta \phi^E_{\{\kappa_E,\vec k\}}(\tau)=0,\quad \varpi^2=4+(\vec k^2-\lambda)/\gamma^2.
\end{align}
The solutions are the associated Legendre functions of second degree and order $\varpi$:
\begin{align}
 \Delta\phi_{\{\kappa_E,\vec k\}}^E(\tau)=\sqrt{N_E(\varpi)} P_2^\varpi(u),
\end{align}
with $N_E(\varpi)$ a normalization constant.

First, one may note that the effective mass of the scalar field in the true vacuum, approached by the kink at $\tau=\pm\infty$, is given by
\begin{align}
\label{eq:mplus}
m^2_\pm=4\gamma^2{\equiv m^2}.  
 \end{align}
Further, there are  two discrete eigenvalues corresponding to $\varpi=1$ (with a positive eigenvalue $\lambda=3\gamma^2$) and $\varpi=2$ (giving a zero mode, associated with time translations)~\cite{Konoplich:1987yd,Garbrecht:2015oea,Garbrecht:2018rqx}. (For the thin-wall problem, there are two discrete modes for
each $\vec k$.) There is no negative
mode because the kink is not a true bounce or tunneling solution (which should tend to the false vacuum both at $\tau\rightarrow\pm\infty$, while the kink only does so only at positive infinity). Further, there is a continuum of modes for imaginary values of $\varpi$.

To relate this spectrum to the discussion of the asymptotic behaviour
of the modes in Section~\ref{sec:continuation}, we use that
for $\varpi\neq1$, we may express the Legendre functions through Jacobi polynomials as
\begin{align}
 P_2^\varpi(u)=\frac{1}{\cos\pi\frac{\varpi}{2}}\left(\frac{u+1}{u-1}\right)^{\frac{\varpi}{2}}(3-\varpi)_\varpi P_2^{-\varpi,\varpi}(u),\quad |u|<1,\,\varpi\neq1.
\end{align}
The asymptotic expansions for $\tau\rightarrow\pm\infty$---corresponding to $u=\pm1$---are
\begin{equation}\label{eq:asympkink}\begin{aligned}
 P_2^\varpi(\tau)\sim&\,\frac{1}{2\cos\pi\frac{\varpi}{2}} (-1)^{\frac{\varpi}{2}}e^{\gamma\varpi\tau}(3-\varpi)_\varpi(1-\varpi) (2-\varpi),\,\,\tau\rightarrow\infty,\\
 P_2^\varpi(\tau)\sim&\,\frac{1}{2\cos\pi\frac{\varpi}{2}} (-1)^{\frac{\varpi}{2}}e^{\gamma\varpi\tau}(3-\varpi)_\varpi(1+\varpi) (2+\varpi),\,\,\tau\rightarrow-\infty.
\end{aligned}\end{equation}
For $\varpi=1$, one has $P_2^1(u)=-3u\sqrt{1-u^2}$, which gives
\begin{equation}
\label{eq:asympkink2}
\begin{aligned}
 P_2^1(\tau)\sim&\,-6e^{-\gamma\tau},\,\,\tau\rightarrow\infty,\\
 P_2^1(\tau)\sim&\,6e^{\gamma\tau},\,\,\tau\rightarrow-\infty.
\end{aligned}
\end{equation}
Based on this, we recover for $\varpi=1,2$ a suppressed behaviour at both ends, in accordance with the fact that we are dealing with discrete modes. There are no other values of $\varpi$ for which 
we obtain a decaying behaviour for both $\tau=\pm\infty$, which confirms that there are only two discrete modes. Furthermore, the continuum spectrum can only come
from oscillating solutions, which demand complex $\varpi$:
\begin{align}
 \label{eq:varpiE}
 \varpi=\frac{i \kappa_E}{\gamma}=\frac{2i\kappa_E}{m}, \kappa_E\in\mathbb{R}.
\end{align}
In this case the continuum eigenvalues are
\begin{align}
\label{eq:contE}
 \lambda=\gamma^2(4-\varpi^2)=m^2+\kappa_E^2,
\end{align}
where we have used Eq.~\eqref{eq:mplus}. This is in accordance with our general result for the Euclidean  eigenvalues in the continuum, Eq.~\eqref{eq:lambdacontE} for $\vec k=0$. The continuum eigenfunctions are then given by
\begin{align}
\label{eq:Realeigenkink}
 \Delta\phi^E_{\kappa_E}(u)=\left.{\cal N}_E(\kappa_E) \,P_2^\varpi(u)\right|_{\varpi=\frac{2i}{m}\kappa_E},
\end{align}
where ${\cal N}_E$ is a normalization constant.
The reciprocal eigenfunctions are obtained by simply changing the sign of $\kappa_E$, as follows from the identity
\begin{align}
 \int_{-1}^1\frac{\D u}{1-u^2}P^{i\xi}_2(u)P^{-i\xi'}_2(u)=\frac{2\sinh \pi\xi}{\xi}\delta(\xi-\xi'),
\end{align}
which implies that our eigenfunctions \eqref{eq:Realeigenkink} satisfy
\begin{align}
\int_{-\infty}^\infty \D\tau  \Delta\phi^{E,-}_{\kappa_E}\phi(\tau) \Delta{\phi}^{E,+}_{\kappa'_E}\phi(\tau)=\frac{m^2}{2\kappa}{\cal N}^2_E(\kappa_E)\sinh\frac{2\kappa_E\pi}{m}\,\delta(\kappa_E-\kappa'_E)\overset{!}{=}2\pi\delta(\kappa_E-\kappa'_E).
\end{align}
Imposing the normalization~(\ref{eq:scalar:product:continuum}) in the last equality
fixes
\begin{align}
\label{eq:NE}
 {\cal N}_E(\kappa_E)=\sqrt{\frac{4\pi \kappa_E}{m^2\sinh\frac{2\pi \kappa_E}{m}}}.
\end{align}

Now, for a rotated time contour, we can rewrite the equation for the fluctuations in terms of a variable
\begin{align}
u_\theta = \tanh [\gamma ie^{-i\theta}(t-t_0)]\,.
\end{align}
The resulting equation is identical to Eq.~\eqref{eq:Legendre}, after substituting $u$ with $u_\theta$. Thus its solutions will be the associated Legendre functions evaluated at $u_\theta$. This is equivalent to the analytic continuation of the Euclidean solutions with the substitution $\tau\rightarrow ie^{-i\theta}t$.
The asymptotic expansions can then be obtained from Eqs.~\eqref{eq:asympkink} and \eqref{eq:asympkink2} by the same analytic continuation, giving for $\varpi\not=1$
\begin{equation}
\label{eq:asympkink3}
\begin{aligned}
 P_2^\varpi(ie^{-i\theta}t)\sim&\,\frac{1}{2\cos\pi\frac{\varpi}{2}} (-1)^{\frac{\varpi}{2}}e^{\gamma\varpi\sin\theta t}e^{i\gamma\varpi\cos\theta t}(3-\varpi)_\varpi(1-\varpi) (2-\varpi),\,\,t\rightarrow\infty,\\
 P_2^\varpi(ie^{-i\theta}t)\sim&\,\frac{1}{2\cos\pi\frac{\varpi}{2}} (-1)^{\frac{\varpi}{2}}e^{\gamma\varpi\sin\theta t}e^{i\gamma\varpi\cos\theta t}(3-\varpi)_\varpi(1+\varpi) (2+\varpi),\,\,t\rightarrow-\infty,\\
 \end{aligned}
 \end{equation}
 and for $\varpi=1$
 \begin{equation}\label{eq:asympkink4}\begin{aligned}
 P_2^1(ie^{-i\theta}t)\sim&\,-6e^{-\gamma\sin\theta t}e^{-i\gamma\cos\theta t},\,\,t\rightarrow\infty,\\
 P_2^1(ie^{-i\theta}t)\sim&\,\,6e^{\gamma\sin\theta t}e^{i\gamma\cos\theta t},\,\,t\rightarrow-\infty.
\end{aligned}\end{equation}
Again, for $\varpi=1,2$ the solutions decay at infinite time for $\theta>0$ and therefore are
legitimately discrete modes. For asymptotically oscillatory solutions, we need the
phase of $\varpi$ to be given by
\begin{align}
\label{varpi:theta}
 e^{-i\theta}\gamma\varpi= \kappa_\theta,\,\kappa_\theta\in\mathbb{R},\Rightarrow \varpi = \frac{e^{i\theta}\kappa_\theta}{\gamma}=\frac{2e^{i\theta}\kappa_\theta}{m}.
\end{align}
These values of $\varpi$ can be obtained from the corresponding Euclidean ones in Eq.~\eqref{eq:varpiE} by an analytic continuation $\kappa_E\rightarrow -ie^{i\theta}\kappa_\theta$, as expected from our general arguments.
The normalized continuum eigenfunctions follow from Eq.~(\ref{continuum:eigenfunction:theta}) as
\begin{align}
\label{Delta:phi:kink:theta}
\Delta \phi_\theta^{\kappa_\theta}(t)={\cal N}_E(-i e^{i\theta}\kappa_\theta)P_2^\varpi(ie^{-i\theta} t).
\end{align}
The continuum eigenvalues can be obtained by applying the same substitution to Eq.~\eqref{eq:lambdacontE}, which gives a result agreeing with Eq.~\eqref{eq:lambdaosc}.

\subsection{Functional determinant of the kink}
\label{sec:det:kink}

Using the eigensystem discussed above,
we now calculate the fluctuation determinant of the kink
for general $\theta$ following the procedure explained in Section~\ref{sec:continuation:determinant}.
First, the two discrete modes associated with $\varpi=1,2$ have the eigenvalues
$\lambda=3\gamma^2,0$, respectively. The zero eigenvalue is dealt with by a volume
integration as in Eq.~(\ref{Zeromodemeasure}). We are hence left with
\begin{align}
\underset{\rm disc}{{\rm tr}'}\log {\cal M}^\theta_K=\log 3\gamma^2=\log \frac32\mu^2,
\end{align}
the prime indicates that we have omitted the zero eigenvalue. We note that according
to our general arguments, the discrete eigenvalues are independent of $\theta$.

The vacuum contribution to the fluctuation determinant is given
by the Coleman-Weinberg form. It cancels
when normalizing with the determinant of the solution
$x^\theta_F(t)=x_+={\rm const.}$ and therefore requires
no further evaluation in the present context.

Substituting the continuum eigenfunctions~(\ref{Delta:phi:kink:theta})
into Eq.~(\ref{beta:bounce}), we obtain the factor
\begin{align}
\label{calB:kink}
{\cal B}^\theta_{\kappa_\theta}(t,\vec x)=\frac{3( u_\theta^2-1)(1+3 u_\theta^2 -\varpi^2)}{(1-\varpi^2)(4-\varpi^2)}
\end{align}
that appears in the integrand of the bounce contribution to the fluctuation determinant and where
$\varpi$ is given by Eq.~(\ref{varpi:theta}).
Now from Eq.~(\ref{tracelog:cont:B}), we know that the part of the determinant arising
from this factor is independent of $\theta$ such that it is then simplest to evaluate
it in Euclidean time $\theta=\pi/2$. The temporal integration
in Eq.~(\ref{tracelog:cont:B}) then yields
\begin{align}
\frac{1}{\gamma}\int\limits_{-1}^1 \frac{\mathrm{d}u}{1-u^2}\frac{3(u^2-1)(1+3  u^2 -\varpi^2)}{(1-\varpi^2)(4-\varpi^2)}=-\frac{6}{\gamma} \frac{2-\varpi^2}{(1-\varpi^2)(4-\varpi^2)}
\end{align}
and the trace over the eigenvalues
\begin{align}
\underset{{\rm cont},{\cal B}}{\rm tr}\,\log {\cal M}^\theta_K=
\int\limits_{-\infty}^\infty\frac{\mathrm{d}\kappa_E}{2\pi}
\left(-\frac{6}{\gamma} \frac{2-\varpi^2}{(1-\varpi^2)(4-\varpi^2)}\right)\log\left(\gamma^2(4-\varpi^2)\right)=-2\left(\log\mu^2 +\log 6\right),
\end{align}
where the Euclidean $\varpi$ is given in Eq.~(\ref{eq:varpiE}).

In total, we arrive at the result
\begin{align}
\label{Eq532}
{\rm tr}'\,\log {{\cal M}}^\theta_K=&
\underset{{\rm disc}}{\rm tr}'\,\log {{\cal M}}^\theta_K+\underset{{\rm cont},B}{\rm tr}\,\log {\cal M}^\theta_K+\underset{{\rm cont},F}{\rm tr}\,\log {\cal M}^\theta_K
\notag\\
=&-\log \mu^2 - \log 24+ie^{-i\theta}T\int_{\rm cont} \frac{\mathrm{d}\kappa_E}{2\pi} \log \lambda_E(\kappa_E,\vec{k}).
\end{align}
Note that the last term is the same as the false-vacuum contribution, such that we have
\begin{align}
\label{Eq533}
{\rm tr}'\log {\cal M}^\theta_{K}-{\rm tr}\log {\cal M}_{F}^\theta=-\log 24\mu^2.
\end{align}
 In Appendix~\ref{app:methods:determinant}, we compare this calculation with a number of
additional methods for computing the functional determinant.
\section{The decay rate from the Minkowski path integral}
\label{sec:Minkowskian:decay:path:integral}


Recalling Eqs.~\eqref{amplitude},~\eqref{eq:FVtheta}, one can relate the partition function $Z^\epsilon_{FB}[T]$ on the integration
cycle through the false vacuum with the transition matrix $M$ of the time-evolution operator ${\cal U}(T) =\exp(i\sigma) \mathbf{1}+iM(T)$:
\begin{align}
 {\cal N}^2 Z^\epsilon_{FB}[T]=e^{i\sigma}\langle {\rm FV}|{\rm FV}\rangle+ \langle{\rm FV}| iM(T) |{\rm FV}\rangle.
\end{align}
Using expression \eqref{eq:ZthetaFBexp}, we may expand $Z^\epsilon_{FB}[T]$ for small $Z^\epsilon_{B,\text{Gau{\ss}ian}}[T]/Z^\epsilon_{F,\text{Gau{\ss}ian}}[T]$
such that
\begin{align}
\label{eq:ZM}
 {\cal N}^2 Z^\epsilon_{F,\text{Gau{\ss}ian}}[T]+{\cal N}^2 Z^\epsilon_{B,\text{Gau{\ss}ian}}[T]\approx e^{i\sigma} \langle{\rm FV}|{\rm FV}\rangle + \langle {\rm FV}| iM (T)|{\rm FV}\rangle.
\end{align}
Since the wave function for $|{\rm FV}\rangle$ is dominated by a contribution corresponding 
to a local ground state at the metastable minimum of the potential, the amplitude ${\cal N}^2 Z^\epsilon_{F,\text{Gau{\ss}ian}}[T]$ is in the  Gau{\ss}ian approximation related with $\langle{\rm FV}|{\rm FV}\rangle$ by a pure phase. Recall that we have isolated this phase in Section~\ref{sec:continuation:determinant}.
Thus, we may identify
\begin{align}
\label{eq:N}
\mathcal{N}^2\;Z_{F,\text{Gau{\ss}ian}}^{\epsilon}[T]=e^{i\sigma}\langle{\rm FV}|{\rm FV}\rangle,
\end{align}
so that Eq.~\eqref{eq:ZM} gives 
\begin{align}
\frac{\langle {\rm FV}| i e^{-i\sigma} M(T) |{\rm FV}\rangle}{\langle{\rm FV}|{\rm FV}\rangle}=\frac{Z_{B,\text{Gau{\ss}ian}}^{\epsilon}[T]}{Z_{F,\text{Gau{\ss}ian}}^{\epsilon}[T]}.
 \end{align}
In accordance to the optical theorem of Eq.~\eqref{opticaltheoremforfalsevacuum2}, the total decay probability is given by
\begin{align}
\label{ImFvtoFv}
p_{FV\rightarrow \text{ all}}[T]=2\,{\rm Im}\frac{\langle \text{FV}|e^{-i\sigma}M(T)|\text{FV}\rangle}{\langle{\rm FV}|{\rm FV}\rangle}=-2\,{\rm Re}\left(\;\frac{{Z}_{B,\text{Gaussian}}^{\epsilon}[T]}{Z_{F,\text{Gaussian}}^{\epsilon}[T]}\right).
\end{align}

As we have shown in Sections~\ref{sec:MinkowskiPathIntegral} and~\ref{sec:continuation}, at NLO, this amplitude can simply be obtained from the Euclidean result through the replacement $\mathcal{T}\rightarrow iT$, and taking into account that the ratio of fluctuation determinants becomes ${\cal T}$-independent (the regulator $\epsilon$ can be taken all the way to zero here). In the quantum-mechanical case we can use the result \eqref{ZFB-Theta2}, while for quantum field theory we may use Eq.~\eqref{ZFB-Theta-QFT}; the result is
\begin{align}
\label{prob1bounce}
p_{FV\rightarrow \text{ all}}[T]=\left\{\begin{array}{c}
T\sqrt{\frac{S_E
[x_B]}{2\pi}}\left|\frac{\det'[-\partial_\tau^2+V''({x_B})]}{\det[-\partial_\tau^2+V''(x_+)]}\right|^{-1/2}\;e^{-S_E[x_B]},\quad \text{QM}\\
TV^{(3)}\left(\frac{S_E
[\phi_B]}{2\pi}\right)^2\left|\frac{\det'[-\partial_\tau^2-\vec{\nabla}^2+V''({\phi_B})]}{\det[-\partial_\tau^2-\vec{\nabla}^2+V''(\phi_+)]}\right|^{-1/2}\;e^{-S_E[\phi_B]},\quad \text{QFT}
                                        \end{array}\right..
\end{align}
From this we find the quantum-mechanical decay rate $\varGamma^{QM}$ and the field-theoretical decay rate per unit volume $\varGamma^{QFT}_V$,
\begin{align}
\label{eq:GammaQM}
\varGamma^{QM}=&\,\sqrt{\frac{S_E
[x_B]}{2\pi}}\left|\frac{\det'[-\partial_\tau^2+V''({x_B})]}{\det[-\partial_\tau^2+V''(x_+)]}\right|^{-1/2}\;e^{-S_E[x_B]},\\
\label{eq:GammaQFT}
\varGamma^{QFT}_V=&
\left(\frac{S_E
[\phi_B]}{2\pi}\right)^2\left|\frac{\det'[-\partial_\tau^2-\vec{\nabla}^2+V''({\phi_B})]}{\det[-\partial_\tau^2-\vec{\nabla}^2+V''(\phi_+)]}\right|^{-1/2}\;e^{-S_E[\phi_B]},
\end{align}
which match the classic result obtained in Euclidean space by Callan and Coleman~\cite{Coleman:1977py,Callan:1977pt}. Note that in Callan's and Coleman's derivation of the decay rate, based on identifying the imaginary part of the energy density in the false-vacuum state (see Appendix~\ref{sec:Callan-Coleman}), the sum over multi-bounce saddle points becomes crucial. This is not the case when calculating the probability of decay through the optical theorem, as the expansion in small $Z^\epsilon_{B,\text{Gau{\ss}ian}}[T]/Z^\epsilon_{F,\text{Gau{\ss}ian}}[T]$ which led to \eqref{eq:ZM} from Eqs.~\eqref{eq:ZthetaFBexp},~\eqref{amplitude} and~\eqref{eq:FVtheta} is equivalent to only considering the single bounce contribution to the false-vacuum partition function.


We emphasize that in order to apply the optical theorem~\eqref{opticaltheoremforfalsevacuum2},
it is necessary that the Minkowskian ratio $Z^\epsilon_{B,\text{Gau{\ss}ian}}[T]/Z^\epsilon_{F,\text{Gau{\ss}ian}}[T]$ is real. Arguably, this follows from the simple analytic continuation of the Euclidean prefactor
$i{\cal T}\to -T$. Nonetheless, this na\"{i}ve substitution is only justified when ignoring the ${\cal T}$-dependence of the classical action and the ratio of determinants;  a $T\to\infty$ limit is implicit here and we have
proved this result in detail at the level of the classical action and of the fluctuation spectrum
at a complex saddle point on the pertaining Lefschetz thimble. This way, it is also further clarified in what sense an instanton describes tunneling in real time.

%
In Eq.~\eqref{prob1bounce} we obtain a probability which is linear in time, while the usual decay behaviour takes the form of an exponential law $p{=}(1-\exp(-\varGamma T))$, as one can derive in the Callan-Coleman formalism. There is, however, no contradiction. The reason for our approximation using the optical theorem being linear in $T$ is that we have accounted only for the single bounce when expanding $Z^\epsilon_{FB}[T]$ for small $Z^\epsilon_{B,\text{Gau{\ss}ian}}[T]/Z^\epsilon_{F,\text{Gau{\ss}ian}}[T]$. Accordingly, one should also expand $\exp(-\varGamma T){\approx} 1-\varGamma T$ such that we have $p {\approx} \varGamma T$ which matches our result. Alternatively, we
may obtain the exponential decay law by following the argument by Callan and Coleman~\cite{Callan:1977pt} based on multi-bounces. The intuitive picture is the following: The particle, initially trapped in the false vacuum, can penetrate into the barrier region between the turning point $\mathbf{p}$ (see Figure~\ref{fig:potential} and $x_+$ due to its quantum nature. Every single complex bounce describes a collision of the particle with the outer boundary of the barrier region. And the probability in Eq.~\eqref{prob1bounce} is the escape probability for the particle to penetrate outside of the barrier for one single collision. Namely, the surviving probability is $1-{p}$ which is the first order expansion of $e^{-p}$. During the history, collisions happen again and again, which leads to an exponential survival probability $e^{-{\varGamma}\,T}$. The strict derivation of this statement can be carried out by repeating the argument in Ref.~\cite{Callan:1977pt} which will be omitted here.

\section{Discussions and conclusions}
\label{sec:conclusions}

In this paper, we have applied Picard-Lefschetz theory to false-vacuum decay in real time and more generally in rotated complex time characterized by a phase angle $\theta$. One motivation is the use of the real-time amplitude
in an optical theorem for tunneling, that leads us to the decay rate
through a route that is alternative to calculating the imaginary part of the
ground-state energy at the false vacuum, or solving for wave functions in the WKB approximation. Using these
different methods in explicit computations of the decay rate for the archetypical example of a quasi-degenerate double-well potential to one-loop accuracy, we illustrate in what way these approaches are related. The interpretation of the real-time false-vacuum to false-vacuum amplitude in terms of the optical theorem provides a new relation between the functional techniques of calculating decay rates
to real-time dynamics. Alternative descriptions in real-time are of course given by the
well-known picture of tunneling through a wave function that penetrates the barrier, that is accounted
in the present paper by the WKB construction in Appendix~\ref{app:sec:wkb}, or by
a distribution of classical paths~\cite{Braden:2018tky} that only leads to approximate results however~\cite{Hertzberg:2019wgx}.

While in Ref.~\cite{Cherman:2014sba} it has been observed that the complex saddle point obtained by Wick rotation of the Euclidean instanton may recover the instanton physics in real time, we focus here on the complex bounce in the vacuum decay problem~\cite{Turok:2013dfa}.
Even though it is not difficult to see that the classical action is invariant under the rotation, proving the equivalence between the Euclidean path integral around the real bounce and the Minkowski integral around its complex continuation is much more difficult than one may expect. Ultimately, this is due to the fact that, even though the Euclidean and Minkowski saddle points can be related through analytic continuation, this is not true for their associated  steepest descent integration contours, because the flow equations that define them  are not holomorphic. In order to prove the equivalence of the path integrals we have made several theoretical developments in this work. First, we have transferred the flow eigenproblem to an eigenproblem in the proper sense, building on the developments in Ref.~\cite{Tanizaki:2014xba}. Based on this relation, we have expressed at the Gau{\ss}ian level the path integral on a Lefschetz thimble through the determinant of the quadratic fluctuation operator continued to rotated time. We have then investigated the
continuation of the fluctuation spectrum, i.e. of the eigenmodes and the
eigenvalues, under rotation of time. It turns out that discrete and
continuum modes behave in a crucially different way and require careful
distinction. Eventually, we have proved that the fluctuation functional determinant around the complex bounce can be obtained from the Euclidean one around the normal bounce via a Wick rotation of the time interval, $\mathcal{T}\rightarrow i e^{-i\theta}T$. Arriving at this result has been less obvious than one would na\"{i}vely expect. In particular, we have made a spectral decomposition of the logarithmic determinant and separated the finite contribution, which is independent of $T$, from a part proportional to the volume of spacetime and hence to $T$. We have observed that the latter piece, which only receives contributions from the continuum spectrum, turns out to be equal to the logarithmic determinant in the asymptotic false vacuum, which is itself related to the usual Coleman-Weinberg potential. This $T$-dependent contribution is then canceled when normalizing the functional determinant by the false-vacuum result. In effect, in the result of the normalized path integral {of the fluctuations about the bounce saddle point,} the Wick rotation only affects the integration over the collective coordinate associated with the spontaneous breakdown of time-translation invariance. This gives a real-time vacuum to vacuum transition amplitude from which one can recover the decay rate using the optical theorem. The result matches the one derived by Callan and Coleman in Euclidean spacetime from considering the imaginary part of the ground-state energy; an interesting difference between the two derivations is that the one using the optical theorem does not rely on summing over multi-bounce configurations. 


To check and illustrate our developments, we have considered the spectrum
of fluctuations in the kink background and its analytic continuation under
rotations of time. We have further computed the functional determinant for the kink in terms of the spectrum, and have compared it with the results from additional methods based on the Gel'fand-Yaglom theorem and on the resolvent of the fluctuation operator. For completeness, we have also reviewed the derivation of the decay rate using the WKB method and shown agreement between all these rather different approaches.

Our work may have applications in the following directions. The transformation between the flow eigenproblem and the proper eigenproblem may turn out to be important in applying Picard-Lefschetz theory to additional problems,
where a real-time description may be of interest, e.g. for QCD instantons.
{Real-time techniques may also be the only way of addressing
vacuum transitions in backgrounds that cannot be Euclideanized, such
as nonequilibrium systems or curved spacetimes~\cite{Brown:2017wpl}. The present work
may serve as the basis for treating such problems to one-loop accuracy
and beyond.} 
Also, by the optical theorem, the decay of the false vacuum is described as the sum of all the possible transitions from the false vacuum to the nucleated field configurations, among which the critical bubble should give the dominant contribution. The precise shape of the probability distribution of the nucleated configurations and its possible phenomenological consequences will be investigated in future work. 

\acknowledgments

WYA is supported by the China Scholarship Council and thanks Juan S. Cruz and Hong Zhang for helpful discussions. BG is grateful to
the Physics Department of the University of Wisconsin-Madison for
generous hospitality during the completion of this work. CT thanks Guillermo Ballesteros and Jos\'e R. Espinosa for valuable input. BG and CT acknowledge support  by the Collaborative Research Centre SFB1258 of the Deutsche
Forschungsgemeinschaft (DFG).

\appendix

\section{Review of the evaluation of the path integral for quantum-mechanical tunneling in Euclidean time}
\label{sec:Callan-Coleman}

\subsection{Euclidean path integral}

In this appendix we review how the path integral is evaluated in the theory of false-vacuum decay
due to Callan and Coleman~\cite{Coleman:1977py,Callan:1977pt}. For simplicity, we focus on the case of quantum mechanics; this serves us to
introduce the notation used in this work and to provide reference formulae. Furthermore, it allows us to contrast the calculation in Euclidean time with the alternative approach where one remains in Minkowski space, the main subject of the present paper.
For this purpose, we also find it useful to recapitulate how the imaginary energy of the false vacuum can emerge and be understood from the evaluation of the path integral based on Picard-Lefschetz theory~\cite{Andreassen:2016cvx}.

We consider the archetypical model from  Ref.~\cite{Callan:1977pt} of quantum-mechanical tunneling in a quartic potential. In Minkowski spacetime, the action is 
\begin{align}
\label{Lagrangian}
S_M=\int \D t\,\mathcal{L}_M=\int \D t\,\left[\frac{1}{2}\left(\frac{\D x}{\D t}\right)^2-V(x)\right]\,,
\end{align}
where the potential $V(x)$ is given in Eq.~(\ref{potential}).
Though we work in quantum mechanics, we still call $x(t)$ field and the ground (resonant) states around $x_-$ and $x_+$ the true vacuum and false vacuum, respectively.

Instead of working with the transition amplitude in Minkowski space that would directly lead to the tunneling rate, Callan and Coleman consider the Euclidean amplitude
\begin{align}
\label{EuclideanPropagator}
D(x_+,\mathcal{T}/2;x_+,-\mathcal{T}/2)=\langle x_+|e^{-H\mathcal{T}}|x_+\rangle=\int \mathcal{D}x(\tau) \,e^{-S_E[x(\tau)]}\equiv Z^E[\mathcal{T}]\,,
\end{align} 
where we define the Euclidean Lagrangian $\mathcal{L}_E=\frac{1}{2}\left(\frac{\D x}{\D\tau}\right)^2+V(x)$ that appears in the action $S_E=\int \D\tau\,\mathcal{L}_E$. All trajectories contributing to the path integral have the boundary conditions $x(-\mathcal{T}/2)=x(\mathcal{T}/2)=x_+$. Note that in the Euclidean Lagrangian, the potential appears upside down compared to the original one. As shown in Eq.~\eqref{relation}, for large $\mathcal{T}$, Eq.~\eqref{EuclideanPropagator} shall give us the information on the lowest-lying energy eigenvalue and its wave function. 

We can evaluate the path integral~\eqref{EuclideanPropagator} using the method of steepest descent. The stationary configurations are given by the equation of motion
\begin{align}
\label{EuclideanEoM}
\frac{\D^2 x(\tau)}{\D\tau^2}-V'(x(\tau))=0
\end{align} 
subject to the Dirichlet boundary conditions $x(-\mathcal{T}/2)=x(\mathcal{T}/2)=x_+$. Here, the prime denotes a derivative with respect to $x$. When we work with real paths $x(\tau)$, this equation describes the motion of a classical particle released at the local maximum $x_+$ of the potential $-V(x)$ at $-\mathcal{T}/2$, and returning to $x_+$ at $\mathcal{T}/2$. In the limit $\mathcal{T}\rightarrow\infty$, we have three types of solutions:\footnote{{\label{fn:Tdependence}Even though  we have taken the limit $\mathcal{T}\rightarrow\infty$ in order to obtain simple expressions for the saddle points, at some other instances in this work, in order to understand the analytic continuation between the Euclidean and Minkowskian configurations, we shall take the view that $\mathcal{T}$ is large but finite. That is, we formally keep the dependence on $\mathcal{T}$ in the quantities that are to be continued analytically. Similar strategies have been employed in the literature \cite{Callan:1977pt}. 
}} the trivial false-vacuum solution $x_F(\tau)\equiv x_+$, the {\it  bounces}  $x_{B_n}(\tau)$\footnote{
The effect of multi-bounces can be accounted for by exponentiating the single-bounce contribution to the path integral~\cite{Callan:1977pt,Plascencia:2015pga,Ai:2018rnh}, as discussed further below in the text.}---which bounce back and forth from the false vacuum $n$ times, with zero initial velocity---and a third one called the {\it shot} in Refs.~\cite{Andreassen:2016cff,Andreassen:2016cvx}, $x_S(\tau)$. The false vacuum and the single bounce ($B_1\equiv B$ and $ x_{B_1}(\tau)\equiv x_B(\tau)$) are well-known from Ref.~\cite{Coleman:1977py}. To understand the shot, note that the particle may be released from  $x_+$ with a nonvanishing initial velocity in such a way that it arrives at the higher maximum $x_-$ with asymptotically vanishing velocity and eventually rolls back to $x_+$. The requirement that the particle must stop exactly at $x_-$ instead of some intermediate point originates from the condition $\mathcal{T}\rightarrow \infty$ for the motion from $x_-$ back to  $x_+$.

Now we can expand the path integral in Eq.~\eqref{EuclideanPropagator} around these three types of saddle points. 
Writing $x(\tau)=x_a(\tau)+\Delta x_a(\tau)$, where $a=F, B_n,S$, we obtain then
\begin{align}
\label{Expandedpathintegral}
&\langle x_+|e^{-H\mathcal{T}}|x_+\rangle\notag\\
&\approx \sum_a\left(e^{-S[x_a]}\int\mathcal{D}
\Delta x_a\, e^{-\int_{-\mathcal{T}/2}^{\mathcal{T}/2}d\tau\left[\Delta x_a(\tau)\left(
-\frac{1}{2}\frac{d^2}{d\tau^2}+\frac{1}{2}V''(x_a(\tau))\right)\Delta x_a(\tau)+\frac{1}{3!}(g+\lambda\,x_a(\tau))\Delta x_a^3(\tau)+\frac{\lambda}{4!}\Delta x_a^4(\tau)\right]}\right)\notag\\
&\equiv Z^E_{F}+{\sum_n Z^E_{B_n}}+Z_S
\end{align}  
It turns out that the bounce fluctuation operator $(-\partial_\tau^2+V''(x_B))$  contains a negative eigenvalue, denoted by $\lambda^B_{0}<0$. {Similar negative eigenvalues are present for the multi-bounce fluctuation operators. (With the multi-bounce given by infinitely separated bounces, there are negative modes corresponding to each of the single bounces, {i.e., there are $n$ negative modes associated with the saddle point $x_{B_n}$}.)} Therefore, a na\"ive Gau{\ss}ian integration in the perturbative expansion of the second line yields an ill-defined result. However, this is not a problem of the
underlying theory but is due to an incorrect application of the method of steepest descent. The directions associated with the negative eigenvalues are in fact {\it not} of steepest descent but rather of steepest {\it ascent}.

\subsection{Contour integration in field space and flow equations}

In order to make appropriate use of the method of steepest descent, we need to complexify the paths $x(\tau)$ to $z(\tau)$ and then perform the path integral on a middle-dimensional contour.\footnote{Specifically, this means that we integrate over a manifold whose dimensionality is one half of the complexified (infinite-dimensional) field space.} For multiple-dimensional integrals as well as their generalization to path integrals this approach is known as Picard-Lefschetz theory (see e.g., Refs.~\cite{Witten:2010cx,Witten:2010zr}). To frame this discussion within a general context, we denote the holomorphic function appearing in the exponential of the integrand as $ -S_E[z]\equiv \mathcal{I}[z]$ and we define the Morse function $h[z]=\mbox{Re}(\mathcal{I}[z])$. The saddle points satisfy the equation of motion $\delta\mathcal{I}[z]=0$ subject to the boundary conditions of interest. For a saddle point $z_a$
of $\mathcal{I}[z]$, one can find a downward flow (the steepest descent path) according to the gradient flow equation~\cite{Witten:2010cx}
\begin{align}
\label{flow}
\frac{\partial z(\tau;u)}{\partial u}=-\overline{\left(\frac{\delta {\mathcal{I}[z(\tau;u)]}}{\delta {z(\tau;u)}}\right)}\;,\  \frac{\partial \overline{z(\tau;u)}}{\partial u}=-\frac{\delta {\mathcal{I}[z(\tau;u)]}}{\delta z(\tau;u)},
\end{align}
{where $u\in\mathbb{R}$ and the boundary condition is $z(\tau;u=-\infty)=z_a{(\tau)}$.}
One can easily check that
\begin{align}
\frac{\partial h}{\partial u}=\frac{1}{2}\left(\frac{\delta\mathcal{I}}{\delta z}\cdot\frac{\partial z}{\partial u}+\frac{\delta\overline{\mathcal{I}}}{\delta \overline{z}}\cdot\frac{\partial\overline{z}}{\partial u}\right)=-\left|\frac{\partial z(\tau;u)}{\partial u}\right|^2\leq 0.
\end{align}
That is, the real part of $\mathcal{I}[z]$ is decreasing when we move away from the saddle point along the contour given by $z(\tau;u)$. Further, one can show that $\partial {\rm Im}\mathcal{I}[z(\tau;u)]/\partial u=0$, meaning that the phase is constant on that contour. All the steepest descent flows generated from a saddle point $z_a$ constitute the so-called Lefschetz thimble, denoted by $\mathcal{J}_a$~\cite{Witten:2010cx}.

Substituting the Euclidean action into Eq.~\eqref{flow}, we have
\begin{align}
\label{Euclideanflow}
\frac{\partial z(\tau;u)}{\partial u}=-\frac{\partial^2\overline{z}(\tau;u)}{\partial\tau^2}+V'(\overline{z}(\tau;u)).
\end{align} 
Expanding $z(\tau;u)$ around the saddle point $z(\tau;u)=z_{a}(\tau)+\Delta z_a(\tau;u)$, one obtains
\begin{align}
\label{Euclideanflowlinearized}
\frac{\partial \Delta{z}_{a}(\tau;u)}{\partial u}=\left(-\frac{\partial^2}{\partial\tau^2}+V''(\overline{z_a}(\tau))\right)\overline{\Delta{z}}_{a}(\tau;u),
\end{align}
subject to the boundary condition $\Delta z_{a}(\tau;u=-\infty)=0$.
In our case, we denote the Lefschetz thimble  associated with $x_F$, $x_{B_n}$, and $x_S$ as $\mathcal{J}_F$, $\mathcal{J}_{B_n}$, and $\mathcal{J}_S$. Every thimble defines a complex integration contour in field space which gives a convergent path integral, as ensured by the decrease of the Morse function along the flow.  Generically, we are looking for a deformation of the original integration contour over the real fields. {If the saddle points are not connected by the flows, i.e., the thimbles end at convergent regions at infinity, this deformed contour $\mathcal{C}$ can be expressed as}
 \begin{align}
\label{Contour}
\mathcal{C}=\displaystyle\sum_{a\in\Sigma}n_a\,\mathcal{J}_{a}
\end{align}
where $\Sigma$ is the moduli space of all the Lefschetz thimbles. The intersection numbers $n_a$ in Eq.~\eqref{Contour} can be either zero or positive integer numbers, and one has an independent perturbative series for each thimble ${\cal J}_a$ near its corresponding saddle-point, with partition function
\begin{align}
Z^E_{a}=e^{\mathcal{I}[z_{a}]}\int\mathcal{D}\Delta z_a e^{\frac{1}{2}\int\D \tau_1\D \tau_2 \left.\Delta z_a(\tau_1)\cdot\frac{\delta^2 \mathcal{I}[z]}{\delta z(\tau_1)\delta z(\tau_2)}\right|_{z_a}\cdot \Delta z_a(\tau_2)+...}.
\end{align} 
One could view every Lefschetz thimble that contributes to the contour $\mathcal{C}$ as a single perturbation theory. Those saddle points then generate the vacua of the theory. 

Two saddle points $z_{a_1}$, $z_{a_2}$ may be connected with each other by the flows when ${\rm Im}\mathcal{I} [z_{a_1}]={\rm Im}\mathcal{I}[z_{a_2}]$, as will happen in our quantum tunneling problem. {In this case some thimbles do not end up at convergent regions at infinity, but rather at other saddle points. Then Eq.~\eqref{Contour} may not be strictly valid; nevertheless, one can still define a basis of paths ending in convergent regions by combining  thimbles or subspaces thereof, and the deformed integration contour will be given by a linear combination of these paths.}
 In this case, the expansion around one saddle point may not be independent of another and one of the saddle points could describe the nonperturbative phenomena relating to different vacua. 

The contour $\mathcal{C}$ is not unique. Suppose we consider a general integral $\int_{\Omega_n} \D\omega$ where $\Omega_n$ is a $n$-dimensional contour in a $2n$-dimensional manifold and $\D\omega$ is a holomorphic differential $n$-form. Then any two contours $\Omega_n^1$, $\Omega_n^2$ that differ by an exact manifold $\Omega_n^3$, i.e., $\Omega_n^3=\partial\Omega_{n+1}$ for some  $n+1-$dimensional manifold $\Omega_{n+1}$ with $\partial$ here denoting the boundary operator, give identical integration result because of the Cauchy theorem. This defines an equivalence relation. To ensure convergence, the integration contour is either compact or its infinite ends lie in the ``convergent regions'' where $h[z]$ is sufficiently small such that the integral is convergent. In this sense, we say that all the contours that ensure a convergent integration are closed and are called integration cycles. Together with their equivalence relations, all the integration cycles give a relative homology group. In our situation, we are just looking for a contour that is homologous to the original one; the thimbles associated with the saddle points {(or, when there are  flows linking saddle points, the cycles obtained from combinations  of subspaces of thimbles),} provide a convenient basis of integration cycles. 

\begin{figure}
  \centering
  \hspace{10pt}
  \includegraphics[scale=0.6]{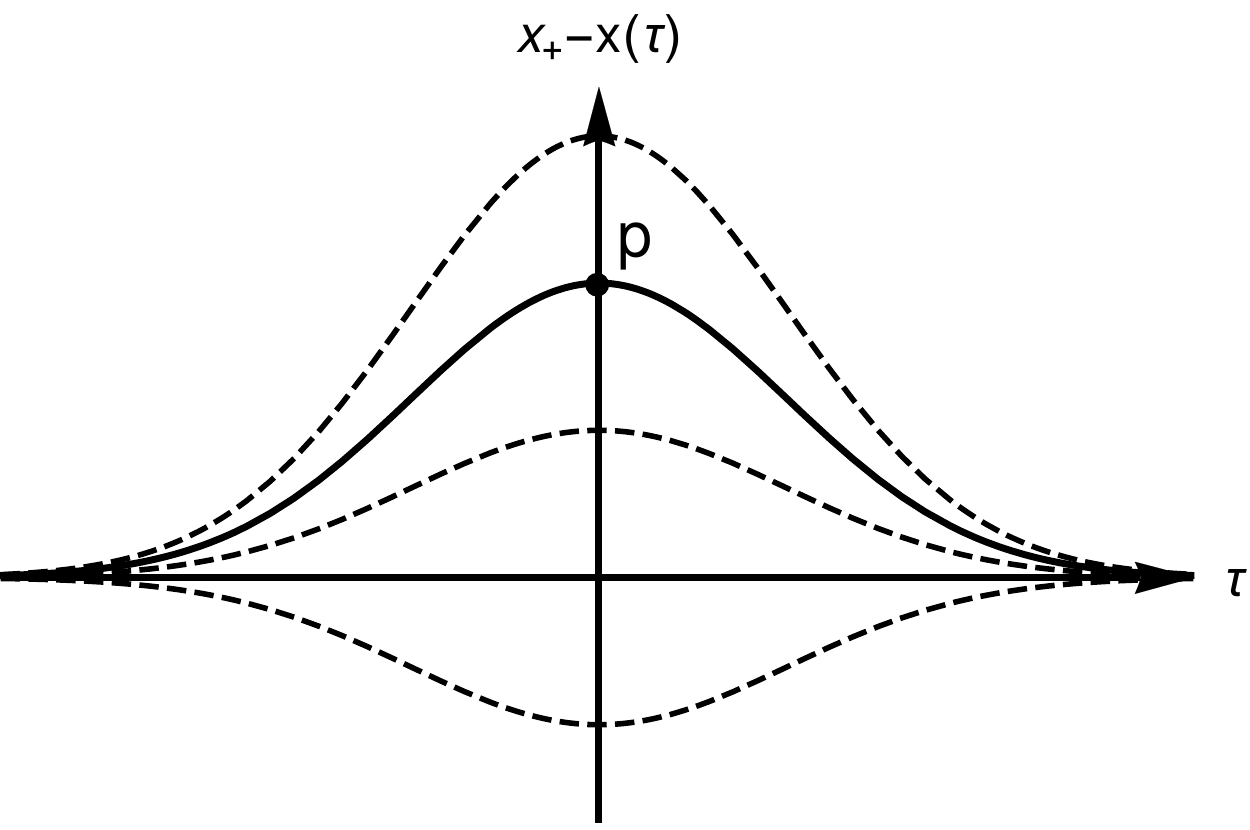}
  \caption{A series of symmetric quantum paths from $x_+$ to itself: $\hat{x}(\tau;\rho)$, parameterized by $\rho$. We take the $\tau$-axis---the trivial false-vacuum solution---as the base point $\hat{x}(\tau;0)$. The path with its maximum marked by $\mathbf{p}$, indicating the turning point, is the bounce $\hat{x}(\tau;b)$, for some number $b$. The paths above the bounce $\hat{x}(\tau;\rho>b)$ are the quantum paths with escape point beyond the turning point $\mathbf{p}$, containing the shot at some point $\rho=s>b$ which we do not show.
  \label{fig:quantumpath}}
\end{figure}

\begin{figure}
  \centering
  \hspace{10pt}
  \includegraphics[scale=0.6]{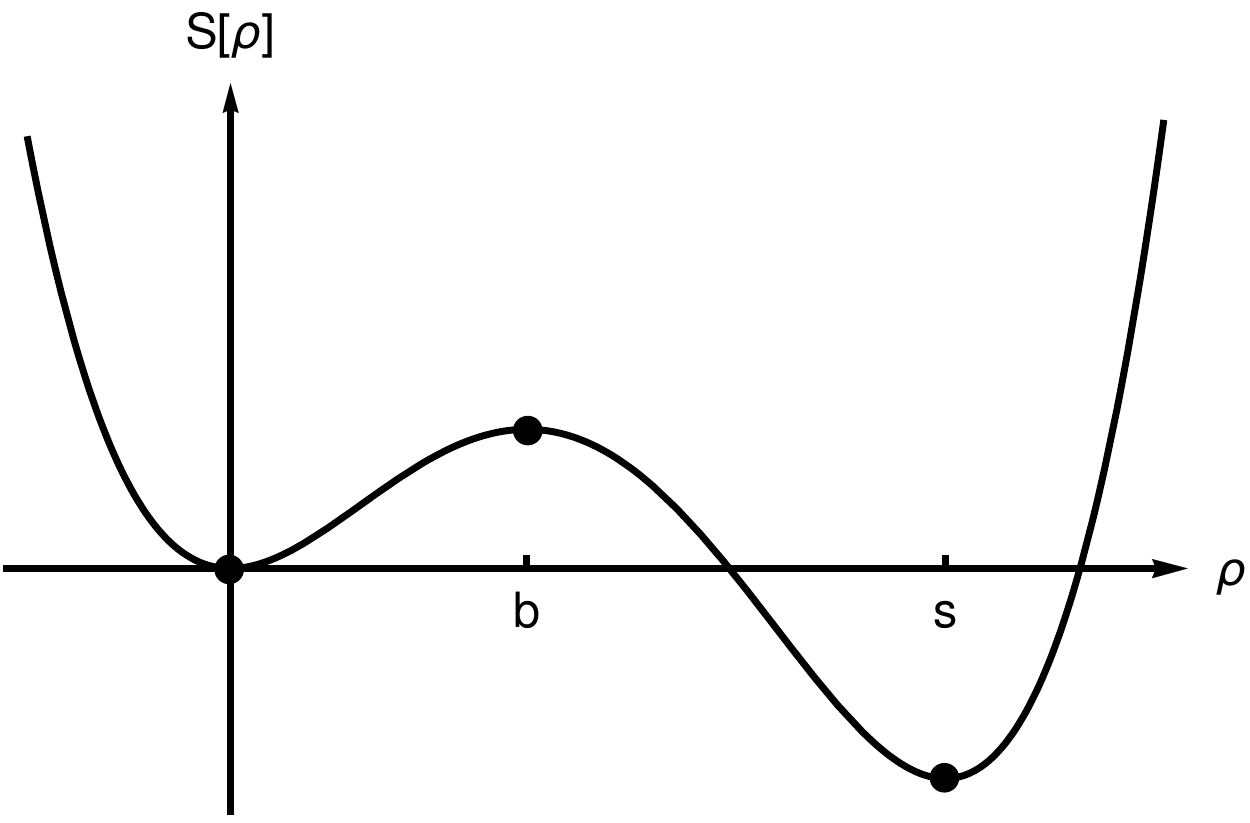}
  \caption{The dependence of the Euclidean action $S_E[\hat{x}(\tau;\rho)]$ on the parameter $\rho$. The saddle points {(marked with dots)} are $x_F(\tau)$, the bounce $x_{B}(\tau)$ and the shot $x_{S}(\tau)$ {from the left to the right, respectively}.
  \label{fig:saddlepoints}}
\end{figure}

Determining all the saddle points and the integers $n_a$ is difficult in general. Fortunately, it is not necessary to do such a complicated analysis for the tunneling problem. {Let us ignore
the multi-bounce saddle-points for the moment, such that one just has the false vacuum, the (single) bounce and the shot.} On the original middle-dimensional contour of real field configurations, all directions except for the one associated with the negative eigenvalue $\lambda_0^B$ at the bounce generate the actual paths of the steepest descent. Along this special direction, the three saddle points are actually related with each other.\footnote{The relation between $x_F$ and the bounce $x_B$ was observed by Callan and Coleman in their original work~\cite{Callan:1977pt}, where they did not discuss the shot however. The full relation between these three saddles is noted in Ref.~\cite{Andreassen:2016cvx} modeled by a toy one-dimensional integral. In that work, it is also pointed out that it is in fact the shot that is essential to understand how an imaginary part for the false-vacuum ground-state energy can emerge from a Euclidean path integral, that is purely real by construction. We will get back to this point later.} To see this, we consider a family of paths $\hat{x}(\tau;\rho)$ with $\rho\in \mathbb{R}$ as represented in Figure~\ref{fig:quantumpath}. The dependence of the Euclidean action  on these paths is shown in Figure~\ref{fig:saddlepoints}. When varying $\rho$, the trivial false-vacuum solution and the shot are situated at the local and the global minimum, respectively. The bounce is at the local maximum, giving a negative second order derivative (the negative eigenvalue $\lambda_0^B$) of the action with respect to $\rho$. Thus along the parameter $\rho$, a steepest {\it ascent} path (recall $h[z]=-S_E[z]$) is generated starting from the bounce.

We find the missed direction of steepest descent when we allow the variable $\rho$ to be complex and deform the one-dimensional path as follows. The variable $\rho$ starts from minus infinity and flows along the real axis towards the point $\rho=b$ where it turns upward\footnote{\label{fn:ambiguity}The path can turn either upward or downward. We take the upward direction in order to have a positive imaginary part in the false-vacuum energy.} into the imaginary direction all the way to $b+i\infty$. That is, the first path is given by $\mathcal{J}_{FB}^{\rho}: -\infty\rightarrow b\rightarrow b+i\infty$, as illustrated in Figure~\ref{fig:JFBrho}. After that, the path flows in the imaginary direction from $b+i\infty$ back to the point $\rho=b$ and then rushes along the real axis to the shot $\rho=s$, ending finally at positive infinity on the real axis. That is, we have a second path $\mathcal{J}_{SB}^{\rho}:b+i\infty\rightarrow b\rightarrow \infty$. By this construction, we are still passing through the original three real saddle points but with a one-dimensional subset of the contour deformed from the real axis to $\mathcal{C}_{\rho}\equiv \mathcal{J}_{FB}^{\rho}+\mathcal{J}_{SB}^{\rho}$. Compared with the original contour, $\mathcal{C}_{\rho}$ contains, in addition, the segment $b\rightarrow b+i\infty\rightarrow b$ and is therefore equivalent to the original one. In the language of {steepest-descent flows}, the above analysis shows that there is a flow passing through both the false vacuum and the bounce; the flow branches out at the bounce into the two steepest-descent directions going upwards and downwards in the imaginary $\rho$ direction, while the deformation of the integration contour only picks one of the branches, given by the {flow} ${\cal J}^\rho_{FB}$ (see Figure~\ref{fig:JFBrho}). A similar situation arises with the {flow} passing through the bounce and the shot, and the {flow}  ${\cal J}^\rho_{SB}$ picking one of the branches. 

We expect that the bounce does not give a single perturbative sector but rather describes the nonperturbative phenomena between the false vacuum and the shot (which actually corresponds to the true vacuum). The integral from the one-dimensional contour $\mathcal{C}_{\rho}$ can be decomposed into
\begin{align}
\label{ZC-1}
Z^E_{\rho}&=Z^E_{FB(\rho)}+Z^E_{SB(\rho)}\equiv\int_{\mathcal{J}^{\rho}_{FB}} \D \rho\, e^{\cal I[\rho]}+\int_{\mathcal{J}^{\rho}_{SB}} \D \rho\,e^{\cal I[\rho]}.
\end{align}
Both $Z^E_{FB(\rho)}$ and $Z^E_{SB(\rho)}$ contain an imaginary part but with opposite sign, leading to a {\it purely real} and also finite result, as expected from the reality of the Euclidean action.

Regarding the multi-bounce saddles,  we expect a similar situation in which the downward flows from the false vacuum and the shot reach a given multi-bounce saddle along with the field directions associated with its negative modes. At every multi-bounce saddle, these {special flows} of the false vacuum and shot  meet and branch out in the imaginary directions associated with the negative modes, 
and the deformation of the integration contour that passes  through either the  false vacuum or the shot picks only half of the branches. 

We will denote {as $\mathcal{J}_{FB}$ the integration cycle passing through the false vacuum and multi-bounces, i.e. $\mathcal{J}_{FB}\sim\mathcal{J}_F+\sum_n\mathcal{J}_{B_n}$, but  picking only half of the branches of  the special one-dimensional flows  linking $F$ and $B_n$, as commented above.} In the analogous way, we define {the cycle} $\mathcal{J}_{SB}$. Note that, {despite the abuse of notation,} $\mathcal{J}_{FB}$ or $\mathcal{J}_{SB}$ shall not be understood as {\it one} thimble, {but rather integration cycles constructed from several thimbles}. We denote the path integral on $\mathcal{J}_{FB}$ as $Z^E_{FB}$ and accordingly for $Z^E_{SB}$.

In Ref.~\cite{Callan:1977pt}, only the contribution $Z^E_{FB}$ is picked out in order to derive the Euclidean transition amplitude $\langle x_+|e^{-H\mathcal{T}}|x_+\rangle$,
leading to an imaginary part of the energy of the false-vacuum state. In the context of the present discussion, this can be explained as follows. Energy, as an eigenvalue of the Hermitian Hamiltonian, must be a real number. Indeed, the dominant purely real part in the full amplitude $\langle x_+|e^{-H\mathcal{T}}|x_+\rangle$, residing in $Z^E_{SB}$,
gives the energy of the true ground state and the corresponding wave function. A complex energy can only emerge when restricting to an open subsystem. In this sense, we may think of $Z^E_{FB}$ and $Z^E_{SB}$ as the theories describing two different subsystems or sectors---the false vacuum and the true vacuum---separately. The imaginary parts from both  $Z^E_{FB}$ and $Z^E_{SB}$ indicate that both the false vacuum and the true vacuum are open systems.\footnote{In the context of quantum mechanics, the notion of an ``open system'' may easily be understood since the false vacuum or true vacuum may be defined only for subregions in space---the left well or the right well. In quantum field theory, ``subsystems'' should be understood in the sense of
regions in field space.} Since the whole system is closed, the imaginary parts must cancel between these two open subsystems.  Note that when taking this point of view, the so-called procedure of potential deformation used in Ref.~\cite{Callan:1977pt} is sidestepped. The factor $1/2$ introduced in the former reference when extracting the imaginary part of the Gau{\ss}ian integral  {around the single-bounce saddle} appears naturally when we restrict to the subsystem represented by $\mathcal{J}_{FB}$ {because, as discussed above, this integration cycle only includes half of the branches of the special one-dimensional flows that link the false-vacuum and bounce saddles. Indeed, the}  contribution to the partition function $Z^E_{FB}$ can be approximated by a sum of Gau{\ss}ian contributions around each of the saddles:
\begin{align}
 Z^E_{FB}\approx Z^E_{F,\text{ Gau{\ss}ian}}+\sum_n Z^E_{B_n,\text{ Gau{\ss}ian}}.
\end{align}
{For the above Gau{\ss}ian contributions, the integration domains must be appropriately constrained in accordance with the fact that $Z^E_{FB}$ does not include all the branches of the special flows that connect the false-vacuum and bounce saddle points, as discussed earlier. As for each saddle $B_n$, there are $n$ branching special flows, and as the  Gau{\ss}ian integrand is the same along the chosen and discarded branches, $Z^E_{B_n,\text{Gau{\ss}ian}}$ is given by $(1/2)^n$ times the full Gau{\ss}ian integration.}
Using the fact that for ${\cal T}\rightarrow\infty$ the multi-bounces are made of infinitely separated bounces, it can be seen that the path integral of their fluctuations factorizes ({the factorial factor comes from integrating the positions of every single bounce in $x_{B_n}$}), which ends up giving  \cite{Callan:1977pt}
\begin{align}
\label{eq:ZGaussian}
 Z^E_{FB}\approx Z^E_{F,\text{ Gau{\ss}ian}}\exp\left(\frac{Z^E_{B,\text{ Gau{\ss}ian}}}{Z^E_{F,\text{ Gau{\ss}ian}}}\right),
\end{align}
with $Z^E_{B,\text{ Gau{\ss}ian}}$ denoting the contribution of fluctuations about a single bounce, {equaling to $1/2$ of the unrestricted Gau{\ss}ian integration. In practice, we will include this factor of $1/2$ in the definition of the integration measure for the path integration of the fluctuations around the bounce.} 
%

In Section~\ref{sec:MinkowskiPathIntegral}, we introduce an additional point of view
on why we need to exclude the perturbative expansion about the shot in order to isolate the imaginary part. That is, we consider the Minkowski amplitude for the transition from false-vacuum to false-vacuum and relate it to the decay rate via the optical theorem. The {integration cycle} that passes through the shot contributes  instead to the true-vacuum to true-vacuum transition amplitude.

The contributions $Z^E_{F,\text{Gau{\ss}ian}}$ and $Z^E_{B,\text{Gau{\ss}ian}}$ (and hence $Z^E_{FB}$) can be readily evaluated because only the deformation of the one-dimensional contour $\mathcal{C}_\rho$ needs a particular care. However, along the lines of Ref.~\cite{Tanizaki:2014xba}, we shall give a general analysis on how to evaluate $Z^E_{FB}$ from the point of view of the flow equation. This can be done by solving the linearized flow equation~\eqref{Euclideanflowlinearized} around each of the relevant saddle points $x_a(\tau)$ with $a=F,B$. We make the separation $\Delta z_{a}(\tau;u)=\sum_n g_n^a(u)\chi_n^a(\tau)$ where $g_n^a(u)\in\mathbb{R}$ and the subscript ``$n$'' denotes a specific direction, such that Eq.~\eqref{Euclideanflowlinearized} becomes
\begin{align}
\label{flow-eigenequation}
(-\partial_\tau^2+V''(\overline{x_a}(\tau)))\,\overline{\chi_n^a}(\tau)g_n^a(u)=\chi_n^a(\tau)\frac{\D g_n^a (u)}{\D u}.
\end{align}
Eq.~\eqref{flow-eigenequation} leads to
\begin{align}
(-\partial_\tau^2+V''(\overline{x_a}(\tau)))\overline{\chi_n^a}(\tau)/\chi_n^a(\tau)={\kappa_n^a}=\frac{1}{g_n^a(u)} \frac{\D g_n^a(u)}{\D u},
\end{align}
where ${\kappa_n^a}\in\mathbb{R}$. The first equation following from this separation is
\begin{align}
\label{flow-eigenEq}
(-\partial_\tau^2+V''(\overline{x_a}(\tau)))\,\overline{\chi_n^a}(\tau)={\kappa_n^a }\chi_n^a(\tau)
\end{align}
with Dirichlet boundary conditions $\chi_n^a(\tau=\pm\mathcal{T}/2)=0$. We refer Eq.~\eqref{flow-eigenEq} as the {\it flow eigenequation} to distinguish from the proper eigenequation and ${ \kappa_n^a}$, $\chi_n^a(\tau)$ as the {\it flow eigenvalue} and {\it flow eigenfunction}, respectively.
The complex conjugate of Eq.~\eqref{flow-eigenEq} is
\begin{align}
\label{flow-eigenEq:conjugate}
\left(-\partial^2_\tau+V''({x_a}(\tau))\right){\chi_n^a}(\tau)={\kappa_n^a} \overline{\chi_n^a}(\tau).
\end{align}
Combining both Eqs.~\eqref{flow-eigenEq} and~\eqref{flow-eigenEq:conjugate} by taking the direct product,
we note that $(\chi^a_n,\overline{\chi^a_n})$ satisfy an eigenvalue equation with a Hermitian operator, such that
we can impose the normalization
\begin{align}
\label{norm-relations}
\int_{-\mathcal{T}/2}^{\mathcal{T}/2}\D\tau\ \overline{\chi^a_m}(\tau){\chi^a_n}(\tau)=\delta_{mn}
\end{align}
on the flow eigenfunctions.

One important property for Eq.~\eqref{flow-eigenEq} is that ${ \kappa_n^a}$ is always paired with $-{ \kappa_n^a}$, which is associated with the flow eigenfunction $i\chi^a_n(\tau)$ as can be checked easily. The additional
equation from separating Eq.~\eqref{flow-eigenequation} is $g_n^a(u)=a_n^a \exp({\kappa_n^a} u)$, where $a_n^a\in \mathbb{R}$. Recalling the boundary condition $g_n^a(u=-\infty)=0$, we have ${\kappa_n^a}>0$, i.e. close to the saddles, where the linearized flow equations apply, the directions on the thimble are
those with positive flow eigenvalues.\footnote{It is possible to have a zero mode in the limit $\mathcal{T}\rightarrow\infty$ which has to be handled separately.} 

Now, since in the calculation based on the Euclidean action,
$x_a(\tau)$ is real, we see from Eqs.~\eqref{flow-eigenEq} and~\eqref{flow-eigenEq:conjugate} that ${\chi_n^a}(\tau)$ and $\overline{\chi_n^a}(\tau)$ are the flow eigenfunctions associated with the same flow eigenvalue. Thus one has $\overline{\chi_n^a}(\tau)=\pm \chi^a_n(\tau)$, assuming there is no degeneracy for the nonzero modes as it is the case in general. Therefore, Eq.~\eqref{flow-eigenEq} has purely real or purely imaginary flow-eigenfunctions, and it reduces to the eigenequation in the proper sense
\begin{align}
\label{Euc-eigen}
(-\partial_\tau^2+V''(x_a(\tau)))f_n^a(\tau)=\lambda_n^a f_n^a(\tau).
\end{align} 
For $\lambda_n^a>0$, we simply have $\chi_n^a(\tau)=f_n^a(\tau)$ and ${ \kappa_n^a}=\lambda_n^a$. For the negative mode $f_0^{B}(\tau)$,\footnote{Following Callan's and Coleman's notation, we use the subscript ``0'' to denote the negative mode and ``1'' to denote the zero mode.} we have $\chi_0^{B}(\tau)=i f_0^{B}(\tau)$ in order to have positive $\kappa_0^{B}$. 

Now let us look at the integrand $\exp(\mathcal{I}[z])$ in the path integral by substituting $z(\tau;u)=z_a(\tau)+\sum_n g_n^a(u)\chi_n^a(\tau)$ into $I[z]$. One has up to $\mathcal{O}(\Delta z^2)$
\begin{align}
\label{expan}
{\cal I} [z]&={\cal I}[z_a]-\frac{1}{2}\int\D\tau\,\Delta z_a(\tau;u)\,(-\partial_\tau^2+V''(x_a(\tau)))\,\Delta z_a(\tau;u)\notag\\
&={\cal I}[z_a]-\frac{1}{2}\sum_n{{ \kappa_n^a}}\,(g_n^a(u))^2,
\end{align}
where in the second equality, we have used Eq.~\eqref{flow-eigenEq:conjugate} and the orthonormality relation~\eqref{norm-relations}. Since $g_n^a(u)$ are real and ${\kappa_n^a}$ are real and positive, the last line in Eq.~\eqref{expan} tells us that the saddle-point approximation to the path integral on the Lefschetz thimble $\mathcal{J}_a$ is a Wiener integration at the Gau{\ss}ian level, and is thus convergent.

\subsection{Integration measure and Gau{\ss}ian integration}
\label{sec:measureandzeromodes}

From $\Delta z_a(\tau;u)=\sum_n g_n^a(u)\chi_n^a(\tau)$, we define the measure of the path integral around a given saddle point as
\begin{align}
\mathcal{D}\Delta z_a=J_a\prod_n\frac{1}{\sqrt{2\pi}}\D g^a_n,
\end{align}
where $J_a$ is the Jacobian due to the transformation from the original real basis to the new basis $\{\chi_n^a(\tau)\}$.
At the Gau{\ss}ian level, we have (again, the zero mode will be considered separately)
\begin{align}
\label{Gau}
J_a\prod_n\int\D g_n^a\ \frac{1}{\sqrt{2\pi}}\ e^{-\frac{1}{2}\sum_n{ \kappa_n^a}({g_n^a})^2}=J_a\prod_n\frac{1}{\sqrt{{\kappa_n^a}}}=J_a|\det(-\partial_\tau^2+V''(x_a))|^{-1/2}.
\end{align}
Without the deformation of the contour, the path integral measure is defined from the decomposition of $\Delta z_a(\tau;u)$ into the real eigenfunctions of $-\partial_\tau^2+V''(x_a)$, $\Delta z_a(\tau;u)=\sum_n c_n^a f_n^a(\tau)$, as
\begin{align}
\mathcal{D}\Delta z_a=\prod_n \frac{1}{\sqrt{2\pi}}\D c_n^a.
\end{align}
Since for the saddle point $x_F$, the basis $\{\chi_n^F(\tau)\}$ is the same as $\{f_n^F(\tau)\}$, we have $J_F=1$. For the bounce, since $\chi_0^{B}(\tau)=if_0^{B}(\tau)$, we have $\D c_0^{B}=i\D g_0^{B}$ as can be seen from $\Delta z_{{B},0}(\tau)= g_0^{B} \chi_0^{B}(\tau)=c_0^{B} f_0^{B}(\tau)$. Since for the other modes, $\chi_{n\neq 0}^{B}$ are the same as $f_{n\neq 0}^{B}$, we finally arrive at $J_{B}=i$. Thus the Jacobian can be identified as {the exponential of} {minus} half of the phase of the determinant of $-\partial_\tau^2+V''(x_a)$. This claim as well as the second equality of Eq.~\eqref{Gau} are actually quite general and we give the proof in Section~\ref{app:sec:Flow-Jacobian}.

Recall that the bounce is connected with the false vacuum via the flow $\mathcal{J}^{\rho}_{FB}$, leading the integral over $g^{B}_{0}$ cut by half in $Z^E_{B,\text{Gau{\ss}ian}}$ (see again Figure~\ref{fig:JFBrho}). We thus account for this fact by adding a factor $1/2$ to the path integral measure 
\begin{align}
\label{EucintegralmeasureB}
\mathcal{D}\Delta z_{B}\rightarrow \widetilde{\mathcal{D}\Delta z_{B}}=\frac{1}{2}\prod_{n}\frac{J_{B}}{\sqrt{2\pi}}\D g^{B}_n.
\end{align}
We finally can use Eq.~\eqref{eq:ZGaussian}, with 
\begin{align}
\label{eq:ZBGaussian}
{Z}^E_{B,\text{Gau{\ss}ian}}=e^{-S_E[x_{B}]}\int\widetilde{\mathcal{D}
\Delta{z}_{B}}\,e^{-\int_{-\mathcal{T}/2}^{\mathcal{T}/2}d\tau\left[\Delta{z}_{B}\left(
-\frac{1}{2}\frac{d^2}{d\tau^2}+\frac{1}{2}V''(x_{B})\right)\Delta{z}_{B}
\right]}.
\end{align}

Now let us consider the zero mode $\chi_1^{B}(\tau)=f_1^{B}(\tau)$ which appears in the limit $\mathcal{T}\rightarrow\infty$, corresponding to the spontaneous symmetry breaking of the time-translation symmetry by the bounce solution,
\begin{align}
f^{B}_1(\tau)=S_E[x_{B}]^{-1/2}\,\frac{\D x_{B}(\tau)}{\D \tau}.
\end{align} 
This zero mode can be traded for an integral over the collective coordinate of the bounce as can be seen from
\begin{align}
\D\Delta z_{B}=\frac{\D x_{B}(\tau)}{\D \tau}\D \tau=S_E[x_B]^{1/2}\,f^B_1(\tau)\D \tau=f^B_1(\tau)\D c_1.
\end{align}
Thus $1/\sqrt{2\pi}\,\D c_1$ can be traded for $(S_E[x_B]/2\pi)^{1/2}\D \tau$ and the integration over the zero mode gives us $\mathcal{T}(S_E[x_B]/2\pi)^{1/2}$. After the Gau{\ss}ian integration of \eqref{eq:ZBGaussian} with the proper treatment of the zero mode, we obtain from Eq.~\eqref{eq:ZGaussian} at NLO,
\begin{align}
\label{ZFB-Euc}
\begin{aligned}
\frac{Z^E_{FB}}{Z^E_{F}}&\approx{\exp}\left(\frac{Z^E_{B,\text{Gau{\ss}ian}}}{Z^E_{F,\text{Gau{\ss}ian}}}\right)\\ 
&=\,{\exp\left(\frac{\mathcal{T}}{2}\sqrt{\frac{S_E[x_B]}{2\pi}}e^{-S_E[x_B]}\left(\frac{{\det}'[-\partial_\tau^2+V''(x_B)]}{\det[-\partial_\tau^2+V''(x_F)]}\right)^{-1/2}\right)}\\
&=\,{\exp}\left(\frac{i\mathcal{T}}{2}\sqrt{\frac{S_E[x_B]}{2\pi}}e^{-S_E[x_B]}\left|\frac{{\det}'[-\partial_\tau^2+V''(x_B)]}{\det[-\partial_\tau^2+V''(x_F)]}\right|^{-1/2}\right),
\end{aligned}\end{align}
where ${\det}'$ indicates that the zero eigenvalue is to be omitted when computing the determinant.
The decay rate of the false vacuum is then obtained from the imaginary part of the false-vacuum energy after using $Z^E_{FB}/Z^E_{F}\sim e^{-E_0 {\cal T}}$, which yields a decay rate 
\begin{align}%
\label{CC-deCayformula}
\varGamma=-2{\rm Im}E_0=\frac{2}{\mathcal T}\,\left|{\rm Im} \log \left(\frac{Z_{FB}^E}{Z_F^E}\right)\right|=\frac{2}{\mathcal T}\,\left|{\rm Im}\, \frac{Z_{B,\text{Gau{\ss}ian}}^E}{Z_{F,\text{Gau{\ss}ian}}^E}\right|.
\end{align}
Substituting Eq.~\eqref{ZFB-Euc} into Eq.~\eqref{CC-deCayformula} gives the formula~\eqref{eq:GammaQM} for the decay rate.

\section{Different methods of evaluating the functional determinant}
\label{app:methods:determinant}

In this appendix, we review two additional ways of obtaining the functional determinant of differential operators, the Gel'fand-Yaglom method and the calculation based on the resolvent generalizing the Green's function. We compare these with the method used in the main text, i.e. the direct integration over the eigenvalues, see  Section~\ref{sec:det:kink}. As for the result for the decay rate, these methods should
also be compared with the WKB approximation presented in Appendix~\ref{app:sec:wkb}. We provide the discussion for the Euclidean fluctuation operators since the functional determinants in the Minkowski formalism or for general complex time can be obtained by analytic continuation of the Euclidean results, as we have shown in the main text. In Appendix~\ref{app:sec:GY-Green}, we take the archetypical model of particle tunneling in a quasi-degenerate quartic potential as an application and compare the results from the two methods discussed in this appendix with the direct evaluation of the logarithmic determinant from Eq.~\eqref{Eq533}.

All of these methods of calculating determinants, that we work out
here for quantum-mechanical tunneling,  can also be applied to false-vacuum decay because the $O(4)$-symmetry of the background allows the decomposition of a four-dimensional partial differential operator into a hyperradial operator and the Laplace-Beltrami operator. The angular spectrum can be exactly solved. Thus, the evaluation of the determinant of a four-dimensional hyperspherically symmetric partial differential operator can be essentially reduced to evaluating the determinant of a hyperradial ordinary differential operator. In Appendix~\ref{app:utility:of:methods}, we
make some remarks on the advantages and disadvantages of the various methods of calculating functional determinants about solitons that can be found in this paper.

To briefly summarize how all these approaches fit into the calculations
for false-vacuum decay or tunneling, we note that
the decay rate can be either obtained from the imaginary part
of the ground-state energy of the false vacuum, from the outgoing
flux or from the imaginary part of the false-vacuum to false-vacuum (forward scattering) amplitude
via the optical theorem. The WKB method can be used to compute
either the ground-state energy or the flux (cf. Appendix~\ref{app:sec:wkb}), whereas the functional determinant
leads to the ground-state energy (cf. Appendix~\ref{sec:Callan-Coleman}) or the false-vacuum to false-vacuum amplitude. The determinant can be calculated either by direct integration
over the spectrum (cf. Section~\ref{sec:det:kink}), by using the Gel'fand Yaglom theorem  (cf.~Appendix~\ref{sec:GY}) or integration
of the resolvent  (cf. Appendix~\ref{sec:resolvent}).

\subsection{Gel'fand Yaglom method}
\label{sec:GY}

The Gel'fand-Yaglom method is based on a powerful theorem of the same name~\cite{Gelfand:1959nq}. It is widely employed in calculations for
tunneling in theoretical as well as phenomenological models, see e.g. Refs.~\cite{Isidori:2001bm,Dunne:2005rt,Andreassen:2017rzq,Chigusa:2017dux}. In this section, we closely follow Ref.~\cite{Coleman:1988}.

\subsubsection{Gel'fand-Yaglom theorem}
Consider the equation
\begin{align}
\label{pde}
(-\partial_\tau^2+W(\tau))\,\psi(\tau)=\lambda\,\psi(\tau),
\end{align}
where $W(\tau)$ is a bounded function of $\tau\in [-\mathcal{T}/2,\mathcal{T}/2]$. The functions $\psi_{\lambda}(\tau)$ are the solutions of Eq.~\eqref{pde} satisfying the boundary conditions
\begin{align}
\label{boundaryconditions}
\psi_{\lambda}(-\mathcal{T}/2)=0, \ \ \left.\partial_\tau\psi_{\lambda}(\tau)\right|_{\tau=-\mathcal{T}/2}=1.
\end{align}
The determinant of the operator $-\partial_\tau^2+W(\tau)$ is defined as
\begin{align}
\det(-\partial_\tau^2+W(\tau))=\prod_n\lambda_n,
\end{align}
where the $\lambda_n$ satisfy  
\begin{align}
\label{ODE}
(-\partial_\tau^2+W(\tau))\,\psi_{\lambda_n}(\tau)=\lambda_n\,\psi_{\lambda_n}(\tau),
\end{align}
with boundary conditions $\psi_{\lambda_n}(-\mathcal{T}/2)=\psi_{\lambda_n}(\mathcal{T}/2)=0$.

The Gel'fand-Yaglom theorem states that
\begin{align}
\label{app:G-Y}
\frac{\det[-\partial_\tau^2+W^{(1)}(\tau)-\lambda]}{\det[-\partial_\tau^2+W^{(2)}(\tau)-\lambda]}=\frac{\psi^{(1)}_{\lambda}(\mathcal{T}/2)}{\psi^{(2)}_{\lambda}(\mathcal{T}/2)}.
\end{align}
Applying the above formula to the case $\lambda=0$ and taking the limit $\mathcal{T}\rightarrow\infty$, we obtain the ratio of determinants that appears e.g. through Eq.~\eqref{ZFB-Euc} in the formula for the decay rate~\eqref{CC-deCayformula}.

\subsubsection{Evaluating the ratio of the functional determinants}
\label{sec:GY:evaluation}

The ratio of functional determinants in the expression for the tunneling rate now can be readily evaluated. We first consider the fluctuation operator at the false vacuum where $W^{(1)}(\tau)=V''(x_+)\equiv m^2$. The solution to Eq.~\eqref{pde} with the boundary conditions~\eqref{boundaryconditions} is
\begin{align}
\psi^{(1)}_0(\tau)=\frac{1}{m}\sinh[m(\tau+\mathcal{T}/2)],
\end{align}
and thus, $\psi^{(1)}_0(\mathcal{T}/2)=e^{m\mathcal{T}}/2m$ for large $\mathcal{T}$. 

Next, we look at fluctuations about the bounce, where $W^{(2)}(\tau)=V''(x_B(\tau))$. We have to evaluate the primed determinant, i.e. the zero eigenvalue associated
with time translations is to be taken out. Following Coleman, we can do this by evaluating the full determinant on a finite interval $[-\mathcal{T}/2,\mathcal{T}/2]$, dividing it by its smallest, nonnegative, {\it finite} eigenvalue near zero, $\lambda_0$ (to be distinguished from the negative eigenvalue $\lambda_0^B$), and eventually letting $\mathcal{T}$ go to infinity. The function $\psi^{(2)}_0(\tau)$ can be constructed from an arbitrary basis of solutions. Actually, it is sufficient to know its asymptotic behavior at $\pm \mathcal{T}/2$ in order to apply the formula~\eqref{app:G-Y}. Consider therefore the equation
\begin{align}
\label{Eq69}
[-\partial^2_\tau+V''(x_B(\tau))]\,\psi(\tau)=0.
\end{align}
One of the basis solutions can be chosen to be
\begin{align}
\label{asymptotic}
x_1(\tau)=B^{-1/2}\,\frac{\D x_B(\tau)}{\D\tau}\rightarrow \pm\, \frac{A}{\sqrt{m}}\,e^{-m|\tau|},\ {\rm as}\ \tau\rightarrow\pm\infty,
\end{align}
where $A$ is determined by the asymptotic behaviour of $x_1(\tau)$ (cf. Eq.~\eqref{expressionofA}). Note that $\psi_0^{(2)}(\tau)$ cannot be $x_1(\tau)$ because $x_1(\tau)$ does not satisfy the particular boundary conditions given below Eq.~\eqref{ODE}.

We also note here that for the classical bounce, there is the constant
of motion
\begin{align}
\frac{1}{2}\left(\frac{\D x_B(\tau)}{\D\tau}\right)^2-V(x_B(\tau))=0.
\end{align} 
Therefore, $\D x_B(\tau)/\D\tau=\sqrt{2V(x_B(\tau))}$, which leads to
\begin{align}
\tau=\int_{x_{\mathbf p}}^x \D x\,\frac{1}{\sqrt{2V(x)}}.
\end{align}
Using the asymptotic behaviour from Eq.~\eqref{asymptotic}, one obtains
\begin{align}
\label{expressionofA}
m\tau\equiv m\int_{x_{\mathbf p}}^x \D x\,\frac{1}{\sqrt{2V(x)}}=-\log\left[{B}^{-1/2}\, m^{3/2}\,A^{-1}\,(x_+-x)\right]+\mathcal{O}(x_+-x).
\end{align}
This equation will be used in Appendix~\ref{app:sec:wkb}.

Next, we consider another independent solution to Eq.~\eqref{Eq69} that
we denote as $x_2(\tau)$. One can choose the normalization for $x_2(\tau)$ such that 
\begin{align}
\label{app:normalization}
x_1\,\partial_\tau x_2-x_2\,\partial_\tau x_1=2\,A^2.
\end{align}
Therefore, we can deduce its asymptotic behaviour
\begin{align}
\label{asymptotic2}
x_2(\tau)\rightarrow \frac{A}{\sqrt{m}}\, e^{m|\tau|},\ {\rm as}\ \tau\rightarrow\pm\infty.
\end{align} 
According to the boundary conditions~\eqref{boundaryconditions}, one can construct $\psi^{(2)}_0(\tau)$ as
\begin{align}
\label{app:psi2}
\psi^{(2)}_0(\tau)=-\frac{1}{2{\sqrt{m}}A}\left(e^{m \mathcal{T}/2}\,x_1(\tau)+e^{-m \mathcal{T}/2}\,x_2(\tau)\right),
\end{align}
leading to $\psi^{(2)}_0(\mathcal{T}/2)=-1/m$.

Now let us subtract the smallest positive eigenvalue $\lambda_0$. Since $\lambda_0$ is small, we can expand $\psi_{\lambda_0}(\tau)=\psi^{(2)}_0(\tau)+\delta\psi_{\lambda_0}(\tau)$ in the eigenequation. Hence, one has
\begin{align}
(-\partial_\tau^2+V''(x_B(\tau)))\,\delta\psi_{\lambda_0}(\tau)=\lambda_0\,\psi^{(2)}_0(\tau),
\end{align}
which is solved by
\begin{align}
\psi_{\lambda_0}(\tau)=\psi^{(2)}_0(\tau)-\frac{\lambda_0}{2\,A^2}\int_{-\mathcal{T}/2}^\tau \D \tau'\,[x_2(\tau)x_1(\tau')-x_1(\tau)x_2(\tau')]\,\psi^{(2)}_0(\tau'),
\end{align}
such that
\begin{align}
\psi_{\lambda_0}(\mathcal{T}/2)={-\frac{1}{m}+}\frac{\lambda_0}{4\,{m}A^2}\int_{-\mathcal{T}/2}^{\mathcal{T}/2} \D\tau'\,[e^{m\mathcal{T}}x_1^2(\tau')-e^{-m\mathcal{T}}x_2^2(\tau')].
\end{align}
Since $x_1(\tau)$ is normalized, we arrive at
\begin{align}
\psi_{\lambda_0}(\mathcal{T}/2){\approx -\frac{1}{m}+}\frac{\lambda_0}{4\,m A^2}\,e^{m\mathcal{T}}.
\end{align}
By requiring the boundary condition $\psi_{\lambda_0}(\mathcal{T}/2)=0$, we obtain $\lambda_0=4\,A^2/e^{m\mathcal{T}}$. In total, we have
\begin{align}
\label{app:det'}
\frac{\det'[-\partial_\tau^2+V''(x_B)]}{\det[-\partial_\tau^2+V''(x_+)]}=\frac{\psi^{(2)}_0(\mathcal{T}/2)}{\lambda_0\,\psi^{(1)}_0(\mathcal{T}/2)}=-\frac{1}{2\,A^2}.
\end{align}
Note that this is a {\it negative} number, indicating the existence of a negative eigenvalue in the eigenspectrum of the operator $-\partial_\tau^2+V''(x_B(\tau))$. Had we used a kink solution $\bar{x}(\tau)$ (see Appendix~\ref{app:sec:GY-Green}) instead of the bounce $x_B(\tau)$, the asymptotic behaviour in Eq.~\eqref{asymptotic} would be different and lead to a positive result in Eq.~\eqref{app:det'}~\cite{Coleman:1988}. Substituting the above result into Eqs.~\eqref{ZFB-Euc},~\eqref{CC-deCayformula}, we have
\begin{align}
\label{decayrateappx}
\varGamma=\sqrt{\frac{B}{\pi\hbar}}\,e^{-B/\hbar}\,A,
\end{align}
where we have inserted $\hbar$ explicitly in view of the comparison with the WKB method in Appendix~\ref{app:sec:wkb}.

\subsection{Integration over the resolvent}
\label{sec:resolvent}

The method for calculating the fluctuation determinants based on the resolvent has been applied to tunneling problems in Refs.~\cite{Baacke:1993jr,Baacke:1993aj,Baacke:1994ix,Baacke:2008zx,Garbrecht:2015yza}.
We consider the following eigenvalue equations
\begin{subequations}
\begin{align}
&G_1^{-1}\psi^{(1)}_n(\tau)\equiv (-\partial_\tau^2+W^{(1)}(\tau))\psi^{(1)}_n(\tau)=\lambda_n^{(1)}\psi^{(1)}_n(\tau),\\
&G_2^{-1}\psi^{(1)}_n(\tau)\equiv (-\partial_\tau^2+W^{(2)}(\tau))\psi^{(2)}_n(\tau)=\lambda_n^{(2)}\psi^{(2)}_n(\tau),
\end{align}
\end{subequations}
and the pertaining ratio
\begin{align}
\label{A23}
Q\equiv\log\frac{\det[-\partial_\tau^2+W^{(1)}(\tau)]}{\det[-\partial_\tau^2+W^{(2)}(\tau)]}=\sum_{n} \log\frac{\lambda^{(1)}_n}{\lambda^{(2)}_n}.
\end{align}

In order to obtain an expression for the fluctuation determinant in terms of the Green's functions, we consider the operator
\begin{align}
G_i^{-1}(s)=G_i^{-1}+s,
\end{align}
where $i=1,2$ and $s\in \mathbb{R}$ is an auxiliary parameter. Its inverse,
satisfying
\begin{align}
\label{eq:resolvent}
[G_i^{-1}(\tau)+s] G_i(s)=\delta(\tau-\tau^\prime),
\end{align}
is called the resolvent and is a generalization of the Green's function
that can be written in the spectral decomposition as
\begin{align}
G_i(\tau,\tau';s)=\sum_n\frac{\overline{\psi_n^{(i)}}(\tau)\,\psi_n^{(i)}(\tau')}{\lambda_n^{(i)}+s}.
\end{align}
Integrating $G_i(\tau,\tau;s)$ over $\tau$, we obtain
\begin{align}
\int\D \tau\; G_i(\tau,\tau;s)=\sum_n\frac{1}{\lambda_n^{(i)}+s}
\end{align} 
by virtue of the orthonormality of the eigenfunctions.

Further, we integrate over $s$ up to some large cutoff $\Lambda$, giving
\begin{align}
\label{eq:integral:s}
\int_0^{\Lambda^2}\D s\int\D\tau\; G_i(\tau,\tau;s)=-\sum_n\log \frac{\lambda_n^{(i)}}{\lambda_n^{(i)}+\Lambda}.
\end{align}
Comparing this with Eq.~\eqref{A23}, we finally get
\begin{align}
\label{eq:logdet:resolvent}
\log\frac{\det[-\partial_\tau^2+W^{(1)}(\tau)]}{\det[-\partial_\tau^2+W^{(2)}(\tau)]}=\lim_{\Lambda\rightarrow\infty}-\int_0^{\Lambda}\D s\int\D\tau\; \Big(G_1(\tau,\tau,s)-G_2(\tau,\tau,s)\Big).
\end{align}

\subsection{Application to the kink background}
\label{app:sec:GY-Green}

In Section~\ref{sec:det:kink}, we have calculated the logarithmic determinant
directly by integration over the spectrum, that is known analytically for
the kink operator. We now compare this explicitly with what one gets
from the Gel'fand-Yaglom as well as the resolvent method. These answers
for the kink soliton directly lead to the determinant of the bounce in the archetypical thin-wall model that corresponds to a kink--antikink pair.
The kink instanton is given by 
\begin{align}
\bar{x}(\tau)=v\tanh\left(\frac{\mu}{\sqrt{2}}\tau\right),
\end{align} 
which gives $B\equiv S_E[\bar{x}]=4\sqrt{2}\mu^3/\lambda$. Here we have set the position of the kink centre to be at $\tau_0=0$ for simplicity.
Note that we still use $B$ here, while being one half of the bounce result, to denote the kink action. 

The kink is different from the bounce solution in the degenerate limit of the double-well model because the kink solution approaches different vacua at $\tau\rightarrow\pm\infty$, whereas the bounce, being a kink--antikink pair, approaches the false vacuum in both limits. 
Therefore, the function $x_1(\tau)$, that appears in the calculation of Appendix~\ref{sec:GY:evaluation} based on
the Gel'fand-Yaglom method, now 
has the following asymptotic behaviour
\begin{align}
\label{B30}
x_1(\tau)={B}^{-1/2}\,\frac{\D \bar{x}(\tau)}{\D\tau}\rightarrow \, \frac{A}{\sqrt{m}}\,e^{-m|\tau|},\ {\rm as}\ \tau\rightarrow\pm\infty,
\end{align}
where $A=2\sqrt{3}\mu$, $m=\sqrt{2}\mu$. The normalization condition~\eqref{app:normalization} then gives us
\begin{align}
x_2(\tau)\rightarrow \pm\frac{A}{\sqrt{m}}e^{m|\tau|}.
\end{align}
These asymptotics introduce a relative minus sign in $\psi_0^{(2)}(\tau)$ when compared to the corresponding result~\eqref{app:psi2} in the background of the bounce
and hence in the formula~\eqref{app:det'}. Finally, we obtain
\begin{align}
\label{eq:app:determinant}
\frac{\det'(-\partial_\tau^2+V''(\bar{x}))}{\det(-\partial_\tau^2+V''(x_+))}=\frac{1}{24\mu^2},
\end{align}
in agreement with what follows from the direct integration over the spectrum
in Eq.~\eqref{Eq533}.

On the other hand, the ratio of functional determinants can be calculated via the Green's function method as in Eq.~\eqref{eq:logdet:resolvent},
\begin{align}
\label{Green-det}
\log\frac{\det^\prime(-\partial_\tau^2+V''(\bar{x}))}{\det(-\partial_\tau^2+V''(x_+))}=-\int_{-\infty}^{\infty}\D\tau\int_0^\infty\D s\ \Big(G^{\prime}(\bar{x};\tau,\tau,s)-G(x_+;\tau,\tau,s)\Big).
\end{align}
Again, the prime on the determinant indicates the omission of the
zero mode. Correspondingly, $G^{\prime}(\bar x;\tau,\tau^\prime,s)$ stands for the resolvent from which the zero mode is subtracted.

We first solve for the resolvent in the kink background. 
Defining $u\equiv \tanh(\mu\tau/\sqrt{2})$, Eq.~\eqref{eq:resolvent}
turns into
\begin{align}
\label{eq:Green:kink}
\left(\frac{\D}{\D u}(1-u^2)\frac{\D}{\D u}-\frac{\varpi^2}{1-u^2}+6\right)G(\bar{x};u,u',s)=-\left(\frac{\sqrt{2}}{\mu}\right)\delta(u-u'),
\end{align}
where $\varpi^2=4+2s/\mu^2$. This
equation has been solved analytically in Refs.~\cite{Garbrecht:2015oea,Garbrecht:2018rqx}. 
Moreover, since the spectrum of the kink is known, as discussed in
Section~\ref{sec:flukes:kink}, the Green's function can be decomposed into contributions from the discrete and continuum spectrum, respectively~\cite{Garbrecht:2018rqx}. The part from the discrete spectrum is
\begin{align}
G_{d}(\bar{x};u,u',s)=\frac{\sqrt{2}}{\mu}\left(-\frac{3}{2}\frac{u u'}{1-\varpi^2}\sqrt{1-u^2}\sqrt{1-u'^2}-\frac{3}{4}\frac{1}{4-\varpi^2}(1-u^2)(1-u'^2)\right),
\end{align}
where the second term is from the time-translational zero mode that needs to be subtracted. The piece from the continuum spectrum is
\begin{align}
&G_{c}(\bar{x};u,u',s)=\frac{\sqrt{2}}{\mu}\left\{\frac{3}{2}\frac{u u'}{1-\varpi^2}\sqrt{1-u^2}\sqrt{1-u'^2}+\frac{3}{4}\frac{1}{4-\varpi^2}(1-u^2)(1-u'^2)\right.\notag\\
&+\left.\left(\frac{1}{2\varpi}\theta(u-u')\left(\frac{1-u}{1+u}\right)^{\frac{\varpi}{2}}\left(\frac{1+u'}{1-u'}\right)^{\frac{\varpi}{2}}\frac{3u^2+3u\varpi+\varpi^2-1}{(1+\varpi)(2+\varpi)}\frac{3u'^2-3u'\varpi+\varpi^2-1}{(1-\varpi)(2-\varpi)}\right.\right.\notag\\
&+(u\leftrightarrow u')\bigg)\bigg\}.
\end{align}
We are thus able to directly subtract the translational zero mode
from the Green's function. In case the spectral decomposition is unknown,
one can alternatively project out the zero-mode contributions from the Green's
functions, as discussed e.g. in Ref.~\cite{Garbrecht:2018rqx}. Further,
the resolvent in the false vacuum is given by
\begin{align}
G(x_+;u,u^\prime,s)=
\frac{\sqrt{2}}{\mu}\left(\frac{1}{2\varpi}\theta(u-u')\left(\frac{1-u}{1+u}\right)^{\frac{\varpi}{2}}\left(\frac{1+u'}{1-u'}\right)^{\frac{\varpi}{2}}+(u\leftrightarrow u')\right).
\end{align}

Taking the coincident limit of the resolvent, with the
zero mode and the false-vacuum part subtracted, we obtain
\begin{align}
G^{\prime}(\bar{x};u,u,s)-G(x_+;u,u,s)\equiv& G_{d}(\bar{x};u,u,s)+G_{c}(\bar{x};u,u,s)
+\frac{\sqrt{2}}{\mu}\left(\frac{3}{4}\frac{1}{4-\varpi^2}(1-u^2)^2\right)\notag\\-&G(x_+;u,u,s)
\notag\\
=&-\frac{\sqrt 2}{\mu}\frac 34\frac{(1-u^2)(1+3 u^2 +2 u^2\varpi-\varpi^2+u^2\varpi^2)}{\varpi(1-\varpi)^2(2+\varpi)}\,.
\end{align}
Doing the integral in Eq.~\eqref{Green-det}, we get
\begin{align}
-\int_{-\infty}^{\infty}\D\tau\int_0^\infty\D s\ \left(G^{\prime}(\bar{x};\tau,\tau,s)
-G(x_+;\tau,\tau,s)\right)=-\log (24\mu^2)+\log\Lambda.
\end{align}
The term $\log\Lambda$ appears because
we have deleted the zero mode, such that one of the logarithms of
$\Lambda$ is not cancelled, as can be seen from Eqs.~\eqref{eq:integral:s}
and~\eqref{eq:logdet:resolvent}, and it is to be discarded.
Therefore, we finally arrive at the same result as in Section~\ref{sec:det:kink}. We thus obtain agreement with Eq.~\eqref{Eq533} from two additional methods.

\subsection{Utility of the different approaches}
\label{app:utility:of:methods}

In many cases, the functional determinant cannot be calculated analytically.
In such situations, the numerical effort requested by implementing the
Gel'fand-Yaglom or the resolvent method appears to be comparable as both approaches amount
to solving ordinary differential equations. The main advantage of the resolvent
method therefore appears to be that the Green's function can readily be
employed in order to compute e.g. corrections to the bounce~\cite{Garbrecht:2015oea,Garbrecht:2015yza,Ai:2018guc} and other
one-loop resummed~\cite{Garbrecht:2018rqx} or higher-order quantities~\cite{Bezuglov:2018qpq,Bezuglov:2019uxg}. Both approaches
avoid the direct solution for the spectrum that has been used in
Section~\ref{sec:det:kink}. While the spectrum may yield interesting
insights into a particular problem, solving for it numerically, in particular to a precision sufficient to compute a renormalized determinant, appears to be substantially more
difficult.

\section{Decay rate from the WKB method}
\label{app:sec:wkb}

In this appendix, we rederive the decay rate~\eqref{decayrateappx} from solving the static Schr\"{o}dinger equation using the WKB expansion. This derivation closely follows the calculation of the ground-state energy in a symmetric double-well potential in Ref.~\cite{Coleman:1988} but modifies
it to be applicable to vacuum decay.

Inside the potential barrier, for $x_{\mathbf p}<x<x_+$ (see Figure~\ref{fig:potential}), we have the following WKB wave function
\begin{align}
\label{A21}
\psi_{\rm WKB}(x)=\frac{c_1}{\sqrt{\kappa(x)}}\,e^{\frac{1}{\hbar}\int_{x_{\mathbf p}}^x \D x'\,\kappa(x')}+\frac{c_2}{\sqrt{\kappa(x)}}\,e^{-\frac{1}{\hbar}\int_{x_{\mathbf p}}^x \D x'\,\kappa(x')},
\end{align}
where $\kappa(x)=\sqrt{2(V(x)-E)}$. We are going to match this wave function with those near the turning points $x_{\mathbf p}$ and $x_+$. Let us first consider the region around $x_+$, where the potential is $V(x)\approx m^2 (x_+-x)^2/2$. We expect the wave function of the ground state to be approximated by the solution to this harmonic-oscillator potential.
For the false-vacuum bound state, we consider the zero-point energy written as $E=\hbar m\,(1/2+\varepsilon)$, where $\varepsilon$ denotes a small correction. We expand next $\kappa(x)$ as
\begin{align}
\kappa(x)=\sqrt{2V(x)}{\left(1-\frac{E}{2V(x)}\right),}
\end{align}
and substitute this into Eq.~\eqref{A21}. Using
\begin{align}
\int_{x_{\mathbf p}}^x \D x'\sqrt{2V(x')}&=\int_{x_{\mathbf p}}^{x_+} \D x'\sqrt{2V(x')}+\int_{x_+}^x \D x'\sqrt{2V(x')}\notag\\
&=\frac{B}{2}-\frac{1}{2} m (x_+-x)^2,
\end{align}
we obtain
\begin{align}
\psi_{\rm WKB}(x)&=\frac{c_1}{\sqrt{m(x_+-x)}}\,e^{\frac{1}{\hbar}\left(\frac{B}{2}-\frac{1}{2}m(x_+-x)^2+E m^{-1}\log(B^{-1/2}{m^{3/2}}A^{-1}(x_+-x))\right)}\notag\\
&+\frac{c_2}{\sqrt{m(x_+-x)}}\,e^{-\frac{1}{\hbar}\left(\frac{B}{2}-\frac{1}{2}m(x_+-x)^2+E m^{-1}\log(B^{-1/2}{ m^{3/2}}A^{-1}(x_+-x))\right)},
\end{align}
where we have used Eq.~\eqref{expressionofA}.
Substituting $E=\hbar m\,(1/2+\varepsilon)$ into the above expression, we finally have
\begin{align}
\label{Eq86}
\psi_{\rm WKB}(x)=&\left(c_1\,e^{B/2\hbar}
B^{-1/4}A^{-1/2}m^{1/4} e^{-\frac{m}{2\hbar}(x_+-x)^2}\right.\notag\\
&\left.+\frac{c_2}{m^{5/4}(x_+-x)}\,e^{-B/2\hbar}B^{1/4}A^{1/2}e^{\frac{m}{2\hbar}(x_+-x)^2}\right)\ \times[1+\mathcal{O}(\varepsilon)].
\end{align}

To fix the coefficients, we need to match $\psi_{\rm WKB}(x)$ to the solutions of the Schr\"{o}dinger equation beyond the turning points. First, we consider the false-vacuum region that is approximately described by the following equation
\begin{align}
\label{seq:harmosc}
-\frac{\hbar^2}{2}\,\partial_x^2\psi(x)+\frac{1}{2}m^2\,(x-x_+)^2\,\psi(x)=E\,\psi(x),
\end{align}
where, for the purpose of matching, we look for approximate solutions valid
for $(x-x_+)^2\gg \hbar/m$. Since $\varepsilon$ is a small number, we can solve this problem perturbatively around $\varepsilon=0$. For $\varepsilon=0$, there are two solutions
\begin{align}
\label{C11}
\psi_1(x)=m^{1/4}e^{-m(x_+-x)^2/2\hbar},
\end{align}
and 
\begin{align}
\label{C12}
\psi_2(x)=\frac{1}{m^{1/4}(x_+-x)}\,e^{m(x_+-x)^2/2\hbar},
\end{align}
where the latter is valid for $(x-x_+)^2\gg \hbar/m$.
The Wronskian for these solutions is
\begin{align}
\psi_1(x)\partial_x\psi_2(x)-\psi_2(x)\partial_x\psi_1(x)=-\frac{2m}{\hbar}+{\cal O}\left(\frac{1}{(x-x_+)^2}\right).
\end{align}
For nonvanishing $\varepsilon$, writing $\psi(x)=\psi_1(x)+\delta\psi(x)$,
the perturbation to the Schr\"odinger equation~\eqref{seq:harmosc} is
\begin{align}
-\frac{\hbar^2}{2}\,\partial_x^2\delta\psi(x)+\frac{1}{2}\,m^2(x-x_+)^2\,\delta\psi(x)=(\hbar m)\varepsilon\,\psi_1(x).
\end{align}
The solution is given by
\begin{align}
\psi(x)=\psi_1(x)-\varepsilon\int_x^{\infty}\D x'\, \psi_1(x')\,[\psi_1(x')\psi_2(x)-\psi_2(x')\psi_1(x)],
\end{align}
where $\psi(x)$ vanishes for $x\rightarrow\infty$. This automatically takes care of vanishing boundary conditions for $(x-x_+)\gg \sqrt{\hbar/m}$. To match at $(x_+-x)\gg\sqrt{\hbar/m}$, we can use the following approximate relation
\begin{align}
\int_x^{\infty}\D x'\,\psi_1^2(x')\approx
\int_{-\infty}^{\infty}\D x'\,\psi_1^2(x')=\sqrt{\pi\hbar}
\end{align}
to obtain 
\begin{align}
\label{Eq93}
\psi(x)=N\left[{ m^{1/4}}e^{-m(x_+-x)^2/2\hbar}\,[1+\mathcal{O}(\varepsilon)]-\frac{\varepsilon\,\sqrt{\pi\hbar}}{m^{1/4}(x_+-x)}\,e^{m(x_+-x)^2/2\hbar}\right],
\end{align}
where we have included a normalization factor. Comparing Eq.~\eqref{Eq93} with Eq.~\eqref{Eq86}, we have
\begin{align}
m\varepsilon=-\frac{c_2}{c_1}\,\sqrt{\frac{B}{\pi\hbar}}\,e^{-B/\hbar}\,A.
\end{align}

The ratio $c_2/c_1$ can be determined by matching around $x_{\mathbf p}$. In this region, $V(x)=V'(x_{\mathbf p})(x-x_{\mathbf p})$, and we neglect
the zero-point energy of the false vacuum, taking $E=0$. Hence, we are dealing with the following Schr\"{o}dinger equation:
\begin{align}
-\frac{\hbar^2}{2}\,\partial_x^2\psi(x)+V'(x_{\mathbf p})(x-x_{\mathbf p})\,\psi(x)=0.
\end{align}
Defining $y=x-x_{\mathbf p}$ and $y=(\hbar^2/(2V'(x_{\mathbf p})))^{1/3}z$, we have 
\begin{align}
\partial_z^2\psi(z)-z\psi(z)=0.
\end{align}
The solutions to this equation are Airy functions with the well-known
asymptotic forms
\begin{align}
&{\rm Ai}(z)\rightarrow\frac{1}{2\sqrt{\pi}}\,z^{-1/4}\,\exp\left(-\frac{2}{3}\,z^{3/2}\right)\ {\rm for}\ z\rightarrow+\infty\,,\\
&{\rm Ai}(z)\rightarrow\frac{1}{\sqrt{\pi}}\,|z|^{-1/4}\,\sin\left(\frac{2}{3}\,|z|^{3/2}+\frac{\pi}{4}\right)\ {\rm for}\ z\rightarrow -\infty,
\end{align}
and 
\begin{align}
&{\rm Bi}(z)\rightarrow\frac{1}{\sqrt{\pi}}\,z^{-1/4}\,\exp\left(\frac{2}{3}\,z^{3/2}\right)\ {\rm for}\ z\rightarrow+\infty\,,\\
&{\rm Bi}(z)\rightarrow\frac{1}{\sqrt{\pi}}\,|z|^{-1/4}\,\cos\left(\frac{2}{3}\,|z|^{3/2}+\frac{\pi}{4}\right)\ {\rm for}\ z\rightarrow -\infty.
\end{align}
This gives the following matching formul\ae:
If for $x>x_{\mathbf p}$, we have 
\begin{align}
\frac{c_1}{\sqrt{\kappa(x)}}\,\exp\left[{\frac{1}{\hbar}}\int_{x_{\mathbf p}}^x \D x'\, \kappa(x')\right]+\frac{c_2}{\sqrt{\kappa(x)}}\,\exp\left[-{\frac{1}{\hbar}}\int_{x_{\mathbf p}}^x \D x'\,\kappa(x')\right],
\end{align}
then the solution for $x<x_{\mathbf p}$ takes the form
\begin{align}
\label{eq:psi:out}
\frac{2\,c_2}{\sqrt{k(x)}}\,\sin\left[{\frac{1}{\hbar}}\int_{x}^{x_{\mathbf p}} \D x'\, k(x')+\frac{\pi}{4}\right]+\frac{c_1}{\sqrt{k(x)}}\,\cos\left[{\frac{1}{\hbar}}\int_x^{x_{\mathbf p}} \D x'\, k(x')+ \frac{\pi}{4}\right],
\end{align}
where $k(x)=\sqrt{-2V(x)}$ and we have used $\kappa(x)\sim \sqrt{z}$ for $z>0$ and $k(x)\sim \sqrt{|z|}$ for $z<0$.

In order to describe tunneling, the wave function beyond $x_{\mathbf p}$
must be of the form of a purely outgoing wave {$\sim\exp \left(-\frac{i}{\hbar}\left[\int_{x}^{x_{\mathbf p}} \D x'\, k(x')+\frac{\pi}{4}\right]\right)$} (note the outgoing wave moves toward the negative $x$-direction). To satisfy this condition for $x<x_{\mathbf p}$, we set $c_1=i\,2c_2$.
This gives us
\begin{align}
m\varepsilon=\frac{i}{2}\,\sqrt{\frac{B}{\pi\hbar}}\,e^{-B/\hbar}\,A,
\end{align}
which is imaginary. Finally, we obtain the decay rate
\begin{align}
\varGamma=\frac{2}{\hbar}\,{\rm Im} E=\,\sqrt{\frac{B}{\pi\hbar}}\,e^{-B/\hbar}\,A,
\end{align}
in agreement with the result~\eqref{decayrateappx} derived from the path integral.

As an alternative to inferring the decay rate from the imaginary part
of the zero-point energy of the false vacuum, we can also obtain
it as the ratio $-j/P$ of the flux $j$ that enters the region around the true
vacuum and the probability $P$ to find the particle around the false vacuum.
The flux into the true-vacuum region is given by
\begin{align}
j=\frac{\hbar}{2i}\left(\psi^*(x)\partial_x \psi(x) - \psi(x)\partial_x \psi^*(x)\right)
=-|c_1|^2,
\end{align}
where for $\psi(x)$, we have substituted Eq.~(\ref{eq:psi:out}) with
$c_2=-(i/2) c_1$. Note that the outgoing flux is negative here because it goes
toward the negative $x$-direction. For $B/\hbar \gg 1$, Eq.~\eqref{Eq86}
is dominated by the first contribution, i.e. the Gau{\ss}ian piece.
We thus obtain for the probability
\begin{align}
P=|c_1|^2 e^{B/\hbar} B^{-1/2} A^{-1} \int \D x\, e^{-\frac{m}{\hbar}(x_+-x)^2}
=|c_1|^2 e^{B/\hbar} B^{-1/2} A^{-1}\sqrt\frac{\pi \hbar}{m},
\end{align}
and once again recover the decay rate as
\begin{align}
\varGamma=-\frac{j}{P}=\sqrt\frac{B}{\pi\hbar}e^{-B/\hbar} A.
\end{align}

\end{document}